\documentclass[fleqn,usenatbib]{mnras}

\usepackage{newtxtext,newtxmath}

\usepackage[T1]{fontenc}

\DeclareRobustCommand{\VAN}[3]{#2}
\let\VANthebibliography\thebibliography
\def\thebibliography{\DeclareRobustCommand{\VAN}[3]{##3}\VANthebibliography}

\usepackage{graphicx}	
\usepackage{amsmath}	
\graphicspath{{Fig/}} 

\title[Tidally induced volcanism]{Can tidal evolution lead to close-in planetary bodies around white dwarfs II: volcanism and transits}

\author[Yuqi Li et al.]{
Yuqi Li,$^{1}$\thanks{E-mail: yl817@cam.ac.uk}
Amy Bonsor,$^{1}$
and Oliver Shorttle$^{1,2}$
\\
$^{1}$Institute of Astronomy, University of Cambridge, Madingley Road, Cambridge, CB3 0HA, UK\\
$^{2}$Department of Earth Sciences, University of Cambridge, Downing Street, Cambridge, CB2 3EQ, UK \\
}

\date{Accepted XXX. Received YYY; in original form ZZZ}

\pubyear{2025}

\begin{document}
\label{firstpage}
\pagerange{\pageref{firstpage}--\pageref{lastpage}}
\maketitle

\begin{abstract}
Planetary material accreted by white dwarfs provides unique insights regarding exoplanetary composition. The evolutionary pathways of planetary bodies around white dwarfs are crucial to understanding the presence of close-in planetary material, observed in the form of pollutants in the atmospheres of white dwarfs and planetary material transiting white dwarfs. Periodic transits around white dwarfs potentially reveal the existence of close-in planetary bodies undergoing dust production. Tidal interactions can bring planetesimals that have been gravitationally perturbed onto long-period highly eccentric orbits around white dwarfs towards shorter orbital periods and smaller eccentricities. Tidal interactions may also induce melting and volcanism in these planetesimals, potentially being a mechanism for dust production, the result of which may be seen in transit. Tidally induced volcanism may be triggered in a wide parameter space: for a 100\,km-sized rocky planetesimals perturbed to a pericentre distance $\lesssim 0.01\,\rm AU$ ($\gtrsim$ twice its Roche limit), both on long-period ($\sim 100\,\rm day$) highly eccentric orbits and short-period ($\sim 10\,\rm hr$) near circular orbits. We comment on the potential link between the resultant volcanic ejecta and observed optical transits.
\end{abstract}




\section{Introduction}

Around 10\%--50\% of white dwarfs are contaminated with planetary material \citep{2003ApJ...596..477Z,2010ApJ...722..725Z,2014A&A...566A..34K,2019MNRAS.487..133W,2023MNRAS.518.3055O,2024MNRAS.527.8687O,2024MNRAS.531L..27M}, revealing the existence of close-in planetary bodies and offering insight into the composition of these bodies.. 

The leading theory of the accretion planetary material onto the white dwarf is tidal disruption of gravitationally perturbed planetesimals. As the host star loses a large fraction of its mass towards the white dwarf phase, the planetary system orbiting the host star becomes less gravitationally stable due to the weakening gravitational attraction from the host star and hence an expanding gravitationally instability zone around planetary bodies \citep{2016RSOS....350571V}. Planetary bodies that are initially stable may begin to alter each other's orbits via mechanisms such as scattering, mean motion resonances and secular chaos \citep{2012MNRAS.420.2990B,2012A&A...548A.104B,2018MNRAS.476.3939M,2022MNRAS.513.4178O,2023MNRAS.518.4537V,2024MNRAS.527.11664}. During (potentially successive) perturbations, some of the planetary bodies are ejected to the outer planetary system and some of them enter the Roche limit, where they are stripped by differential tidal force \citep{2014MNRAS.445.2244V}. After tidal disruption, the planetary debris can be accreted onto the white dwarf under various mechanisms, for instance, Poynting-Robertson (PR) effect, gas drag, and the Yarkovsky effect \citep{2011MNRAS.416L..55R,2011ApJ...732L...3R,2011ApJ...741...36B,2012MNRAS.423..505M,2015MNRAS.451.3453V,2020MNRAS.493.4692V,2021MNRAS.501.3806M,2022MNRAS.510.3379V}.

Accretion can be observed indirectly, through dusty material (infrared excess) and gas (circumstellar emission features) around the white dwarf \citep{1987Natur.330..138Z,2005ApJ...632L.119B,2005ApJ...632L.115K,2006ApJ...646..474K,2007MNRAS.380L..35G,2007ApJ...663.1285J,2009ApJ...694..805F,2009AJ....137.3191J,2010ApJ...714.1386F,2010ApJ...722.1078M,2015ApJ...806L...5X,2016NewAR..71....9F,2017MNRAS.468..154B,2020ApJ...902..127X,2021ApJ...920..156L,2023ApJ...944...23W,2024MNRAS.527.6038R}. 

On the other hand, there is a growing number of systems with planetary materials occultating white dwarfs being detected \citep{2021ApJ...912..125G}, some of which reveal periodicity in the light curves:

\begin{itemize}
    \item the $4.5\rm\,hour$ periodic signals around white dwarf WD\,1145+017 \citep{2015Natur.526..546V},
    \item the 107.2\,day period around ZTF\,J0139+5245 \citep{2020ApJ...897..171V},
    \item the 9.937\,hr and 11.2\,hr periods around ZTF\,J0328-1219  \citep{2021ApJ...917...41V},
    \item the 25.02\,hr period around WD\,1054-226 \citep{2022MNRAS.511.1647F}.
\end{itemize}

The transiting features of the most well-studied system WD\,1145+017 are summarized below \citep{2015Natur.526..546V,2016ApJ...818L...7G,2016MNRAS.458.3904R,2017ApJ...836...82C,2017MNRAS.465.3267G}:

\begin{itemize}
    \item Long-lasting primary periodic signal $\sim4.5\,\rm hr$.
    \item Shorter-lived secondary periodic signals (4.490--4.495\,hr) near the primary.
    \item Variable wavelength-independent transit depths, up to $\sim60\%$.
    \item Asymmetric and noisy transit light curves.
    \item Longer transit duration ($\sim 5\,\rm min$) relative to that of an intact planetesimal-sized body.
\end{itemize}

These features are consistent with a disintegrating body (due to, e.g., sublimation, partial tidal disruption) transiting the white dwarf \citep{2013ApJ...776L...6K,2014A&A...572A..76V,2014ApJ...784...40R,2015ApJ...812..112S,2015Natur.526..546V,2016A&A...596A..32V,2016MNRAS.458.3904R, 2017MNRAS.465.1008V, 2020ApJ...893..166D}. A potential orbital evolution pathway of such a close-in body could be tidal evolution \citep{2019MNRAS.486.3831V,2019MNRAS.489.2941V,2020MNRAS.492.6059V,2020MNRAS.498.4005O}.

The classical view of tidal interaction is that differential tidal force exerted by the primary raises tidal bulges on the secondary (and vice versa). The tidal bulges lag behind the equipotential surface due to internal friction, perturbing the orbital evolution of the system \citep{1981A&A....99..126H,2007A&A...462L...5L,2010A&A...516A..64L,2010ApJ...725.1995M,2011A&A...535A..94B,2011A&A...528A..27H,2012ApJ...751..119B,2012ApJ...757....6H,2022ApJ...931...11G,2022ApJ...931...10R,2023ApJ...948...41L}. 

Tidal evolution is usually accompanied with dissipation of orbital energy as heat, which is referred to as tidal heating. Tidal heating is a plausible cause for the active volcanism on Io in the Solar System \citep{doi:10.1126/science.281.5373.87,2002GeoRL..29.1443R,doi:10.1126/science.1147621,LOPES2015747,2015ApJS..218...22T,2018AJ....156..207R,2022Icar..37314737K,2024NatAs...8...94D,2024ApJ...961...22S}, and is a potential explanation for the formation of hot/inflated Jupiters \citep{2009ApJ...702.1413M,2010ApJ...713..751I,2010A&A...516A..64L,2021ApJ...920L..16D,2022ApJ...931...10R,2022ApJ...931...11G,2023ApJ...943L..13V}. Meanwhile, \citealp{2024ApJ...961...22S} investigates the mechanism where runaway melting of exoplanetary bodies undergoing tidal heating can trigger volcanism. Considering the enormous amount of orbital energy dissipated during the evolution process of a planetesimal perturbed onto highly eccentric orbits and subsequently circularized onto a close-in orbit around white dwarfs, it is reasonable to question whether tidal circularization of planetesimals around white dwarfs is accompanied with active volcanism. 

In our paper I \citep{10.1093/mnras/staf182}, we model the orbital period of planetesimals undergoing tidal evolution around white dwarfs, showing that the resultant orbital period distribution will peak at short-period (nearly) circularized orbits ($\sim10\,\rm hour$--$1\,\rm day$), with a rising tail towards long-period highly eccentric orbits ($\sim 100\,\rm day$) and a probability valley in-between. 

In this paper, we qualitatively investigate a scenario where planetesimals barely perturbed outside the Roche limit undergo rapid tidal evolution, shrinking their orbital semi-major axis and eccentricity. The orbital decay is accompanied by tidal heating that triggers melting and volcanism. Dust ejected from volcanically active planetesimals is initially sufficiently opaque to block the observed star light and leads to transits seen in the optical. Similar to the disk formed under tidal disruption, volcanic dust can be accreted by the white dwarf (e.g., under PR drag). In exploring this model, this paper:: 
\begin{itemize}
    \item predicts the planetesimals and their orbital parameters where tidally induced volcanism could be active;
    \item compares predictions for the features of transits originating from tidally induced volcanism to observations, focusing on the range of secondary drift periods, transit duration, transit depth and phase shift;
    \item compares the frequency of tidally induced volcanism and its contributions to white dwarf pollutants to those of tidal disruption.
\end{itemize}

The paper starts by summarizing the model for tidal heating (Section \ref{tidal heating method}) and the conditions required for volcanism (Section \ref{conditions of volcanism method}). By predicting the ejection speed for volcanic ejecta (Section \ref{volcanism ejection speed method}), the subsequent orbit of volcanically ejected material is computed and used to predict features of the transits (Section \ref{period dispersion method}). Our results highlight that which planetesimals are volcanically active and the epochs during their evolution when this activity occurs \ref{volcanism para results}. The fate of the volcanic ejecta, focusing on the range of orbital periods and orbital inclinations, is presented in Section \ref{period dispersion results}. We then discuss the limitations in our model in Section \ref{model limitation discussion}, including the tidal, thermal modeling (Section \ref{tidal model discussion}, Section \ref{surface sublimation discussion}), Section \ref{planetesimal properties discussion} and Section \ref{planetesimal properties discussion 1}) and the orbital evolution of the ejecta (Section \ref{orbital perturbation discussion} and Section \ref{dispersion eccentric orbits discussion}). In Section \ref{observation discussion}, we compare the model predictions to observations, focusing on WD\,1145+017. Finally, we discuss other potential transit-related scenarios accompanied with tidally induced volcanism in Section \ref{other potential scenarios discussion}, together with other models proposed to explaining transits (Section \ref{other models discussion}) and in Section \ref{conclusions}, we summarize our results.

\section{Method}

In this study, we qualitatively study a scenario in which volcanism is induced by tidal interactions after planetesimals are gravitationally perturbed onto highly eccentric orbits around the white dwarf. These planetesimals narrowly avoid tidal disruption, but experience strong tidal interactions. During tidal evolution, the orbital semi-major axis of the planetesimal decays, with the decaying orbital energy deposited as heat inside the planetesimal. Under intense tidal heating, the planetesimals may melt and undergo volcanism. The erupted volcanic dust blocks the white dwarf and leads to transit.

We present our model that is divided into two parts, 1. tidal evolution, volcanism of the planetesimal and 2. the subsequent orbit dispersion of the ejecta, in this section. For the planetesimal, we: 1. follow the orbital decay of the planetesimal under the constant time lag (CTL) model to obtain the accompanied tidal heating (Section \ref{tidal heating method}), and 2. compare between heating and cooling of the planetesimal undergoing tidal evolution to deduce whether and for what parameters volcanism can be triggered and maintained (Section \ref{conditions of volcanism method}). For the volcanic ejecta, we estimate the ejection speed range (Section \ref{volcanism ejection speed method}), from which we deduce the orbit dispersion of the ejecta relative to that of the planetesimal (Section \ref{period dispersion method}).

The definitions of symbols and the corresponding fiducial values (if any) are listed in Table \ref{parameter table}.

\begin{table*}
\begin{center}
\begin{tabular}{|c|c|c| } 
 \hline
 
Symbols&Definition&fiducial value\\ 
\hline

$M_*$& white dwarf mass &0.6\,$M_{\odot}$  \\

$R_*$& white dwarf radius & 0.01\,$R_{\odot}$\\

$L_*$ & white dwarf luminosity & \\

$t_0$ & cooling age of the white dwarf at the start of tidal evolution & 100\,Myr \\

$M_p$& planetesimal mass\\

$R_p$ & planetesimal radius & $100\,\rm km$\\

$\rho_p$& planetesimal bulk density& $3000\,\rm kg/m^3$\\

$k_2$ & potential love number of degree 2\\

$\Delta t$& constant time lag\\

$K_p$& $K_p= 3k_2\Delta t$& 1000\,s\\

$T_p$ & $T_p=\frac{K_p(M_p+M_*)M_*R_p^5}{M_p}\approx \frac{K_pM_*^2R_p^5}{M_p}$\\

$I$ & moment of inertia\\

$\sigma_{\mathit{+}}$ & ultimate tensile strength & 0.1\,MPa\\

$C_p$ & $C_p= \frac{I}{M_pR_p^2}$& $\frac{2}{5}$\\

\hline

$t$ & tidal evolution time\\

$e$& eccentricity\\

$a$& semi-major axis\\

$q$ & pericentre distance\\

$q_0$ & initial pericentre distance\\

$Q$ & apocentre distance\\

$Q_0$ & initial apocentre distance & 3\,AU \\

$r$ & radial distance from the host star\\

$r_{\mathit{Roche}}$ & Roche limit\\

$\tau$ & orbital period\\

$\omega_p$ & planetesimal spin\\

$\bar{v}_{\mathit{orb}}$ & orbital velocity\\

$v_k$ & local keplerian speed of circular orbit\\

$n$ & mean motion, $n=\sqrt{\frac{G(M_p+M_*)}{a^3}}\approx \sqrt{\frac{GM_*}{a^3}}$\\

$J$ & specific orbital angular momentum\\

$\epsilon $ & specific orbital energy\\

$f$ & true anomaly\\

$E$ & eccentric anomaly\\

$\gamma$ & angle between velocity and radial vector\\

$\omega$ & argument of pericentre\\

\hline
$c_p$ & planetesimal specific heat capacity& 1000\,$\rm J/(K\cdot kg)$\\
$\kappa$ & thermal conductivity & 3\,$\rm W/(m\cdot K)$\\
$\alpha_d$ & thermal diffusivity\\
$f_m$ & radial position of magma reservoir in the unit of $R_p$ & 0.9\\
$\epsilon_e$ & emissivity& 1\\
$A_B$ & albedo & 0\\
$\mathrm{Nu}$ & Nusselt number& 1\\
$E_t$& tidal energy\\
$T$& Temperature\\
$T_s$ & (time-averaged) equilibrium surface temperature\\
$T_c$ & critical temperature of melting & 2000\,K\\
$\Delta T_c$ & temperature representation of melting energy & 3000\,K\\
$Q_l$ & heat loss\\
\hline
$n_0$ & mass fraction of volcanic gas\\
$\mu_m$ & mean molecular weight\\
$\bar{u}_{\mathit{eject}}$ & volcanic ejection velocity \\
$\bar{v}_{\mathit{tot}}$ & total velocity of ejecta\\
$v_{\mathit{esc}}$ & escape speed\\
$\theta$ & polar angle\\
$\phi$ & azimuthal angle\\
\hline
\end{tabular}
\end{center}
\caption{Definitions of the symbols and the choice of fiducial values. $M_*=0.6\,M_{\odot}$ is the peak of the observed white dwarf mass distribution \citep{2007MNRAS.375.1315K,2016MNRAS.461.2100T,2017ASPC..509..421K}. $t_0$ quantifies the time taken to perturb planetesimals onto highly eccentric orbits \citep{2018MNRAS.476.3939M}. $c_p$ and $\kappa$ are identical to those in \citealp{2009E&PSL.284..144R,2021Sci...371..365L}. $T_c$ is chosen to be the silicates liquidus temperature \citep{2015aste.book..533S,2023A&A...671A..74J}. $\sigma_{\mathit{+}}$ is chosen to be the lower bound for meteorites samples \citep{2020M&PS...55..962P}. $K_p$ is converted from tidal dissipation for typical rocky bodies \citep{1977RSPTA.287..545L,1997A&A...318..975N,2011A&A...535A..94B,2014Icar..241...26N,2015A&A...584A..60C,2017CeMDA.129..509B,2020JGRE..12506312R,2020A&A...635A.117B,2020A&A...637A..78S, 2022AdGeo..63..231B}.}
\label{parameter table}
\end{table*}

\subsection{Tidal heating}\label{tidal heating method}

As a planetesimal on a highly eccentric orbit around a white dwarf undergoes tidal evolution, its semi-major axis decays, and the damped orbital energy is deposited inside the body. The tidal energy deposition rate $\frac{dE_t}{dt}$ and semi-major axis decay rate ($\frac{da}{dt}$) are related by:

\begin{equation}\label{tidal dissipation equation}
\frac{dE_t}{dt}=-\frac{GM_pM_*}{2a^2}\frac{da}{dt},
\end{equation}

\noindent where we neglect the rotational energy variations, which are orders of magnitude smaller than their counterparts of orbital energy, and we only consider the contributions from planetesimal tide induced by the white dwarf, which dominates the orbital evolution of the system. The inputs required for computing $\frac{da}{dt}$ are the properties of the white dwarf-planetesimal system (e.g., tidal response $T_p$ in Table \ref{parameter table}) and the initial orbit (initial pericentre $q_0$, initial apocentre $Q_0$).

To obtain the tidal heating rate $\frac{dE_t}{dt}$, the key is to compute the orbital evolution rate of the planetesimal. Under the CTL model, where the planetesimal around the white dwarf reaches pseudo-synchronization and spin-orbit alignment rapidly and that the orbital angular momentum is conserved during tidal evolution, the coupled tidal evolutionary equations can be reduced to those regarding semi-major axis ($\frac{da}{dt}$) and eccentricity ($\frac{de}{dt}$) \citep{1981A&A....99..126H,2007A&A...462L...5L,2010A&A...516A..64L,2010ApJ...725.1995M,2011A&A...535A..94B,2011A&A...528A..27H,2012ApJ...751..119B,2012ApJ...757....6H,2022ApJ...931...11G,2022ApJ...931...10R,2023ApJ...948...41L}:

\begin{equation}\label{tidal a general equation}
\begin{aligned}
\frac{da}{dt}&=-7GT_p\left(\frac{2q_0Q_0}{q_0+Q_0}\right)^{-7} e^2(1-e^2)^{-\frac{1}{2}} F(e).
\end{aligned}
\end{equation}

\begin{equation}\label{tidal e general equation}
\begin{aligned}
\frac{de}{dt}&=-\frac{7}{2}GT_p\left(\frac{2q_0Q_0}{q_0+Q_0}\right)^{-8}e(1-e^2)^{\frac{3}{2}}F(e),
\end{aligned}
\end{equation}

\noindent where $T_p$ is a white dwarf-planetesimal system relevant constant (see Table \ref{parameter table}) modulating the tidal evolution rate, $F(e)$ is:

\begin{equation}
\begin{aligned}
F(e)=\frac{1+\frac{45}{14}e^2+8e^4+\frac{685}{224}e^6+\frac{255}{448}e^8+\frac{25}{1792}e^{10}}{1+3e^2+\frac{3}{8}e^4}.   
\end{aligned}   
\end{equation}

We can further eliminate the terms regarding semi-major axis $a$ using conservation of orbital angular momentum:

\begin{equation}\label{conservation of L}
a(1-e^2)=\frac{2q_0Q_0}{q_0+Q_0},   
\end{equation}

\noindent such that $\frac{dE_t}{dt}$ (Eq.\ref{tidal dissipation equation}) under the CTL model can be reduced to an equation of single variable, eccentricity $e$:

\begin{equation}\label{tidal power reduced}
\frac{dE_t}{dt}=-{GM_pM_*}e\left(\frac{2q_0Q_0}{q_0+Q_0}\right)^{-1}\frac{de}{dt},    
\end{equation}

\noindent whose maximum is at $e\approx 0.735$ whose presence is a natural consequence of the competition between the increase in the term $\frac{1}{a^2}$ and the decrease in the term $\left|\frac{da}{dt}\right|$ during tidal evolution in Eq.\ref{tidal dissipation equation}. 

Eq.\ref{tidal a general equation} and Eq.\ref{tidal power reduced} indicate that a planetesimal starting at a smaller $q_0$ evolves much faster, and hence, is heated up much more rapidly under the CTL model. However, a planetesimal undergoing long term tidal evolution and volcanic activities must avoid tidal disruption, setting the minimum pericentre distance a planetesimal can reach--the Roche limit ($r_{\mathit{Roche}}$). Under tidal interactions, $r_{\mathit{Roche}}$ for a pseudo-synchronously rotating planetesimal that is spherical and rigid in the limit of $r_{\mathit{Roche}}\ll Q_0$ can be approximated as \citep{10.1093/mnras/staf182}:

\begin{equation}\label{roche limit approximated equation}
\begin{aligned}
&r_{\mathit{Roche}}\approx \left(\frac{3.36125GM_*}{\frac{4}{3}\pi G\rho_p+\frac{3\sigma_{\mathit{+}}}{4 \rho_pR_p^2}}\right)^{\frac{1}{3}}.
\end{aligned}
\end{equation}

\subsection{Volcanism}

With the tidal heating rate quantified, in this section, we propose two simple criteria for volcanism (Section \ref{conditions of volcanism method}) and estimate the possible range of volcanic ejection speed (Section \ref{volcanism ejection speed method}).

\subsubsection{Conditions for volcanism} \label{conditions of volcanism method}

Based on the tidal evolution track and the accompanied tidal heating, we investigate whether and under what conditions volcanism can be triggered and maintained. We acknowledge that accurate modeling of tidal heating and volcanism is nontrivial and requires proper coupling among tidal, thermal and rheological evolution (see Section \ref{model limitation discussion}). However, for the purpose of qualitative study, we incorporate two simple conditions for volcanism:

\begin{itemize}
    \item melting happening faster than cooling (initiation),
    \item heating rate being faster than cooling rate (maintenance).
\end{itemize}

We quantitatively investigate the first condition (initiation) by comparing the time taken for melting ($t_{\mathit{melt}}$) to the time taken for cooling ($t_{\mathit{cool}}$) at the start of tidal evolution, $t=0$. The melting timescale, $t_{\mathit{melt}}$, is given by:

\begin{equation}
t_{\mathit{melt}}\sim \frac{c_pM_p\Delta T_c}{\dot{E}_{\mathit{t,0}}},   
\end{equation}

\noindent where we use the fact that for a highly eccentric orbit, $\frac{dE_t}{dt}(t=t_{\mathit{melt}})\approx \frac{dE_t}{dt}(t=0)$ (see Appendix A). The energy required for melting is given by the multiplication of specific heat capacity ($c_p$), planetesimal mass ($M_p$) and an temperature representation of melting energy ($\Delta T_c$), $c_p M_p \Delta T_c$. We choose $\Delta T_c=3000\,\mathrm{K}$ as the fiducial value (see Appendix B). 

The cooling timescale, $t_{\mathit{cool}}$, is approximated by the diffusion timescale scaled by the Nusselt number ($\mathrm{Nu}\geq 1$), the ratio of total heat loss rate (excluding heat loss due to volcanism) to conductive heat loss rate:

\begin{equation}
t_{\mathit{cool}}\sim \frac{R_p^2\rho_pc_p}{\pi^2\kappa \mathrm{Nu}},  
\end{equation}

\noindent with $\kappa$ the thermal conductivity. In this study, we focus on the limiting case where heat loss only occurs via conduction as a planetesimal undergoing tidally induced volcanism should usually preserve a solid mantle (see Section \ref{surface sublimation discussion}) where conduction is the dominant heat transport mechanism ($\mathrm{Nu}=1$; an estimation of $\mathrm{Nu}$ for convection through a solid-like crust with melt is presented in Appendix C, where we show that $\mathrm{Nu}\lesssim 10$ usually holds).

The ratio of melting timescale to cooling timescale, $\frac{t_{\mathit{melt}}}{t_{\mathit{cool}}}$, can be approximated as (see Appendix A):

\begin{equation}\label{scaling relation ratio}
\begin{aligned}
\frac{t_{\mathit{melt}}}{t_{\mathit{cool}}}&\sim0.5\mathrm{Nu}\frac{1000\,\mathrm{s}}{K_p}\frac{\kappa}{3\,\mathrm{W/(m\cdot K)}}\frac{\Delta T_{\mathit{c}}}{3000\,\mathrm{K}}\\&\times \left(\frac{0.6\,M_{\mathit{\odot}}}{M_*}\right)^3\left(\frac{100\,\mathrm{km}}{R_p}\right)^4\left(\frac{q_0}{0.01\,\mathrm{AU}}\right)^{7.5}\left(\frac{Q_0}{3\,\mathrm{AU}}\right)^{1.5}.      
\end{aligned}
\end{equation}

We assume that melting occurs for $\frac{t_{\mathit{melt}}}{t_{\mathit{cool}}}<1$ (melting being faster than cooling), with the critical parameters determined by the boundary $\frac{t_{\mathit{melt}}}{t_{\mathit{cool}}}=1$. Most importantly, due to the rapid decay of tidal heating rate with the increase in the initial pericentre distance $q_0$ (see Section \ref{tidal heating method}), the critical initial peicentre distance above which melting is slower than cooling ($q_{\mathit{0,crit}}$) is:

\begin{equation}\label{q0crit equation}
\begin{aligned}
q_{\mathit{0,crit}}&\sim 0.011\,\mathrm{AU}\\ &\times \left(\frac{1}{\mathrm{Nu}} \frac{K_p}{1000\,\mathrm{s}} \frac{3\,\mathrm{W/(m\cdot K)}}{\kappa} \frac{3
000\,\mathrm{K}}{\Delta T_{\mathit{c}}}\right)^{\frac{2}{15}}\\&\times \left(\frac{M_*}{M_{\odot}}\right)^{\frac{2}{5}}\left(\frac{R_p}{100\,\mathrm{km}}\right)^{\frac{8}{15}}\left(\frac{3\,\mathrm{AU}}{Q_0}\right)^{\frac{1}{5}}.   
\end{aligned} 
\end{equation}

The strong dependence of tidal heating rate on $q_0$ indicates that for a fixed characteristic cooling time, a small decrease in $q_0$ will move from a regime where heating and cooling are comparable to heating dominant regime, making the details of cooling model less important (see Appendix D for alternative models with consistent results).

To quantify the second condition (maintenance), we compare the heat loss rate $\dot{Q}_l$ to the tidal power $\dot{E}_t$. $\dot{Q}_l$ is approximated as heat loss via steady-state conduction scaled by the Nusselt number $\mathrm{\mathrm{Nu'}}$ (see Appendix E):

\begin{equation}\label{steady state cooling equation}
\dot{Q}_l=4\pi \kappa \mathrm{Nu'} \frac{f_m}{1-f_m}R_p (T_{\mathit{c}}-T_s), 
\end{equation}

\noindent where $f_m$ is the position of the melt reservoir in the unit of planetesimal size (above which conduction is the dominant heat transport mechanism). We apply a fiducial value of 0.9 when necessary, while keeping it a free parameter in equations, considering the large uncertainties (see Appendix F). 
We use $\mathrm{Nu'}$ as a reminder that heat loss after melting quantified by $\mathrm{Nu'}$ is not necessarily the same as its counterpart during melting, quantified by $\mathrm{Nu}$ in Eq.\ref{scaling relation ratio}. Similar to the melting process, we consider the limiting case where $\mathrm{Nu'}=1$. $T(x=f_mR_{\mathit{p}})=T_{\mathit{c}}$ is the critical temperature of melting, $T(x=R_p)=T_s$ is the time-averaged surface temperature of the planetesimal given by \citep{2022RNAAS...6...56Q}:

\begin{equation}
\label{time averaged equilibrium temperature equation}
\begin{aligned}
T_s=\left[\frac{L_*(1-A_B)}{16\pi\epsilon_e\sigma a^2}\right]^{\frac{1}{4}}\frac{2\sqrt{1+e}}{\pi}E_2\left(\frac{2e}{1+e}\right),
\end{aligned}
\end{equation}

\noindent where $E_2$ is the complete elliptic integral of the second kind, and $E_2\left(\frac{2e}{1+e}\right)\rightarrow 1$, $T_s\rightarrow \frac{2\sqrt{2}}{\pi}\left[\frac{L_*(1-A_B)}{16\pi\epsilon\sigma a^2}\right]^{\frac{1}{4}}$ as $e\rightarrow 1$.

$T_s$ is a function of white dwarf luminosity, which we will approximate as \citep{1952MNRAS.112..583M}:

\begin{equation}\label{white dwarf luminosity equation}
L_*(t)\approx 3.26L_{\odot}\frac{M_*}{0.6M_{\odot}}\left(0.1+\frac{t_0+t}{\mathrm{Myr}}\right)^{-1.18}, 
\end{equation}

\noindent where $t_{\mathit{0}}$ is the cooling age of the white dwarf at the start of tidal evolution ($t=0$).

Similarly, by setting the ratio of tidal power to heat loss rate $\frac{\dot{E}_t}{\dot{Q}_l}=1$, one can constrain the parameter space where tidal power exceeds heat loss of cooling and hence volcanism can occur. A conservative estimation (due to the initially increasing tidal power along the tidal evolution track) for the critical initial pericentre distance below which melting can be maintained $q_{\mathit{0,crit}}'$ is (see Appendix E):

\begin{equation}\label{scaling relation 1}
\begin{aligned}
q_{\mathit{0,crit}}'&\sim 0.014\,\mathrm{AU} \\&\times  \left(\frac{1}{\mathrm{Nu'}} \frac{K_p}{1000\,\mathrm{s}} \frac{3\,\mathrm{W/(m\cdot K)}}{\kappa} \frac{2000\,\mathrm{K}}{T_{\mathit{c}}}\frac{1-f_m}{f_m}\right)^{\frac{2}{15}}\\&\times \left(\frac{M_*}{0.6\,M_{\odot}}\right)^{\frac{2}{5}}\left(\frac{R_p}{100\,\mathrm{km}}\right)^{\frac{8}{15}}\left(\frac{3\,\mathrm{AU}}{Q_0}\right)^{\frac{1}{5}}\\&\sim 1.2 q_{\mathit{0,crit}}\left(\frac{\mathrm{Nu}}{\mathrm{Nu'}}\frac{1-f_m}{f_m}\frac{\Delta T_c}{T_c}\right)^{\frac{2}{15}},     
\end{aligned}
\end{equation}

\noindent where we use the fact that $\left(T_c-T_{\mathit{s,0}}\right)^{-\frac{2}{15}}\approx T_c^{-\frac{2}{15}}$ usually applies ($T_{\mathit{s,0}}\approx T_s(a=\frac{Q_0}{2},e=1)$). Meanwhile, there is no simple scaling relation when $T_s(t)$ is non-negligible due to its dependence on other free parameters (for instance, $q_0$ and $e$). However, one can obtain the corresponding upper/lower bounds of the critical values (which is a conservative estimation of the parameter space) by using the inequality $T_c>T_c-T_s$. For instance, the upper bound of the critical eccentricity below which volcanism can no longer be maintained is (see Appendix E):

\begin{equation}\label{scaling relation 2}
\begin{aligned}
e_{\mathit{crit}}'&\lesssim 0.001 \\&\times  \left(\mathrm{Nu'} \frac{1000\,\mathrm{s}}{K_p} \frac{\kappa} {3\,\mathrm{W/(m\cdot K)}}\frac{T_{\mathit{c}}}{2000\,\mathrm{K}}\frac{f_m}{1-f_m}\right)^{\frac{1}{2}}\\&\times \left(\frac{M_{\odot}}{M_*}\right)^{\frac{3}{2}}\left(\frac{100\,\mathrm{km}}{R_p}\right)^{2}\left(\frac{q_0}{0.005\,\mathrm{AU}}\right)^{\frac{9}{2}}.   
\end{aligned}
\end{equation}

\subsubsection{Volcanic ejection speed} \label{volcanism ejection speed method}

In the tidally induced volcanism scenario we consider in this study, the volcanic ejecta is considered to cause the transits. To understand whether the transit features originate from tidally induced volcanism, one needs to know the orbit of the ejecta, which is dependent on the ejection velocity. We approximate the volcanic ejection speed ($u_{\mathit{ejection}}$) using the conduit model \citep{WOODS1995189,COLUCCI201498,MACEDONIO2005153} as (see Appendix G):

\begin{equation}
u_{\mathit{eject}}\sim\sqrt{n_0RT}\approx 3000\sqrt{\frac{n_0}{\mu_m}\frac{T}{1000\,\mathrm{K}}}\,\mathrm{m/s},    
\end{equation}

\noindent where $n_0$ is the mass fraction of gas in the flow with specific gas constant $R$ and mean molecular wright $\mu_m$, $T$ is the temperature of flow. When substituting $n_0\sim 0.01$--0.1 \citep{WOODS1995189}, $\mu \sim 10$ (typical rocky planet volcanic gas \citealp{RENGGLI2017296,2023JGRE..12807528L}) and $T\sim 1000\,\rm K$ (critical temperature of melting \citealp{LESHER2015113,2015aste.book..533S,2023A&A...671A..74J}), $u_{\mathit{eject}}$ is of the order of magnitude $100\,\rm m/s$, broadly consistent with the numerical simulations in \citealp{WOODS1995189,MACEDONIO2005153,COLUCCI201498} and the observations \citep{STEINBERG197889,1999Sci...285..870G}. Accounting for the uncertainties in the analytical model and free parameters, we choose $u_{\mathit{ejection}}=10\,\rm m/s$--$1000\,\rm m/s$, with the upper limit matching the observations of Io's volcanism \citep{MCEWEN1983191,LOPES2015747}. 

\subsection{Orbit dispersion of volcanic ejecta }\label{period dispersion method}

Volcanically active planetesimals spew out dust and gas. If sufficient dust is ejected, it will initially be opaque and produce optical transits of the white dwarf. The subsequent orbits of the ejecta after ejection is, thus, crucial to the predicted features of the transits, including dispersion in orbital periods, as well as transit depth and duration. 

In this section, we focus solely on the dust ejected and leaving the planetesimal (ejection speed greater than escape speed of the planetesimal $u_{\mathit{eject}}>v_{\mathit{esc}}$), assuming that initially the dust evolves purely dynamically and dominated by the gravity of the white dwarf, with an initial velocity kick induced as it is released. We note here that this ignores other processes that may be important, including mutual interactions, gravitational perturbations due to the planetesimal, radiation and magnetic forces (see Section \ref{orbital perturbation discussion}).  

In this section, we outline a method to compute the orbital parameters of the ejecta as a function of the orbit of the planetesimal, where the ejecta is launched on the surface of the planetesimal, and the ejection velocity. The assumption is that when dust leaves the planetesimal its future trajectory is only influenced by the gravity of the star, such that its initial conditions determine its trajectory, which deviates from the planetesimal's orbit. We also neglect the back reaction of volcanic ejection on the planetesimal's orbit (see Appendix I).

\subsubsection{Computations of orbital parameters of the ejecta}

\begin{figure}
\includegraphics[width=0.45\textwidth]{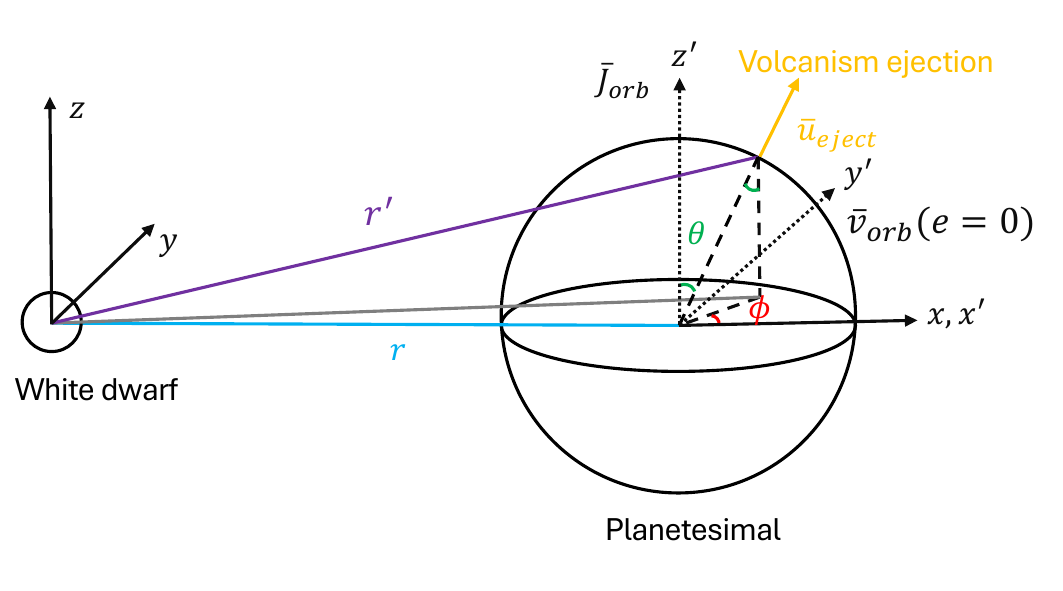}
\caption{The coordinate system used to solve for the final orbit of the ejecta. The sketch is not to scale.}
\label{coordinate setup plot}
\end{figure}

First, based on the variation of specific energy, we compute the variation of orbital semi-major axis/period of the ejecta relative to the planetesimal. In the heliocentric frame of the white dwarf, the specific energy of the planetesimal is given by:

\begin{equation}
\epsilon_p=-\frac{GM_*}{2a}=-\frac{GM_*}{r}+\frac{1}{2}v_{\mathit{orb}}^2,   
\end{equation}

\noindent while the specific energy of the ejecta is:

\begin{equation}\label{epsilon equation}
\epsilon'=-\frac{GM_*}{2a'}=-\frac{GM_*}{r'}-\frac{GM_p}{R_p}+\frac{1}{2}{v_{\mathit{tot}}}^2,  
\end{equation}

\noindent where $v_{\mathit{orb}}=\sqrt{GM_*\left(\frac{2}{r}-\frac{1}{a}\right)}$ is the orbital speed of the planetesimal with an orbital semi-major axis $a$ at distance $r$ from the white dwarf, $r'$ is the distance between the ejecta at eruption to the white dwarf and $v_{\mathit{tot}}$ is the total speed of the ejecta (a combination of orbital velocity, spin velocity and ejection velocity).

The relation between the semi-major axis of the ejecta $a'$ and its counterpart of the planetesimal $a$ is hence:

\begin{equation}\label{a drift general equation}
\begin{aligned}
\frac{1}{a'}&=\frac{1}{a}-2\left(\frac{1}{r}-\frac{1}{r'}\right)+\frac{2M_p}{M_*R_p}-\frac{v_{\mathit{tot}}^2-v_{\mathit{orb}}^2}{GM  _*}\\&=\frac{2}{r'}+\frac{2M_p}{M_*R_p}-\frac{v_{\mathit{tot}}^2}{GM  _*}.        
\end{aligned}
\end{equation}

To solve for $a'$, we define a heliocentric Cartesian coordinate system where (see Fig.\ref{coordinate setup plot}):

\begin{itemize}
    \item the x-axis aligns with the radial vector from the white dwarf to the planetesimal,
    \item the y-axis aligns with the velocity vector the planetesimal would have if it is on a circular orbit,
    \item the z-axis aligns with the orbital angular momentum vector of the planetesimal,
\end{itemize}

We further define a spherical polar coordinate system centred at the planetesimal with respect to a new Cartesian coordinate system $(x',y',z')$, obtained by shifting the origin of the Cartesian coordinate $(x,y,z)$ defined above to $(r,0,0)$ (such that $x'=x-r$, $y'=y$, $z'=z$). We use the polar angle $\theta$ and azimuthal angle $\phi$ to define the position/direction of volcanic ejection where (see Fig.\ref{coordinate setup plot}):

\begin{itemize}
    \item $\theta$ is the polar angle from the $z'$ axis to the ejecta,
    \item $\phi$ is the azimuthal angle from the $x'$ axis to the ejecta's projection onto the $x'$--$y'$ (orbital/equatorial) plane of the planetesimal.
\end{itemize}

Hence, the orbital velocity of the planetesimal is:

\begin{equation}
\begin{aligned}
\bar{v}_{\mathit{orb}}=v_{\mathit{orb}}\begin{bmatrix}\cos{\gamma}\\\sin{\gamma}\\0\end{bmatrix},
\end{aligned}    
\end{equation}

\noindent with $\gamma$ the angle between the radial vector from the white dwarf to the planetesimal and the velocity vector of the planetesimal such that
$\sin\gamma=\sqrt{\frac{a^2(1-e^2)}{r(2a-r)}}$,      
$\cos\gamma=\sqrt{1-\sin^2\gamma}\frac{\sin f}{|\sin f|}$.

The velocity given to the ejected mass due to the spin of the planetesimal is:

\begin{equation}
\begin{aligned}
\bar{v}_{\mathit{spin}}=\omega_{\mathrm{p}}R_p\begin{bmatrix}-\sin{\theta}\sin{\phi}\\\sin{\theta}\cos{\phi}\\0\end{bmatrix},
\end{aligned}    
\end{equation}

\noindent where $\omega_p$ is the spin of the planetesimal, which is given by the pseudo-synchronous spin rate:

\begin{equation}\label{equilibrium spin zero obliquity}
\begin{aligned}
\omega_{\mathit{p}}=S\frac{1+\frac{15}{2}e^2+\frac{45}{8}e^4+\frac{5}{16}e^6}{\left(1+3e^2+\frac{3}{8}e^4\right)}\sqrt{GM_*}\left(\frac{2q_0Q_0}{q_0+Q_0}\right)^{-\frac{3}{2}},     
\end{aligned}
\end{equation}

\noindent with $S$ quantifying the direction of the planetesimal's spin relative to its orbit, which is 1 for prograde orbits and -1 for retrograde orbits.

The ejection velocity, which is assumed to be parallel to the radial vector from the centre of the planetesimal to the position of ejection on the surface of the planetesimal is:

\begin{equation}
\begin{aligned}
\bar{u}_{\mathit{eject}}=u_{\mathit{eject}}\begin{bmatrix}\sin\theta\cos{\phi}\\\sin{\theta}\sin{\phi}\\\cos{\theta}\end{bmatrix},    
\end{aligned}    
\end{equation}

\noindent where $\phi$ and $\theta$ are the azimuthal angle and polar angle quantifying the position of mass ejection on the surface of the planetesimal (where $\theta=\frac{\pi}{2}$ is the equator of the planetesimal/orbital plane, $\phi=0$ aligns with the x-axis). The positional vector of the ejecta in the white dwarf frame $\bar{r}'$ is: 

\begin{equation}
\begin{aligned}
\bar{r}'= \begin{bmatrix}r+R_p\sin\theta \cos \phi \\ R_p\sin{\theta}\sin{\phi}\\R_p\cos \theta\end{bmatrix}.  
\end{aligned}
\end{equation}

The velocity of the ejecta $\bar{v}_{\mathit{tot}}=\bar{v}_{\mathit{orb}}+ \bar{v}_{\mathit{spin}}+\bar{u}_{\mathit{eject}}$ is:

\begin{equation}
\begin{aligned}
\bar{v}_{\mathit{tot}}&=\begin{bmatrix}v_{\mathit{orb}}\cos\gamma-\omega_pR_p\sin\theta\sin\phi+u_{\mathit{eject}}\sin\theta\cos\phi\\ v_{\mathit{orb}}\sin\gamma+\omega_pR_p\sin\theta\cos\phi+u_{\mathit{eject}}\sin\theta\sin\phi\\u_{\mathit{eject}}\cos \theta\end{bmatrix}, 
\end{aligned}
\end{equation}  

We can now solve for the orbital semi-major axis of the ejecta using Eq.\ref{a drift general equation}.

Second, based on the orbital angular momentum of the ejecta $\bar{J}'\approx \bar{r}'\times \bar{v}_{\mathit{tot}}$, one can solve for the orbital inclination $i'$ of the ejecta:

\begin{equation}\label{inclination general equation}
i'=\arctan\left(\frac{\sqrt{J_x'^2+J_y'^2}}{J_z'}\right)\frac{\cos\theta}{|\cos\theta|}, 
\end{equation}

\noindent where the last term $\frac{\cos\theta}{|\cos\theta|}$ defines the direction of the tilt.

Finally, we approximate the orbital eccentricity vector (whose magnitude is the eccentricity) of the ejecta via: 

\begin{equation}
\bar{e}'=\frac{{v}_{\mathit{tot}}^2\bar{r}'-(\bar{v}_{\mathit{tot}}\cdot \bar{r}')\bar{v}_{\mathit{tot}}}{GM_*} -\frac{\bar{r}'}{r'}.
\end{equation}

\subsubsection{Near circular orbit approximations and extremes}\label{circular orbit method}

We show in our paper I that if the transits around WD\,1145+017 (4.5\,hr period) and ZTF\,J0328-1219 (9.937\,hr and 11.2\,hr period) originate from planetesimals undergoing tidal evolution, the planetesimals are most likely on near circular orbits. Furthermore, we show that tidal evolution may intrinsically lead to a pile-up of planetesimals on short-period near circular orbits, which together with a potentially higher transit probability, making them more likely to be identified by observations. These short-period planetesimals are also ideal targets for periodicity measurements (easier to constrain orbital periods using dedicated high-speed photometry) and dynamical studies (more constrained initial pericentre distance and tidal circularization timescale) \citep{10.1093/mnras/staf182}. Therefore, it is valuable to consider the special case where the transiting volcanically active planetesimal is on a near circular orbit (see Appendix H for generalization to moderately eccentric orbit).

In terms of the observations, we are mainly interested in the orbital period ($\Delta \tau$)/semi-major axis ($\Delta a$) dispersion and orbital plane dispersion (inclination $i'$) of the ejecta relative to the planetesimal, which translate to the range of drift periods/transit duration and transit depth, respectively. 

We note that $\Delta \tau$ and $i'$ are effectively equal to their circular-orbit counterparts regardless of the true anomaly for a near circular orbit ($e\lesssim 0.01$) and it is reasonable to approximate the geometry as a circular orbit (see Appendix H). 

The calculations of $\Delta a$ and $i'$ can be greatly simplified assuming $e\rightarrow 0$ by using the facts that:

\begin{itemize}
    \item $\frac{v_{\mathit{spin}}}{v_{\mathit{orb}}}\ll 1$,
    \item $\frac{u_{\mathit{eject}}}{v_{\mathit{orb}}} \ll 1$
    \item $\frac{R_p}{a}\ll 1$,
    \item $\frac{2M_pa}{M_*R_p}=\frac{v_{\mathit{esc}}^2}{v_{\mathit{orb}}^2} <\frac{u_{\mathit{eject}}^2}{v_{\mathit{orb}}^2}$.
\end{itemize}

As a result, we may limit the calculations to the first-order terms in $\frac{v_{\mathit{spin}}}{v_{\mathit{orb}}}$, $\frac{u_{\mathit{eject}}}{v_{\mathit{orb}}}$ and $\frac{R_p}{a}$. Note that due to the multiplications with the trigonometry functions, the magnitude of the first-order terms can be reduced to that of their second-order counterparts for some $\theta$ and $\phi$. We will neglect these second-order effects in the derivations. 

The first order approximation for $\Delta a=a'-a$ is:

\begin{equation}\label{a drift equation}
\begin{aligned}
\frac{\Delta a}{a} &\approx 2\sin\theta\left[(1+S)\frac{R_p}{a}\cos\phi +\frac{u_{\mathit{eject}}}{v_{\mathit{orb}}}\sin\phi\right].
\end{aligned}    
\end{equation}

The corresponding orbital period dispersion of the ejecta relative to the orbital period of the planetesimal $\Delta \tau$ is:

\begin{equation}
\begin{aligned}
\frac{\Delta \tau}{\tau}\approx \frac{3}{2}\frac{\Delta a}{a}.
\end{aligned}
\end{equation}

Eq.\ref{a drift equation} approaches 0 (but not exactly 0 due to the higher-order contributions we neglect) at $\sin\theta=0$ (poles). The extremes of $\frac{\Delta \tau}{\tau}$ present at $\theta=\frac{\pi}{2}$ (equator) and $\phi=\arctan{\frac{\frac{u_{\mathit{eject}}}{v_{\mathit{orb}}}}{\frac{R_p}{a}+\frac{\omega_pR_p}{v_{\mathit{orb}}}}}=\arctan{\frac{u_{\mathit{eject}}a^{\frac{3}{2}}}{\sqrt{GM_*}(1+S)R_p}}$, with the extremes being:

\begin{equation}\label{extreme equation}
\begin{aligned}
\frac{\Delta \tau_\pm}{\tau}&\approx \pm 3\sqrt{\frac{u_{\mathit{eject}}^2}{v_{\mathit{orb}}^2}+(1+S)^2\frac{R_p^2}{a^2}},
\end{aligned}    
\end{equation}

\noindent which gives the maximum period drift/maximum phase-shift rate of the ejecta relative to the planetesimal.

The first-order approximation for the orbital angular momentum of the ejecta $\bar{J}'$ is:

\begin{equation}
\begin{aligned}
\bar{J'}&\approx -R_pv_{\mathit{orb}}\cos\theta \hat{i}-u_{\mathit{eject}}a\cos\theta\hat{j}\\&+(av_{\mathit{orb}}+\omega_pR_pa\sin\theta\cos\phi+au_{\mathit{eject}}\sin\theta\sin\phi\\&+R_pv_{\mathit{orb}}\sin\theta\cos\phi)\hat{k},       
\end{aligned}
\end{equation}

\noindent from which one can approximate the tilt of the ejecta's orbital plane to the planetesimal's orbital plane, orbital inclination $i'$ as:

\begin{equation}\label{inclination equation}
i'\approx \cos\theta\sqrt{\frac{u_{\mathit{eject}}^2}{v_{\mathit{orb}}^2}+\frac{R_p^2}{a^2}},    
\end{equation}

\noindent which approaches 0 at $\cos\theta=0$ (equator) and whose extremes present at $\cos\theta=\pm1$ (poles), with the extremes being:

\begin{equation}\label{maximum inclination}
i'_{\pm}\approx \pm \sqrt{\frac{u_{\mathit{eject}}^2}{v_{\mathit{orb}}^2}+\frac{R_p^2}{a^2}},
\end{equation}

\noindent which gives the maximum extent of the ejecta perpendicular to the orbital plane of the planetesimal.

Although there seems to be no dependence on planetesimal density in the equations, one should be cautious that the analytical model assumes that the ejecta is instantaneously unbound to the planetesimal/bound to the white dwarf such that the ejecta's orbit is torqued negligibly by the planetesimal (see Section \ref{orbital perturbation discussion} for the viability of this assumption). Therefore, the analytical model is expected to converge to the real case at higher ejection speed for a planetesimal with a higher density. 

\section{Results}\label{results}

\subsection{Is volcanism active?}\label{volcanism para results}

\begin{figure*}
\includegraphics[width=0.8\textwidth]{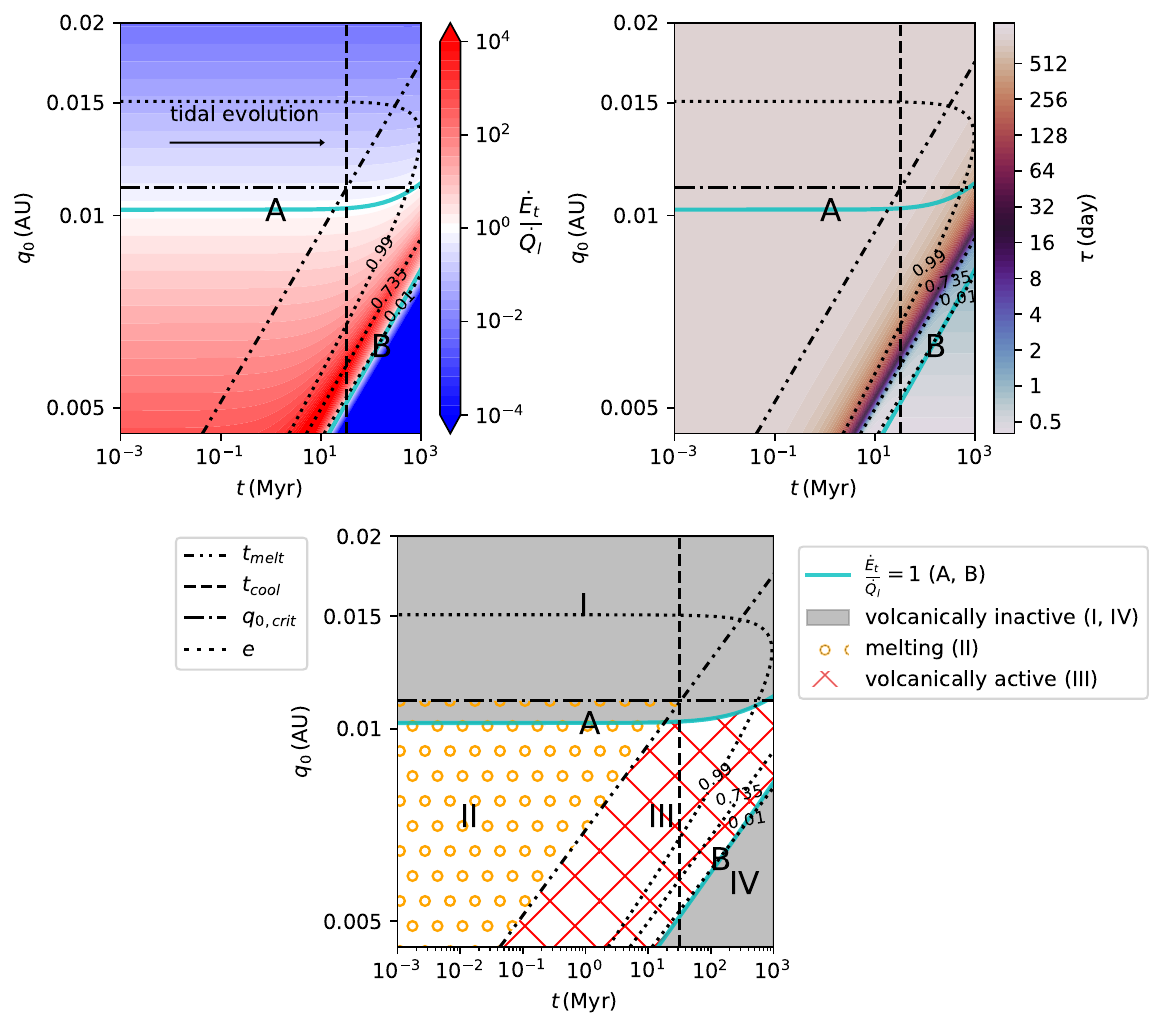}
\caption{The contour plot of the ratio of tidal power to steady state heat loss rate (upper-left panel), orbital period (upper-right panel), and a highlight of which planetesimal ($q_0$) and when ($t$, $e$) are volcanically active (lower panel). The dashdotdotted, dashed and dashdot lines are the melting timescale, cooling timescale and the critical initial pericentre distance of melting. The dotted contour lines are the orbital eccentricities. Other free parameters are identical to those listed in Table \ref{parameter table}.}
\label{volc parameter space plot}
\end{figure*}

\begin{figure}
\includegraphics[width=0.45\textwidth]{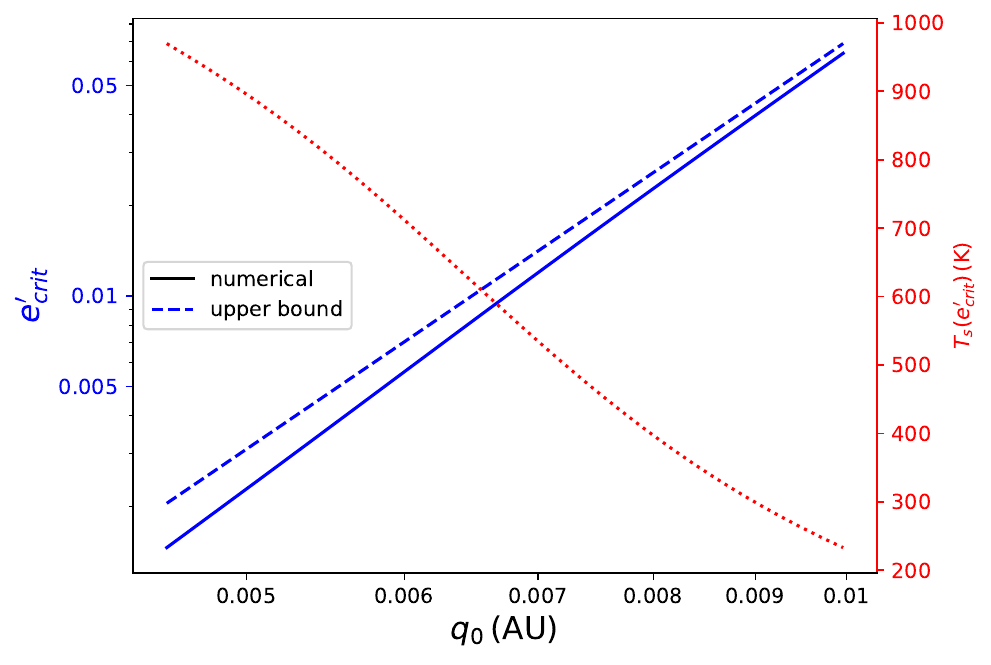}
\caption{The critical eccentricity ($e_{\mathit{crit}}'$) below which volcanism cannot be maintained computed numerically (using Eq.\ref{tidal power reduced} and Eq.\ref{steady state cooling equation}, solid line, left axis) together with the time-averaged equilibrium temperature at $e=e_{\mathit{crit}}'$ (dotted line, right axis), and the analytical upper bound of $e=e_{\mathit{crit}}'$ (Eq.\ref{scaling relation 2}, dashed line, left axis). The free parameters are identical to those in Table \ref{parameter table}.}
\label{ecrit plot}
\end{figure}

This section determines whether or not planetesimals perturbed close to a white dwarf become volcanically active. We consider planetesimals sufficiently close to suffer strong tidal interactions, but insufficiently close to be tidally disrupted. Based on the model presented in Section \ref{conditions of volcanism method}, the initial pericentre, $q_0$ of the planetesimal is the key parameter. 

First, in Fig.\ref{volc parameter space plot} (all 3 panels have identical axes), we follow the tidal evolution tracks of planetesimals starting at a common apocentre distance (parameters listed in Table \ref{parameter table}) and different pericentre distances ranging from 0.00456\,AU (Roche limit) to 0.02\,AU. Note that Fig.\ref{volc parameter space plot} is a snapshot in parameter space where all the free parameters except for $q_0$ (the dominant factor) are fixed. Each sample planetesimal evolves horizontally across the figure (as is indicated by the arrow in the upper-left panel, representing the increase in time elapsed after the planetesimal arrives at $q_0$, $t$), decreasing orbital eccentricity ($e$, the dotted contour lines) and orbital period ($\tau$, contours in the upper-right panel). Meanwhile, orbital energy is dissipated as heat inside the planetesimal, until the planetesimal's orbit circularizes, where tidal evolution ceases. Contours on the upper two panels plot the ratio of tidal power to cooling rate ($\frac{\dot{E}_t}{\dot{Q}_l}$, upper-left panel) and orbital period of the planetesimal ($\tau$, upper-right panel). The lower panels highlight which planetesimal ($q_0$) undergoes active volcanism and when ($t$, $e$). 

The physical meanings of the lines in Fig.\ref{volc parameter space plot} are summarized below:

\begin{itemize}
    \item Dashdotdotted line: melting time ($t_{\mathit{melt}}$), time required for melting from the start of tidal evolution.
    \item Dashed line: cooling time ($t_{\mathit{cool}}$), time required to transport the tidal energy outside the planetesimal.
    \item Dashdot line: the critical initial pericentre distance for melting ($q_{\mathit{0,crit}}$, Eq.\ref{q0crit equation}), where melting time is equal to the cooling time ($t_{\mathit{melt}}=t_{\mathit{cool}}$), above which the planetesimal cannot melt under tidal heating.
    \item Solid (cyan) line A: the contour line where tidal power equal to cooling rate at high $e$ ($\frac{\dot{E}_t}{\dot{Q}_l}=1$, high $e$), above which the melt cannot be maintained due to the rapidly decaying tidal evolution rate arising from large $q_0$. The intersection of A with the y-axis ($t=0$) defines the critical initial pericentre distance for melt maintenance, below which tidal power exceeds cooling rate at $t=0$, the start of tidal evolution ($q_{\mathit{0,crit}}'$, Eq.\ref{scaling relation 1}).
    \item Solid (cyan) line B: the contour line where tidal power equal to cooling rate at low eccentricity ($\frac{\dot{E}_t}{\dot{Q}_l}=1$, low $e$), below which the melt cannot be maintained due to the diminishing tidal heating rate when the planetesimal approaches tidal circularization state. An optimistic estimation of the orbital eccentricity along B, $e_{\mathit{crit}}'$ is given by Eq.\ref{scaling relation 2}.
\end{itemize}

The potential volcanically active regime is region III (cross-hatched) in the lower panel of Fig.\ref{volc parameter space plot}, spanning from long-period highly eccentric orbit to short-period near circular orbit. The physical processes in regions I--IV in the lower panel of Fig.\ref{volc parameter space plot} are summarized below:

\begin{itemize}
    \item I: volcanically inactive ($\tau \gtrsim 1\,\rm yr$, $e\sim 1$), above A ($\sim q_{\mathit{0,crit}}'$) or $q_{\mathit{0,crit}}$, whichever is lower, volcanism is inactive as the planetesimal is not close enough to the white dwarf for strong tidal interactions to induce/maintain melt.
    \item II: melting ($\tau \gtrsim 1\,\rm yr$, $e\sim 1$), bounded by $q_{0,\mathit{crit}}\sim 0.011\,\rm AU$ and $t_{\mathit{melt}}$, the planetesimal potentially undergoes melting.
    \item III: volcanically active ($\tau\sim 10\,\rm hour$--$1\,\rm yr$, $e\sim 0.001$--1), bounded by A or $q_{\mathit{0,crit}}$, whichever is lower, $t_{\mathit{melt}}$ and B, the planetesimal is potentially melted and volcanism is active.
    \item IV: volcanically inactive ($\tau\sim 10\,\rm hour$--$1\,\rm day$, $e\sim 0$--0.01), below B, volcanism is inactive due to the slowing down tidal heating being insufficient to keep the melt from crystallization as the planetesimal approaches tidal circularization.
\end{itemize}

In the remainder of this section, we take a detailed look at the conditions required for volcanism including the parameter dependence.

\subsubsection{Conditions for melting}\label{conditions for melting results}

First, we investigate the conditions required for melting: melting being faster than cooling ($t_{\mathit{melt}}<t_{\mathit{cool}}$). In Fig.\ref{volc parameter space plot}, melting is considered possible if a horizontal line representing the time evolution of a planetesimal intersects the solid line $t_{\mathit{melt}}$ before it intersects the dashed line $t_{\mathit{cool}}$. Due to the rapidly decreasing tidal evolution rate with $q_0$, $t_{\mathit{melt}}$ increases rapidly with $q_0$ ($t_{\mathit{melt}}\propto q_0^{7.5}$), such that a planetesimal starting at a larger $q_0$ melts much slower. The maximum $q_0$ where melting is possible is given by the $q_0$ where $t_{\mathit{melt}}=t_{\mathit{cool}}$, which is represented by the dashdot line. The numerically computed $q_{\mathit{0,crit}}\approx 0.011\,\rm AU$ is in agreement with its analytical counterpart Eq.\ref{q0crit equation}. Based on Eq.\ref{q0crit equation} and setting $q_{\mathit{0,crit}}$ equal to the Roche limit ($r_{\mathit{Roche}}=0.00456\,\rm AU$) where the planetesimal undergoes the most intense tidal heating, we can deduce the conditions that rule out tidally induced volcanism:

\begin{equation}
\begin{aligned}
&\mathrm{Nu}\frac{1000\,\mathrm{s}}{K_p}\frac{\kappa}{3\,\mathrm{W/(m\cdot K)}}\frac{\Delta T_{\mathit{c}}}{3000\,\mathrm{K}}\\&\times \left(\frac{0.6\,M_{\mathit{\odot}}}{M_*}\right)^3\left(\frac{100\,\mathrm{km}}{R_p}\right)^4\left(\frac{Q_0}{3\,\mathrm{AU}}\right)^{1.5}\gtrsim 700,
\end{aligned}    
\end{equation}

\noindent which, when varying single parameter, translates to:

\begin{itemize}
    \item $K_p\lesssim 1\,\rm s$,
    \item $R_p\lesssim 20\,\rm km$,
    \item $Q_0\gtrsim 200\,\rm AU$,
\end{itemize}

\noindent while other conditions (e.g., $M_*\lesssim 0.07\, M_{\odot}$, $\kappa\gtrsim 2000 \,\rm W/(m\cdot K)$, $\Delta T_c \gtrsim 2\times 10^{6} \,\rm K$, $\mathrm{Nu}\gtrsim 700$) are likely non-physical. Note that the obtained size constraint above (and that in the following section) is a tidal heating constraint rather than a geological constraint. We conduct similar analysis for different cooling models in Appendix D and find very similar constraints (except for a much larger $Q_0$ required to rule out tidally induced melting). We conclude that there exists a parameter space for tidally induced melting and it is possible for a planetesimal to form a magma ocean, serving as the reservoir of volcanism.

\subsubsection{Conditions for melt maintenance}

Second, once the planetesimal is melted under tidal heating, we investigate whether the tidal power, despite maintaining this melt, can further supply volcanic eruption. We do this by comparing the tidal power ($\dot{E}_t$) to the cooling rate through a solid crust above the melt ($\dot{Q}_l$), where $\dot{Q}_l>\dot{E}_t$ means that volcanism is suppressed due to insufficient tidal power.

We plot the ratio of the tidal power to the cooling rate $\frac{\dot{E}_t}{\dot{Q}_l}$ in the upper-left panel of Fig.\ref{volc parameter space plot}. For each planetesimal evolving horizontally across the figure, $\frac{\dot{E}_t}{\dot{Q}_l}$ initially increases with time, $t$, reaching its maximum around $e\sim 0.7$ ($\tau\sim 10\,\rm day$, not exactly at $e=0.735$ due to the dependence of $\dot{Q}_l$ on $e$), and then decreases with time $t$, towards 0 as the planetesimal approaches circularization. At identical orbital eccentricity (along the dotted contour lines), a planetesimal with smaller $q_0$ tends to undergo more intense heating. We stress that due to the complexities in tidal-thermal-rheological coupling, $\frac{\dot{E}_t}{\dot{Q}_l}$ in Fig.\ref{volc parameter space plot} only serves as a condition for volcanism for a planetesimal at a given snapshot in time instead of being an accurate prediction for the orbital parameters where volcanism is most active (see Section \ref{tidal model discussion}).
 
The regime where $\frac{\dot{E}_t}{\dot{Q}_l}>1$ is bounded by A (tidal power equal to cooling rate high $e$) and B (tidal power equal to cooling rate low $e$), where the tidal power drops below the steady-state cooling rate. 

Bound A starts horizontal, while shifts towards larger $q_0$ at later times, indicating that planetesimals that started on orbits with relatively large pericentres ($q_0$) and do not fulfill the condition for volcanism may become volcanically active at a later tidal evolution stage due to the initial increasing trend of tidal power with tidal evolution ($e\gtrsim 0.735$). A conservative estimation for A and for the $q_0$ required to maintain melt is $q_0\lesssim q_{\mathit{0,crit}}'$ (Eq.\ref{scaling relation 1}), with $q_{\mathit{0,crit}}'$ around 0.010\,AU under the fiducial parameters. Similarly, using Eq.\ref{scaling relation 1} and placing the planetesimal at its Roche limit ($r_{\mathit{Roche}}=0.00456\,\rm AU$), we deduce that tidally induced volcanism is impossible if:

\begin{equation}
\begin{aligned}
&\mathrm{Nu'}\frac{1000\,\mathrm{s}}{K_p}\frac{\kappa}{3\,\mathrm{W/(m\cdot K)}}\frac{T_{\mathit{c}}}{2000\,\mathrm{K}}\frac{f_m}{1-f_m}\\&\times \left(\frac{0.6\,M_{\mathit{\odot}}}{M_*}\right)^3\left(\frac{100\,\mathrm{km}}{R_p}\right)^4\left(\frac{Q_0}{3\,\mathrm{AU}}\right)^{1.5}\gtrsim 3700,
\end{aligned}    
\end{equation}

\noindent where $\frac{f_m}{1-f_m}=9$ under the fiducial parameter $f_m=0.9$.

Assuming that only one free parameter is varied with others kept at their fiducial values, tidally induced-volcanism is ruled out if:

\begin{itemize}
    \item $K_p\lesssim 1\,\rm s$,
    \item $R_p\lesssim 20\,\rm km$,
    \item $Q_0\gtrsim 150\,\rm AU$,
    \item $f_m\gtrsim 0.9997$,
\end{itemize}

\noindent which is similar to the parameter space where melting is impossible. Therefore, we may conclude that tidally induced volcanism is a possible scenario from an energy point of view.

Meanwhile, on a near circular orbit, a planetesimals with $q_0\lesssim 0.007\,\rm AU$ can remain volcanically active (see the overlap of bound B with the contour line $e=0.01$). With the decrease of $q_0$, volcanism tends to cease at smaller eccentricities $e$, which is shown more explicitly in Fig.\ref{ecrit plot}, where the numerically computed $e_{\mathit{crit}}'$ (solid blue line), circularized below which tidal heating is insufficient to support volcanism, increases from $\sim 0.002$ at $q_0\sim 0.005\,\rm AU$ to $\sim 0.07$ at $q_0\sim 0.01\,\rm AU$. The dashed blue line in Fig.\ref{ecrit plot} is the analytical upper bound of (Eq.\ref{scaling relation 2}), which is generally of the same order of magnitude as the numerical results and whose fractional error decreases with the increase in $q_0$/decrease in $T_s(e=e_{\mathit{crit}}')$.

\subsubsection{Where to expect active volcanism}

Tidally induced volcanism is a potential scenario for planetesimals perturbed close to the white dwarf. Volcanism can occur if the planetesimal initially reaches a small pericentre, $q_0\lesssim 0.01\,\rm AU$ (with the parameter dependence given by Eq.\ref{q0crit equation} and Eq.\ref{scaling relation 1}), unless the planetesimal is small ($R_p\lesssim 20\,\rm km$) and/or possesses a weak tidal response ($\gtrsim 1000$ times weaker than the fiducial value). A planetesimal perturbed to a small pericentre distance may become volcanically active at an early tidal evolution stage on long-period highly eccentric orbits ($\tau\sim 1\,\rm yr$, $e\sim 1$). Meanwhile, tidal heating may remain sufficient to support volcanism up to short-period near circular orbits ($\tau\sim 10\,\rm hr$, $e\sim 0$), with the minimum eccentricity where tidal heating exceeds cooling approximated by Eq.\ref{scaling relation 2}. From an energy perspective, a larger planetesimal with a smaller initial pericentre distance undergoes more intense tidal heating, and may remain volcanically active to smaller orbital eccentricities and orbital periods. 

\subsection{Ejecta from a planetesimal on a near circular orbit}\label{period dispersion results}

\begin{figure*}
\includegraphics[width=0.8\textwidth]{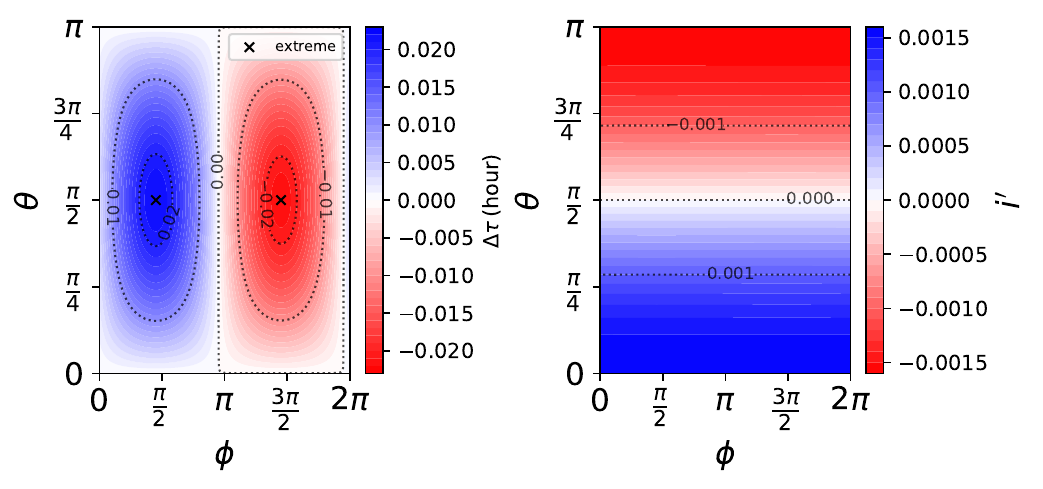}
\caption{The orbital period (left panel) and orbital inclination (right panel) of the volcanic ejecta with ejection speed $500\,\rm m/s$ launched at different positions of the planetesimal surface relative to the 4.5\,hr orbital period of the planetesimal on a (near) circular orbit around a $0.6\,M_{\odot}$ white dwarf. The planetesimal is assumed to have a density of $\rho_p=6000\,\rm kg/m^3$. The crosses in the left panel are the analytically predicted extremes (Section \ref{circular orbit method}).}
\label{period dispersion}
\end{figure*}

\begin{figure*}
\includegraphics[width=0.8\textwidth]{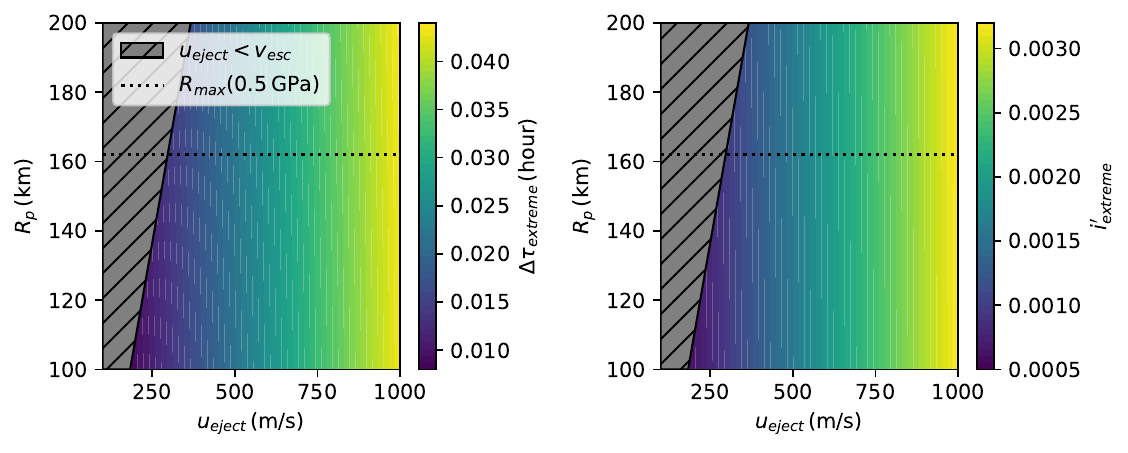}
\caption{The maximum orbital period dispersion ($\Delta \tau_{\mathit{extreme}}$, left panel) and the extent perpendicular to the orbital plane/along the z axis ($\Delta z_{\mathit{extreme}}$ right panel) from the planetesimal on a 4.5\,hr (near) circular orbit around a $0.6\,M_{\odot}$ white dwarf for different ejection speeds and planetesimal sizes. The density of the planetesimal is $\rho_p=6000\,\rm kg/m^3$. The shaded area corresponds to the parameter space where the ejection speed is less than the escape speed. The dotted line is the maximum radius a planetesimal could have without undergoing tidal disruption for a tensile strength of 500\,MPa ($\sim$ upper limit of measured meteorites \citealp{2019P&SS..165..148O,2020M&PS...55..962P}).}
\label{max period dispersion}
\end{figure*}

In this section, we first compute the orbital period dispersion (dispersion in the orbital plane) and inclination (dispersion perpendicular to the orbital plane) of the ejecta based on Eq.\ref{a drift general equation} and Eq.\ref{inclination general equation} for a planetesimal with $R_p=100\,\rm km$ on a 4.5\,hr orbit around WD\,1145+017. A planetesimal circularized to such a short-period orbit without undergoing tidal disruption must have an ultimate tensile strength in the range of iron meteorites ($\gtrsim 100\,\rm MPa$, \citealp{10.1093/mnras/staf182}). Therefore, we assume this planetesimal to be iron-rich and we use $\rho_p=6000\,\rm kg/m^3$ instead of the fiducial value $\rho_p=3000\,\rm kg/m^3$. According to Eq.\ref{scaling relation 2}, this planetesimal may still undergo active volcanism until $e\lesssim 0.001$ even when accounting for the much higher thermal conductivity of iron-rich bodies \citep{OPEIL2010449,https://doi.org/10.1111/maps.13895}. 
In Fig.\ref{period dispersion}, we plot the orbital period dispersion $\Delta \tau$  (left panel) and the orbital inclination $i'$ (right panel) for a fixed ejection speed of 500\,m/s as a function of ejection angle ($\theta$, $\phi$). In Fig.\ref{max period dispersion}, we plot the maximum of $|\Delta \tau|$, $\Delta \tau_{\mathit{extreme}}$, left panel) and the maximum $i'$, $i'_{\mathit{extreme}}$ (right panel) for a range of ejection speed ($u_{\mathit{eject}}=$100--1000\,m/s) and planetesimal size ($R_p=100$--200\,km). The key points are summarized below:

\begin{itemize}
    \item The orbit of the ejecta remains close to that of the planetesimal: the orbital period dispersion remains $\lesssim 0.1\%$ of the orbital period (a phase shift of $\lesssim 0.02\pi$ per orbit relative to the planetesimal) and the orbital inclination of the ejecta is below $\sim 0.003$ (Fig.\ref{max period dispersion}, but note that the maximum z coordinate of the ejecta may reach $\sim 2500\,\mathrm{km}\gg R_p$; see Section \ref{transit depth and duration discussion} for discussion).
    \item The first-order approximations in Section \ref{circular orbit method} are in perfect agreement with the results obtained by solving the complete equations Eq.\ref{a drift general equation} and Eq.\ref{inclination general equation}.
    \item For an ejection speed of 500\,m/s, the extremes of $\Delta \tau$, $\Delta \tau_{\pm}\approx \pm 0.022\,\rm hr$ (phase shift of $\sim 0.01\pi$ per orbit), occurs at $\theta=\frac{\pi}{2}$ (equator), $\phi\approx 0.451\pi$ and $\phi=1.451\pi$ (nearly parallel/anti-parallel to the keplerian orbital velocity), as is indicated by the crosses in the left panel of Fig.\ref{period dispersion}.
    \item $\Delta \tau$ is greatly suppressed as $\theta\rightarrow 0$ and $\pi$ (poles), as is predicted by Eq.\ref{a drift equation} (Fig.\ref{period dispersion}, left panel).
    \item $\Delta \tau$ approaches 0 near $\phi=\pi$ and $2\pi$, where the contributions from the term $\frac{u_{\mathit{eject}}}{\mathit{v_{\mathit{orb}}}}$ is suppressed by the trigonometry function $\cos\phi$ in Eq.\ref{a drift equation}, as $\frac{u_{\mathit{eject}}}{v_{\mathit{orb}}}$ is much larger than $\frac{R_p}{a}$ for the parameters we consider in this section (Fig.\ref{period dispersion}, left panel). For longer-period near circular orbits (decrease in $v_{\mathit{orb}}$ and increase in $a$), $\frac{u_{\mathit{eject}}}{v_{\mathit{orb}}}$ becomes more dominant, making $\Delta \tau$ and $i'$ less sensitive to planetesimal size.
    \item  For an ejection speed of 500\,m/s, it is not rare to reach $|\Delta \tau|\sim 0.01\,\rm hr$, the observed range of drift periods around WD\,1145+017 \citep{2016MNRAS.458.3904R}, as is indicated by the contour lines in the left panel of Fig.\ref{period dispersion}.
    \item The maximum/minimum of $i'\approx \pm 0.0016$ (corresponding to a maximum z coordinate of $\sim 1000\,\rm km$) occurs when the volcanism eruption is at the poles of the planetesimals ($\theta=0,\pi$) and approaches 0 when eruption is at the equator of the planetesimal ($\theta=\frac{\pi}{2}$), as is predicted by Eq.\ref{inclination equation} (Fig.\ref{period dispersion}, right panel).
    \item $i'$ is insensitive to $\phi$ for near circular orbit (Fig.\ref{period dispersion}, right panel).
    \item The maximum of $|i'|$ (ejection at poles) corresponds to the minimum of $|\Delta \tau|$ ($\sim 0$), and vice versa (Fig.\ref{period dispersion}).
    \item The maximum orbital period dispersion and inclination increases with planetesimal size and volcanic ejection speed, with the latter usually being dominant (Fig.\ref{max period dispersion}).
\end{itemize}

According to Fig.\ref{max period dispersion}, although the discrete secondary periods of the order of magnitude $0.01\,\rm hr$ around WD\,1145+017 \citep{2016MNRAS.458.3904R} can be explained by volcanism, volcanism can hardly account for the six orbital periods, A--F (A being the 4.5\,hr period that is dominant/primary), derived from K2 data ranging from 4.5 to 4.9\,hr \citep{2015Natur.526..546V}. Conversely, if the 4.5--4.9\,hr period range originates from orbit dispersion from a single planetesimal, an ejection speed $\gtrsim 10000\,\rm m/s$ is required. Therefore, the existence of orbital periods B, C, D, E and F (if not being false positives) may imply the existence of multiple bodies/more energetic dynamical processes (e.g., collisions, early stage tidal disruption of a large parent body and subsequent tidal circularization of its constituents). With the proof of accuracy for the first-order approximations in Section \ref{circular orbit method}, we can deduce the maximum orbital period dispersion for the 9.937\,hr and 11.2\,hr periods around ZTF\,J0328-1219 with $M_*=0.731\,M_{\odot}$ \citep{2021ApJ...917...41V} assuming (near) circular orbit. We find out that $\Delta \tau \lesssim 0.1\,\rm hr$, which hardly explains the two observed orbital period dispersion $>1\,\rm hr$ using a single planetesimal. Therefore, the two periodic signals around ZTF\,J0328-1219 potentially corresponds to distinct bodies. As eccentricity can promote orbital dispersion (see Section \ref{dispersion eccentric orbits discussion}), we further test the case where $e=0.01$ for WD\,1145+017 and $e=0.2$ for ZTF\,J0328-1219 (based on Fig.11 of \citealp{10.1093/mnras/staf182}). We find out that these eccentricities remain insufficient to explain the 4.5\,hr--4.9\,hr period in K2 data for WD\,1145+017 and the 9.937\,hr and 11.2\,hr periods for ZTF\,J0328-1219, emphasizing the preference for multiple-body scenarios.

\section{Discussion}\label{discussion}

Motivated by the observed periodic transits around a small sub-set of white dwarfs, potentially linked to close-in disintegrating planetary bodies, we explore the fate of the planetary bodies orbiting white dwarfs. We study a scenario where planetesimals perturbed just outside the Roche limit undergo rapid tidal evolution, ending up on various orbits closer to the white dwarf \citep{10.1093/mnras/staf182}. During the orbital decay/circularization, the accompanied tidal heating of the planetesimal potentially triggers volcanism, with the erupted dust being a possible origin of transits. 

In this section, we firstly discuss the limitations of the model regarding the tidally induced volcanism scenario (Section \ref{model limitation discussion}), starting with the tidal and thermal modeling (Section \ref{tidal model discussion}, Section \ref{surface sublimation discussion}), Section \ref{planetesimal properties discussion} and Section \ref{planetesimal properties discussion 1}), followed by additional orbital perturbations to the volcanism ejecta (Section \ref{orbital perturbation discussion}) and tidally induced volcanism on eccentric orbits (Section \ref{dispersion eccentric orbits discussion}). Then, we apply our model to observations in Section \ref{observation discussion}, focusing on WD\,1145+017, via the mass loss rate during volcanism (Section \ref{maximum mass loss discussion}), the resultant transit depth, duration and phase shift of primary transit (Section \ref{transit depth and duration discussion}). Then, we predict the frequency of tidally induced volcanism and its contributions to white dwarf pollutants (Section \ref{frequency of pollutants discussion}). Finally, we briefly discuss other transit-related physical processes accompanied with the tidally induced volcanism scenario (Section \ref{other potential scenarios discussion}) and other existing models explaining transits (Section \ref{other models discussion}).

\subsection{Model limitations} \label{model limitation discussion}

\subsubsection{Tidal, thermal and rheological coupling} \label{tidal model discussion}

In this paper, we only consider the conditions required for volcanism from an energy pointy of view, by comparing tidal heating to cooling at a given snapshot in time. In reality, volcanism and tidal evolution are much more complex. Volcanism of planetesimals may be accompanied with thermal processes such as melting and advection, and hence, alterations of the planetesimal's properties, for instance, rheology, composition, mass and radius. As a result, both the tidal evolution and thermal evolution of the planetesimal are affected.

We utilize the CTL model to simulate the orbital evolution of the planetesimal and the resultant heat deposition. Under the CTL model, the tidal evolution rate is modulated by $T_p=\frac{K_pM_*^2R_p^5}{M_p}\propto \frac{K_pR_p^2}{\rho_p}$ (Table \ref{parameter table}), which is assumed to remain constant along the tidal evolution track. In reality, planetesimal size ($R_p$) and density ($\rho_p$) can both vary under volcanism eruptions. Furthermore, variations in rheology during melting and volcanism can alter the tidal response of the planetesimal ($K_p$). 

On the other hand, the dissipation of planetary bodies under tide (e.g., usually quantified by the tidal quality factor $Q$ or the multiplicity of potential love number and constant time lag $K_p=3k_2\Delta t$ in this study) is poorly constrained and model-dependent. In section \ref{volcanism para results}, we show that unless the tidal evolution is more than 2--3 orders of magnitude slower than the model prediction, there remains a possibility of tidally induced volcanism. In other words, tidally induced volcanism is not rare in terms of the parameter space. However, we are unable to predict the fraction of exoplanetary bodies lying in the volcanically active regime. There may exist a population of planetesimals that disfavour the tidal circularization scenario, due to, for instance, lacking bodies with strong tidal response (e.g., no large-sized bodies, difficulty in raising tidal bulges), gravitational perturbations being too strong such that planetesimals can hardly undergo long-term tidal evolution (e.g., ejection), gravitational perturbation being too weak such that planetesimals are hardly perturbed close enough to the white dwarf for strong tidal interactions. 

Furthermore, the CTL model fixes the rheological dependence of tidal response (proportional to tidal forcing frequency), such that the summed tidal dissipation efficiency of all forcing frequencies can be characterized by a single proportionality constant along the tidal evolution track (e.g., $K_p$ in this study). In reality, tidal dissipation efficiency (and its dependence on the forcing frequency) and the tidal energy distribution inside the body are closely correlated with the rheology of the body, which is altered during melting and the accompanied compositional and structural variations \citep{2005Icar..177..534T,2013Icar..223..308B,2013ApJ...764...27M,2014Icar..241...26N,2014MNRAS.438.1526S,2015E&PSL.427...74Z,2019CeMDA.131...30B,2019MNRAS.486.3831V,2023ApJ...943L..13V,2024ApJ...961...22S}. How tidal interactions are altered during melting/volcanism depends on the applied tidal model and the properties of the planetary body. For instance, under the viscoelastic tidal model, as the thickness of the solid mantle decreases during melting, some planetesimals (given the proper rheology, rigidity, relaxation frequency, etc.) may undergo a runaway process where tidal dissipation efficiency increases before saturation \citep{2024ApJ...961...22S}. Meanwhile, the tidal evolution rate may be further enhanced in the presence of a magma ocean, whose dynamics and rheology differ from that of a solid \citep{2015ApJS..218...22T,2025ApJ...979..133F}. Another important consequence of a rheology dependent tidal response is that a stable pseudo-synchronous rotation state may not persist, adding uncertainties to the Roche limit of the planetesimal \citep{2013ApJ...764...27M,2014MNRAS.438.1526S,2020MNRAS.498.4005O,2020MNRAS.496.3767V}. Moreover, high-eccentricity tidal interactions, which mainly occur near the pericentre, may possess distinct dynamics from their low-eccentricity counterparts (for instance, chaotic and non-chaotic dynamical tides studied for gas giants and stars which potentially lead to more efficient tidal dissipation and hence more rapid tidal heating; see Appendix S for details), thus requiring separate, but self-consistent, modeling \citep{1977ApJ...213..183P,2018ApJ...854...44M,2019MNRAS.489.2941V,2019MNRAS.484.5645V, 2020MNRAS.492.6059V,2022ApJ...931...11G,2022ApJ...931...10R}. 

Despite uncertainties in tidal evolution that may prevent or promote tidally induced volcanism from an energy prospect, long-term volcanic activities may also be disfavoured for certain planetesimals even if sufficient tidal power is provided. For instance, there may exist planetesimals: 1. that undergo thermal disruption (e.g., due to thermal stress), 2. that maintain the equilibrium between heating and cooling without undergoing volcanism (for instance, via more efficient cooling, or suppressing tidal evolution in response to more intense tidal heating), 3. that lack volcanic gas/consume the volcanic gas rapidly to provide excess pressure for eruption and 4. that undergo runaway volcanism, losing a considerable amount of mass, suppressing tidal evolution at an early tidal evolution stage. On the other hand, the conditions for tidally induced volcanism may be much easier to achieve than what we assume, without the need of a shallow magma ocean \citep{park2025io}.

Finally, the orbital parameters where a planetesimal is volcanically active (e.g., the boundaries between melting and volcanism/II and III in Fig.\ref{volc parameter space plot}) are potentially sensitive to the thermal modeling (see Appendix D), including the heat transport mechanisms and tidal energy distribution. For instance, efficient cooling/tidal energy partially damped near the surface and lost rapidly  may slow down the propagation of melt, shifting the volcanically active stage of the planetesimal to a later tidal evolution stage. On the other hand, if tidal energy is preferentially distributed at the position where the shallow magma ocean for volcanism is formed, or a shallow magma ocean is not the necessary condition for volcanism \citep{park2025io}, the volcanically active stage may be brought to an earlier tidal evolution stage.

To properly model the tidally induced volcanism, a model coupling both solid and fluid tidal interactions, thermal and rheological evolution of the planetesimal is required.

\subsubsection{Surface melting/sublimation}\label{surface sublimation discussion}

The deposition and transport of tidal energy results in additional heat flux through the surface, hence increasing the surface temperature of the planetesimal. Furthermore, stellar irradiation may lead to high surface temperature, especially for the day side of tidally locked planetesimals. We find that tidal heating and/or stellar irradiation rarely lead to melting/sublimation of the planetesimal's surface, unless the planetesimal is perturbed sufficiently close-to/within the typical Roche limit of a rocky body ($\sim 0.004\,\rm AU$) and circularize rapidly at an early white dwarf cooling age (see Appendix L). Hence we do not consider planetesimals that undergoes surface melting/sublimation as a common representative of the planetesimal population. Furthermore, volcanism may be greatly suppressed for planetesimals whose surfaces undergo melting/sublimation under intense tidal heating, which requires concentration of tidal energy at the surface and is associated with much more rapid cooling (cooling through sublimation and/or enhanced thermal emission). Therefore, the planetesimal undergoing tidally induced volcanism should usually preserve a solid mantle where cooling through conduction is dominant.

\subsubsection{Thermal properties of planetesimals}\label{planetesimal properties discussion}

The thermal properties of planetesimals is crucial in determining when they melt and how fast they cool. In this work, we parameterize the energy required for melting in terms of the a constant specific heat capacity, $c_p$ and $\Delta T_c$, a temperature representation of melting energy computed using the properties of forsterite (see Appendix B). Meanwhile, the cooling process in the presence of a shallow melt reservoir is parameterized using a constant thermal conductivity $\kappa$. For a range of reasonable values for these parameters (different mineralogy), and accounting for the corresponding temperature dependence, we find that our chosen fiducial values (Table \ref{parameter table}) leads to a conservative estimation for volcanism to occur. Accounting for the uncertainties of the thermal properties, the critical initial pericentre distance of melting $q_{\mathit{0,crit}}$ (Eq.\ref{q0crit equation}) may maximally increase by $\approx 20\%$, and the critical initial pericentre distance for maintaining the melt at high eccentricity $q_{\mathit{0,crit}}'$ (Eq.\ref{scaling relation 1}) may maximally increase by $\approx 10\%$ (see Appendix K). 

On the other hand, additional heat loss channel quantified by $\mathrm{Nu}$, for instance, convection, may suppress the parameter space of volcanism. In the limiting case where $\mathrm{Nu}=10$ for convection in melt-bearing solid layer (see Appendix C), $q_{\mathit{0,crit}}$ and $q_{\mathit{0,crit}}'$ may maximally reduce by $\sim 25\%$, which is comparable to the potential maximum increase above.

Therefore, we do not expect that the thermal properties of the planetesimal and the potential convective cooling can suppress the parameter space of tidally induced volcanism significantly.

\subsubsection{The Roche limit of planetesimals}\label{planetesimal properties discussion 1}

This work assumes that planetesimals are perturbed from an outer planetary belt onto highly eccentric orbits around the white dwarf. If they are perturbed too close at the pericentres, they are torn apart by the differential tidal forces. This work focuses on planetesimals that avoid being torn apart, such that their orbits can evolve under tide. The Roche limit (Eq.\ref{roche limit approximated equation}) provides a useful representation for how close a planetesimal can come to the white dwarf without being tidally disrupted. 

The Roche limit is a function of planetesimal properties such as tensile strength, shape and density. We assume that the planetesimal is spherical and rigid, where it is strongest against tidal disruption. In this case, a strength-less planetesimal with the density similar to that of Solar System ordinary chondrites can circularize to a $\lesssim 10\,\rm hr$ orbit without undergoing tidal disruption, while a planetesimal with $R_p\sim 100\,\rm km$ must be of the tensile strength to Solar System iron meteorites to avoid tidal disruption if it is tidally circularized to a 4.5\,hr orbital period \citep{10.1093/mnras/staf182}. However, as is the case for Solar System asteroids, planetesimals are better described by ellipsoid rather than spheres \citep{2018AJ....156..139M}, potentially making them weaker to tidal disruption \citep{1999Icar..142..525D,10.1093/mnras/staf182}. On the other hand, the tensile strength of planetesimals may weaken with an increase in planetesimal size (scale effect, \citealp{2021AcAau.189..465A}), and the tensile strength may also alter significantly during thermal evolution \citep{2020M&PS...55..962P}. All these effects, together with the spin evolution of the planetesimal under tidal interactions (Section \ref{tidal model discussion}), add uncertainties to the estimation of the Roche limit of a planetesimal, with the deformation and scale effect tending to increase the Roche limit (hence reducing the allowed initial pericentre distance space for tidally induced volcanism) and the effects of the thermal evolution, spin evolution being uncertain.

\subsubsection{Orbital perturbations} \label{orbital perturbation discussion}

In Section \ref{period dispersion method}, we compute the orbital period dispersion of the ejecta analytically, neglecting additional effects such as the gravitational effect of the planetesimal, mutual interactions of the ejecta, general relativistic effect and radiation forces. In this section, we further investigates some of these effects. 

First, we consider the gravitational effect of the planetesimal. The simulations using Rebound (including GR effect, \citealp{2012A&A...537A.128R}) give qualitatively similar results as the analytical expression (fractional error $\lesssim 5\%$ for $u_{\mathit{eject}}=500\,\rm m/s$, with the error decreases with the increase in $u_{\mathit{eject}}$. The exceptions (large fractional error) concentrate in a narrow parameter space of ejection angle ($\theta$, $\phi$) where $\Delta \tau\rightarrow 0$ in Fig.\ref{period dispersion}. Within this narrow parameter space, the ejecta remains close to the planetesimal, continuously being perturbed and with a higher risk of collisions (see Appendix M). Accounting for the very limited parameter space where the analytical model loses accuracy and the consistency in the parameter dependence between the analytical and numerical results, it is generally reasonable to apply the analytical model even when accounting for the planetesimal.

Second, we consider the effect of stellar irradiation on the orbital evolution of the ejecta, with the radiation pressure and Poynting Robertson (PR) drag being the dominant perturbations for dust. 

The radiation pressure effectively leads to a reduced mass of the star, $M_*(1-\beta)$, as seen from the ejecta, with $\beta$ the fraction of radiation to gravitational force \citep{1950ApJ...111..134W}. As a result, the ejecta moves at sub-keplerian speed and has a slightly longer orbital period and lower GR precession rate than its counterpart with keplerian speed at the same orbit (see Appendix M).

On the other hand, the PR drag leads to the loss of angular momentum of the ejecta orbiting the white dwarf. As a result, the ejecta drifts inwards, altering the parameter space where collisions between the ejecta and the planetesimal occur (see Appendix M). The orbital-averaged semi-major axis and eccentricity evolutionary equations under PR drag are \citep{1950ApJ...111..134W,2015MNRAS.451.3453V,2022MNRAS.510.3379V}:

\begin{equation}\label{PR a equation}
\frac{da}{dt}=-\frac{3(Q_{\mathit{abs}}+Q_{\mathit{ref}})(2+3e^2)L_*}{16\pi c^2D\rho a(1-e^2)^{\frac{3}{2}}},    
\end{equation}

\begin{equation}\label{PR e equation}
\frac{de}{dt}=-\frac{15(Q_{\mathit{abs}}+Q_{\mathit{ref}})eL_*}{32\pi c^2 D\rho a^2\sqrt{(1-e^2)}},    
\end{equation}

\noindent where $Q_{\mathit{abs}}$ is the absorption coefficient, $Q_{\mathit{ref}}$ is the reflection coefficient (albedo), $D$ and $\rho$ are radius and density of the particle. We assume that $Q_{\mathit{abs}}+Q_{\mathit{ref}}=1$. 

The PR accretion timescale (timescale of accreting onto the white dwarf) of a particle on a near circular orbit is:

\begin{equation}
\begin{aligned}
t_{\mathit{PR}}&\sim 8\,\mathrm{yr}\\&\times \left(\frac{M_{*}}{0.6\,M_{\odot}}\right)^{\frac{2}{3}}\frac{0.01\,L_{\odot}}{L_*}\frac{D}{1\,\mathrm{\mu m}}\frac{\rho}{3000\,\mathrm{kg/m^3}}\left(\frac{\tau}{4.5\,\mathrm{hr}}\right)^{\frac{4}{3}}\frac{\Delta \tau}{\tau}.    
\end{aligned}
\end{equation}

The long-lived drift periods around WD\,1145+017 is observed to have an orbital decay timescale over 1000\,yr \citep{2016MNRAS.458.3904R}. Assuming that PR drag contributes to the orbital decay, we find out that at least some of the transiting particles are larger than 100\,$\rm \mu m$, not violating the typical size distribution of volcanism ejecta \citep{BUTWIN201999,atmos11060567}. Meanwhile, some of the dips only last for weeks \citep{2016MNRAS.458.3904R}, potentially due to the removal of small particles under PR drag. For instance, for $\frac{\Delta \tau}{\tau}=0.002$ (from the orbital period of the planetesimal to the shortest observed drift period), $t_{\mathit{PR}}$ is $\sim 10\,\rm day$ for micron-sized particles (hence we do not violate the lower bound for the particle size deduced from wavelength-independent transit light curves, $\sim 0.5\,\rm \mu m$ \citealp{2016A&A...589L...6A}). 

Based on the analytical model, we can broadly predict the dispersion of the dust. However, the long-term evolution of the ejecta involves rich dynamics. To properly model the evolution of volcanism ejecta, and hence the transit light curve, one should account for the orbital perturbations due to the planetesimal and radiation, and additionally, the mutual interactions among the volcanism ejecta (e.g., gas-dust interactions) and the magnetic perturbations that may become relevant in the presence of hot plasmas from volcanism/sublimation. 

\subsubsection{Volcanism ejecta of planetesimals on eccentric orbits}\label{dispersion eccentric orbits discussion}

\begin{table}
\begin{center}
\begin{tabular}{|c|c|c|c|c|c| } 
 \hline
System&Mass\,($M_{\odot}$)& Peiord &Predicted eccentricity\\
\hline
ZTF\,J0139+5245&0.52&107.2\,day&$\sim 0.98$\\
\hline
WD\,1054–226 &0.62& 25\,hour & $\sim 0.4$\\
\hline 
\end{tabular}
\end{center}
\caption{A list of the properties of the observed transiting systems potentially correspond to eccentric orbits \citep{2020ApJ...897..171V,2022MNRAS.511.1647F}, together with the predicted orbital eccentricity \citep{10.1093/mnras/staf182}.}
\label{transit table}
\end{table}

\begin{figure}
\includegraphics[width=0.45\textwidth]{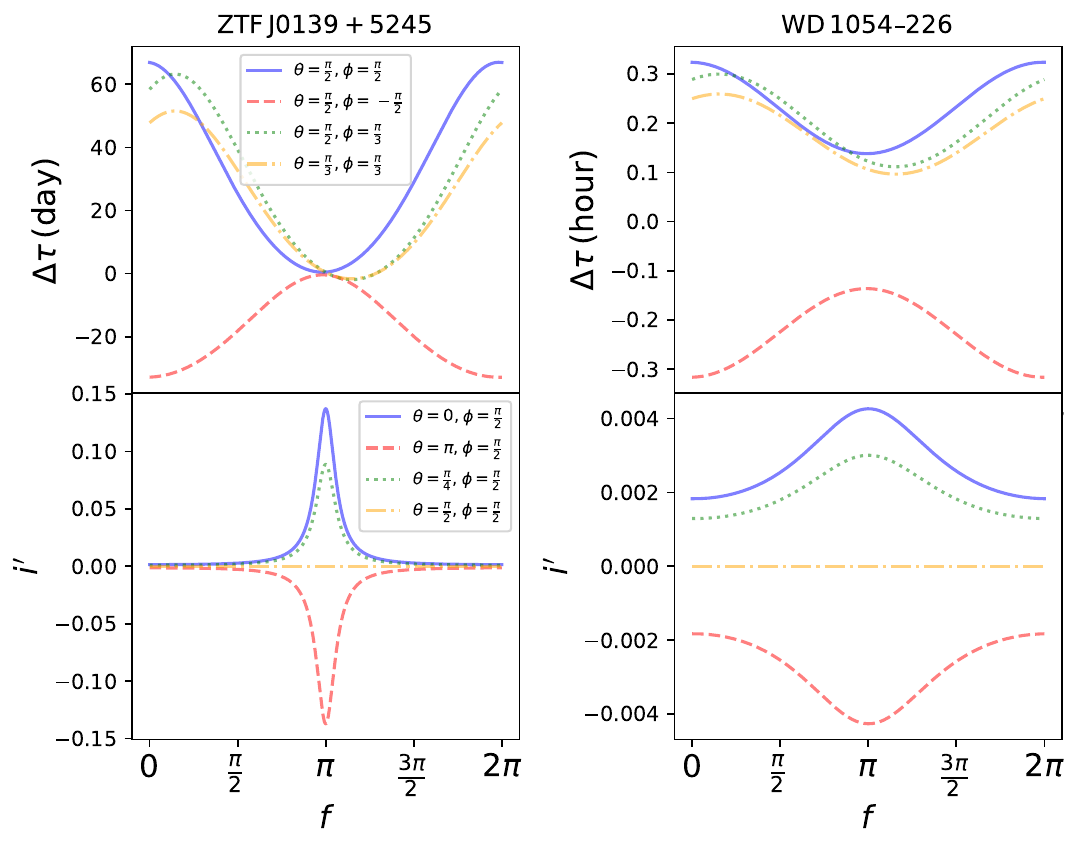}
\caption{The orbit dispersion (upper panel: period, lower panel: inclination) relative to the planetesimal around ZTF\,J0139+5245 (left panel) and WD\,1054-226 (right panel), respectively, as a function of true anomaly for four sets of ejection angles. The ejection speed is assumed to be 500\,m/s with the white dwarf mass and transit periods given in Table \ref{transit table} (other free parameters identical to Table \ref{parameter table}).}
\label{eccentric orbits}
\end{figure}

In Section \ref{period dispersion results}, we mainly focus on volcanism ejection from a planetesimal on a near circular orbit. However, we show that volcanism can equivalently occur on a highly eccentric orbit in Section \ref{volcanism para results}, potentially corresponding to the moderate/long-period transits observed around white dwarfs. In this section, based on the two transiting systems ZTF\,J0139+5245 and WD\,1054-226 (parameters listed in Table \ref{transit table}), we discuss the orbit dispersion of the ejecta from a planetesimal on a eccentric orbit (see Appendix H for analytical approximations for moderately eccentric orbit).

In Fig.\ref{eccentric orbits}, we plot the predicted orbital period dispersion $\Delta \tau$ (upper panel) and inclination $i'$ (lower panel) (computed using the analytical model, which generally agrees with N body simulations) for ZTF\,J0139+5245 (left panel) and WD\,1054-226 (right panel) for ejection at different true anomalies $f$ with a fixed ejection speed of $500\,\rm m/s$. The key points are summarized below:

\begin{itemize}
    \item The orbit dispersion is greatly enhanced by orbital eccentricity, up to $\sim 60\%$ in $\tau$ and up to 0.15 for $i'$ for ZTF\,J0139+5245.
    \item The orbit dispersion is sensitive to the true anomaly for highly eccentric orbits, with the period dispersion more enhanced near the pericentre and inclination more enhanced near the apocentre.
    \item With the decrease in orbital eccentricity, the orbit dispersion gradually converges to the that of the planetesimal. This is the case for WD\,1054-226, with the orbital period dispersion declines to $\lesssim 1\%$ of the orbital period of the planetesimal and inclination decays to $\lesssim 0.004$.
\end{itemize}

For highly eccentric orbits, the orbit of the volcanism ejecta may seem irrelevant to that of the parent body due to the much larger orbit dispersion compared to near circular orbits. Hence, for long-period dust cloud-like transits, unless multiple transits with identical periods are verified, it is hard to tell whether the observed periodic signal is an indication of the orbital period of a planetary body. Although the theoretical maximum of the transit depth is much higher for highly eccentric orbits (see Appendix O), one must be cautious whether the much more dispersed ejecta remains optically thick.

\subsection{Does volcanism lead to transit and pollutants?} \label{observation discussion}

In the previous sections we illustrate that tidally induced volcanism is possible from an energy point of view and can occur in planetesimals with a wide variety of orbits (Section \ref{volcanism para results}). If we directly link the observed periods for the transiting material to volcanically active planetesimals, there is broad agreement  \citep{10.1093/mnras/staf182}.    

In this work, we speculate that dust ejected from volcanically active planetesimals is initially sufficiently opaque to block light from the star. In this section, we discuss the features of the light curves that would result from volcanically active planetesimals orbiting white dwarfs. 

In Section \ref{period dispersion results}, we show that for a planetesimal on (near) circular orbit, volcanism can lead to dust spanning a narrow range of orbital periods and inclinations centered around those of the planetesimal (primary transits correlated with the planetesimal). On the other hand, a large amount of dust launched with small dispersion in ejection velocity may be interpreted as a secondary transit. We show that the predicted orbital period dispersion is able to cover the secondary drift periods around WD\,1145+017 for a ejection speed of $\sim 500\,\rm m/s$ (Section \ref{period dispersion results}).

The remaining questions are: 1. whether volcanism can explain other transiting features (e.g., phase shift of primary transits, variability/dust across the orbit, the observed transit depth and transit duration), and 2. the contributions of volcanism to white dwarf pollutants, which we will discuss in this section.

\subsubsection{Transit features} \label{transit depth and duration discussion}

Due to the distinct orbital parameters the ejecta could have relative to the planetesimal, (a fraction of) the ejecta leave the planetesimal and diffuse away. In this section, we will show that volcanic ejection can broadly explain the transit depth, transit duration, and the observed $180^{\circ}$ phase-shift of the transits for WD\,1145+017 based on the analytical expressions in Section \ref{circular orbit method}.

The maximum transit depth $\frac{\Delta L}{L}$ assuming a rectangular region of dust with infinite width and a height of $2\Delta z$ is given by:

\begin{equation}
\frac{\Delta L}{L}=\frac{2\sin\theta_z\cos\theta_z+2\theta_z}{\pi},    
\end{equation}

\noindent where $\theta_z=\arcsin\left(\frac{\Delta z}{R_*}\right)$. The maximum of $\Delta z$ due to volcanism ejection is a function of maximum orbital inclination, which, for a near circular orbit, is (Eq.\ref{maximum inclination}):

\begin{equation}
\Delta z_{\mathit{max}}=a\sin i'_{\mathit{max}}\approx a\sqrt{\frac{R_p^2}{a^2}+\frac{u_{\mathit{eject}}^2}{v_{\mathit{orb}}^2}}.    
\end{equation}

The orbital period dispersion of the ejecta relative to the planetesimal leads to the formation of leading (shorter-period) and trailing (longer-period) tails, which are phase shifted relative to the planetesimal in the opposite direction, gradually filling the orbit of the planetesimal. As a result, the transit light curve will gradually become noisy over the orbital phases, and the theoretical maximum of the transit duration increases with time. Ultimately, the leading tail will catch up the trailing tail, when the dust spans over the orbit and mutual interactions among the ejecta may become significant (see Appendix O for an illustration of this scenario). The maximum transit duration for volcanism ejecta (before the leading tail catches up the trailing tail) is given by \citep{2005ApJ...627.1011T,2008ApJ...679.1566B}:

\begin{equation}\label{transit duration equation}
\begin{aligned}
t_{\mathit{transit,max}}&\approx \left(\frac{\Delta \tau_{+}}{\tau}-\frac{\Delta \tau_{-}}{\tau}\right)t+\frac{R_*\tau}{\pi a}\\&\approx 6t\sqrt{\frac{u_{\mathit{eject}}^2a}{GM_*}+(1+S)^2\frac{R_p^2}{a^2}}+\frac{R_*\tau}{\pi a},     
\end{aligned}  
\end{equation}

\noindent where $\frac{\Delta \tau_{\pm}}{\tau}$ is given by Eq.\ref{extreme equation} and $t$ is the time elapsed after the ejection. 

On the other hand, the time taken for the leading tail to catch up the trailing tail ($t_{\mathit{catch}}$) can be estimated via:

\begin{equation}
\begin{aligned}
t_{\mathit{catch}}&\approx \tau\left(\frac{\Delta \tau_{+}}{\tau}-\frac{\Delta_{-}}{\tau}\right)^{-1}\\&\approx \frac{\tau}{6\sqrt{\frac{u_{\mathit{eject}}^2a}{GM_*}+(1+S)^2\frac{R_p^2}{a^2}}}.  
\end{aligned}  
\end{equation}

For WD\,1145+017, we have $\left(\frac{\Delta L}{L}\right)_{\mathit{max}}\sim 50\%$ ($\Delta z_{\mathit{max}}\sim 2500\,\rm km$), $t_{\mathit{transit,max}}\sim 0.02t+40\,\rm s$, and $t_{\mathit{catch}}\sim 50\tau\sim 9\,\rm day$ (see Appendix O for numerical simulations).

In comparison to the K2 data, where the transit duration is 4\,min and the transit depth is $8\%$ (undiluted, \citealp{2015Natur.526..546V}), the maximum transit depth of our model is well above the observed value. The minimum time to reach the observed transit duration after volcanism is $\sim 3\,\rm hr$. On the other hand, follow-up ground based observations show much deeper transit depth, up to $50\%$ \citep{2015Natur.526..546V,2016ApJ...818L...7G,2016MNRAS.458.3904R,2018haex.bookE..37V}, reaching the upper bound of our model. According to $t_{\mathit{catch}}$, we predict self-interactions among the volcanism ejecta after $ \sim 9\,\rm day$, potentially leading to observable features out of phase from the planetesimal (starting near $180^{\circ}$, see Appendix O), roughly in agreement with the observed phase shift (appeared $\approx1 \,\rm week$ later, \citealp{2015Natur.526..546V}). If we constrain the volcanism ejection speed based on the time interval for observing transits with $180^{\circ}$ phase shift $u_{\mathit{eject,max}}\gtrsim 1300\,\rm m/s$, we find out $\Delta z_{\mathit{max}}\gtrsim 3000\,\rm km$ and $\left(\frac{\Delta L}{L}\right)_{\mathit{max}}\gtrsim 60\%$. If the maximum ejection speed exceeds $\sim 2500\,\rm m/s$, the dust is able to cover the whole white dwarf ($\Delta z\geq R_*$). 

Tidally induced volcanism can broadly recover the observed transit features around WD\,1145+017, with the observed transit depth close to the theoretical maximum, potentially suggesting an underestimation of the orbit dispersion, either due to an underestimation of volcanic ejection speed or the existence of other dynamical processes (e.g., sublimation, mutual interactions, magnetic forces, production of density waves/vortex \citealp{https://doi.org/10.1029/2021GL092899} in an evolved disk composed of previous volcanic ejecta by new ejections).  
A detailed modeling of the ejecta constrained by the observations of the dust/gas disks (e.g., infrared excess, emission features) is required to better understand its evolution and the corresponding transit features.

\subsubsection{Dust production rate from volcanism}\label{maximum mass loss discussion}

The dust ejected from volcanic activity may not only result in to optical transits, but can also be accreted by the white dwarf. In this section we show that the maximum possible mass loss rate from volcanism, neglecting any geological control, and accretion efficiency, is sufficient to explain the observed accretion rates seen for WD\,1145+017 for a 100\,km-sized planetesimal. Clearly, we do not anticipate that planetesimals lose mass at this maximum rate throughout their lifetime, but instead discuss late in this section the stochastic nature of the process that enables the planetesimal to survive the long-term tidal evolution. 

During volcanism ejection, the thermal energy of the ejecta is lost to the environment (cooling via advection). The tidal heating must be sufficient to sustain this rate of heat loss. The heat loss rate due to ejection $\dot{Q}_{e}$ can be approximated as:

\begin{equation}\label{mass loss equation}
\dot{Q}_{e}=\frac{1}{2}\dot{m}u_{\mathit{eject}}^2+\dot{m}c_p\Delta T_c,  
\end{equation}

The maximum volcanic mass loss rate can be estimated by setting $\dot{Q}_{e}$ equal to the difference between the tidal power ($\dot{E}_t$, Eq.\ref{tidal power reduced}) and the non-volcanic heat loss rate ($\dot{Q}_l$, Eq.\ref{steady state cooling equation}) (the energy required for fracture is usually negligible as is shown in Appendix J):

\begin{equation}
\dot{m}=\frac{\dot{E}_t-\dot{Q}_l}{\frac{1}{2}u_{\mathit{eject}}^2+c_p\Delta T_c}.    
\end{equation}

\noindent where the denominator is usually dominated by the thermal energy term $c_p\Delta T_c$ and for a high dust production rate, the numerator is dominated by $\dot{E}_t$. Hence, we may approximate the maximum mass loss rate via (see Appendix N for numerical results):

\begin{equation}
\begin{aligned}
\dot{m}_{\mathit{max}}&\sim \frac{\dot{E}_t}{c_p\Delta T_c}\\
&\sim 2\times 10^{8}\,\mathrm{g/s}\frac{K_p}{1000\,\mathrm{s}}\left(\frac{M_*}{0.6\,M_{\odot}}\right)^3\left(\frac{R_p}{100\,\mathrm{km}}\right)^5\left(\frac{e}{0.01}\right)^2\\ &\times \left(\frac{0.005\,\mathrm{AU}}{q_0}\right)^9\left(\frac{1000\,\mathrm{J/(K\cdot kg)}}{c_p}\right)\left(\frac{3000\,\mathrm{K}}{\Delta T_c}\right),
\end{aligned}   
\end{equation}

\noindent with the dominant factor being $q_0$ followed by $R_p$. We acknowledge that is nonphysical for $\dot{m}_{\mathit{max}}$ to exceed the ejection-speed limited mass loss rate, $4\pi R_p^2\rho u_{\mathit{eject}}$ (with $\rho$ the vent exit density). For the parameters we consider in this section, $4\pi R_p^2\rho u_{\mathit{eject}}$ is orders of magnitude larger than the range of mass loss rate limited by tidal power, even when taking the minimum ejection speed, $10\,\rm m/s$.

For a planetesimal with a radius of 100\,km on a 4.5\,hr period ($q_0=0.0027\,\rm AU$) around WD\,1145+017 with an eccentricity of 0.01. $\dot{m}_{\mathit{max}}$ is $\sim 5\times 10^{10} \,\rm g/s$, one order of magnitude larger than the predicted value based on transits, $8\times10^{9}\,\rm g/s$ (note that this value is uncertain and model-dependent \citealp{2015Natur.526..546V}) and slightly larger than the accretion rate of WD\,1145+017, $4.3\times 10^{10}\,\rm g/s$ \citep{2016ApJ...816L..22X}. Setting $\dot{m}_{\mathit{max}}=8\times 10^{9}\,\rm g/s$, we find a lower limit of planetesimal radius $\sim 70\,\rm km$ to explain the transit around WD\,1145+017 via tidally induced volcanism. Under the mass loss rate of $8\times10^{9}\,\rm g/s$ and $4.3\times 10^{10}\,\rm g/s$, the planetesimal with $R_p=100\,\rm km$ and $\rho_p=6000\,\rm kg/m^3$ will lose $10\%$ of its mass in $\sim 10^4\,\rm yr$ and $\sim 2\times 10^3\,\rm yr$, respectively, much longer than the observation span of this system. Hence, volcanism is a possible explanation for transits, and can potentially contribute to white dwarf pollutants.

Conversely, for a strength-less rocky planetesimal ($\rho_p\lesssim 4000\,\rm kg/m^3$ \citealp{2020M&PS...55..962P} and hence Roche limit $\gtrsim0.00415\,\rm AU$) to reach a dust production rate of $8\times10^{9}\,\rm g/s$, the planetesimal must be larger than $\sim 150\,\rm km$. For a planetesimal with a radius of 800\,km (above which the escape speed exceeds the maximum volcanic ejection speed), the planetesimal must be perturbed within $\sim 0.01\,\rm AU$ to reach a dust production rate of $8\times10^{9}\,\rm g/s$.

If volcanism is the origin of transits corresponding to short-period near circular orbit, problems remain on whether the planetesimal still have sufficient amount of gas and silicate melts to supply volcanism eruption at the end of its tidal evolution stage. To explain tidally induced volcanism that leads to transits on near circular orbits, either mass loss during tidally induced volcanism is much less efficient than the theoretical maximum, or tidally induced volcanism preferentially occurs at a late tidal evolution stage for some planetesimals. Some potential scenarios include: 

\begin{itemize}
    \item High speed volcanic ejection (beyond the escape speed) with a large amount of material erupted is of low possibility, such that high mass loss rate cadences are discrete and of limited lifetime. The planetesimal can recycle a large fraction of its volcanic products back to its interior and re-triggering volcanism eruption after the corresponding geological timescale. The true accumulated mass loss of the planetesimal is much smaller compared to its counterpart integrated using the theoretical maximum mass loss rate.
    \item The planetesimal starts massive and only the extremely high speed ejection can contribute to mass loss. Mass loss starts slow as only the extremely high speed eruption can produce volcanic products escaping from the planetesimal. High mass loss rate may only become relevant at a late tidal evolution stage when the planetesimal loses a fraction of its mass, reducing its escape speed to a certain degree.
    \item Cooling may be efficient enough to delay the the propagation of melt reservoir towards the surface such that volcanism from a shallow melt reservoir only occurs at a late tidal evolution stage (see Section \ref{model limitation discussion} and Appendix D).
\end{itemize}

\subsubsection{Frequency of tidally induced volcanism and signatures in white dwarf pollutants} \label{frequency of pollutants discussion}

\begin{figure}
\includegraphics[width=0.45\textwidth]{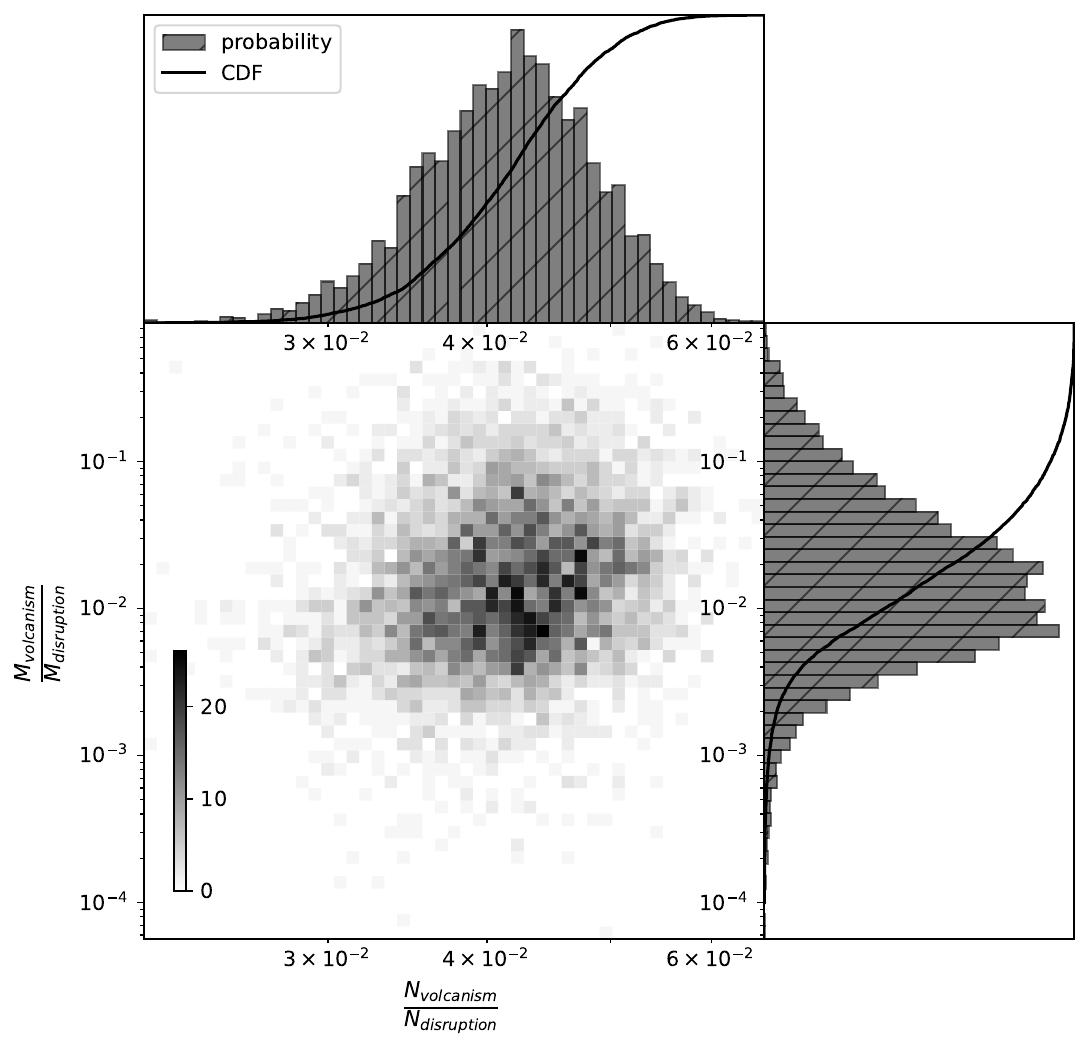}
\caption{The simulated distribution for the frequency (x-axis) and accreted mass (y-axis) ratio of tidally induced volcanism relative to tidal disruption for $\alpha=1$ and $\beta=-4$ and 5000 systems (each system contains 200000 sample planetesimals). Middle panel: 2D density map in frequency ratio-mass ratio space. Upper panel: probability distribution (bar chart) and the cumulative distribution (CDF, line) for frequency ratio. Right panel: same as upper panel, but for accrerted mass ratio. The probability distributions and the density map are plotted on the logarithmic scale.}
\label{ratio plot}
\end{figure}

In this study, we focus on a population of planetesimals gravitationally perturbed close to the white dwarf (just outside the Roche limit). Meanwhile, planetesimals can be perturbed within the Roche limit, where they undergo tidal disruption and subsequent accretion onto the white dwarf, which is the leading theory for white dwarf pollution. Both processes can occur simultaneously. Similar to the debris formed from tidal disruption, volcanic ejecta can be accreted onto the white dwarf, via, for instance, PR drag (Section \ref{orbital perturbation discussion}). As a result, tidally induced volcanism provides an additional channel of white dwarf pollutants. In this section, we compare between the frequency and mass contributions of tidal disruption and tidally induced volcanism ($\frac{N_{\mathit{volcanism}}}{N_{\mathit{disruption}}}$, $\frac{M_{\mathit{volcanism}}}{M_{\mathit{disruption}}}$). 

We investigate a simple model based on the following assumptions:

\begin{itemize}

    \item A size distribution of planetesimals with a probability density function $P(R_p)\propto R_p^{\beta}$, from $R_{\mathit{p,min}}=10\,\rm km$ to $R_{\mathit{p,max}}=\,\rm 1000\,\rm km$ \citep{1969JGR....74.2531D,2012ApJ...754...74G,2012ApJ...747..113P,2016ApJ...822...55S,2017ApJ...847L..12S}.

    \item An uniform planetesimal density distribution between $3000\,\rm kg/m^3$ and $6000\,\rm kg/m^3$.

    \item An initial pericentre distribution with a probability density function $P(q_0)\propto q_0^{\alpha}$.

    \item A $K_p$ ($3k_2\Delta t$ which quantifies the tidal dissipation efficiency) distribution where $\log_{10}\frac{K_p}{1\,\rm s}$ uniformly distributes between 0 and 4 (the lower limit is given by the measured value of gas giant \citealp{2023A&A...669A...2F} and the upper limit is arbitrarily chosen to be an order of magnitude larger than the fiducial value).
    
    \item A Roche limit ($r_{\mathit{Roche}}$) distribution that is uniform between $0.003\,\rm AU$ and $0.005\,\rm AU$ (estimated using the property range of Solar System ordinary chondrites \citealp{2020M&PS...55..962P}). 
    
    \item A planetesimal potentially undergoes volcanism when $r_{\mathit{Roche}}<q_0<q_{\mathit{0,crit}}'$, with $q_{\mathit{0,crit}}'=0.01\,\mathrm{AU} \left(\frac{K_p}{1000\,\mathrm{s}}\right)^{\frac{2}{15}}\left(\frac{R_p}{100\,\rm km}\right)^{\frac{8}{15}}$ and when the tidal escaping timescale $\tau_{\mathit{e,tide}}= 5\frac{r_HM_pq_0^{7.5}}{GK_pM_*^2R_p^5Q_0^{0.5}}$ (with $r_H$ the Hill radius of a 10 Earth mass planet at 3\,AU, Appendix P) is smaller than the 'outward' perturbation timescale $\tau_p$ such that $\log_{10}\frac{\tau_p}{1\,\rm Myr}$ uniformly distributed between between -3 to 3.    

    \item A distribution of the fraction of planetesimal mass accreted by the white dwarf during tidally induced volcanism $f_\mathit{m,v}\frac{\mathrm{max}(0,1000\,\mathrm{m/s}-v_{\mathit{esc}})}{1000\,\mathrm{m/s}-10\,\mathrm{m/s}}$ (accounting for the lower probability of mass loss for a higher escape speed) where $\log_{10}f_\mathit{m,v}$ uniformly distributes between -4 and -1.

    \item An accretion efficiency of tidal disruption 0.5 (arbitrary choice, accounting for the fact that the debris disk can be continuously perturbed by the planet that perturbed the planetesimal in the first instance, with some of the fragments ejected and some stable against accretion).

\end{itemize}

We simulate the frequency and pollutant mass ratio of tidally induced volcanism to that to tidal disruption, assuming $\alpha=1$ (an arbitrary choice accounting for the lower possibility of perturb a planetesimal closer to the white dwarf \citealp{2024MNRAS.527.11664}), with a larger $\alpha$ making tidally induced volcanism more preferred) and $\beta=-4$ \citep{1969JGR....74.2531D,2012ApJ...754...74G,2012ApJ...747..113P}, with a larger $\beta$ making tidally induced volcanism more preferred. We plot the simulated distribution of $\frac{N_{\mathit{volcanism}}}{N_{\mathit{disruption}}}$ and $\frac{M_{\mathit{volcanism}}}{M_{\mathit{disruption}}}$ on the logarithmic scale (order of magnitude comparison) in Fig.\ref{ratio plot}. Under our model, tidal disruption is usually more than 10 times more frequent than tidally induced volcanism. Compared to the frequency ratio, the ratio of mass delivery of volcanism to that of tidal disruption spans a much wider order of magnitude ($10^{-4}$--1). 

Note that we minimize the free parameters required for simulations and hence neglect various complexities in the tidal disruption-tidally induced volcanism scenario, for instance, the correlation between planetesimal size/density and tensile strength, volcanic ejection speed, the geological constraints on volcanism, the dynamics of tidally disrupted disks. Furthermore, many of the free parameters/distributions remain poorly constrained. The computations of mass ratio $\frac{M_{\mathit{volcanism}}}{M_{\mathit{disruption}}}$ involves more assumptions and are hence less certain compared to the frequency ratio $\frac{N_{\mathit{volcanism}}}{N_{\mathit{disruption}}}$. The occurrence of tidally induced volcanism is not a sufficient condition for transits, as other conditions, for instance, transit probability, dust production rate, must be accounted for.

The volcanic activities we are familiar with, for instance, the volcanism on Io and Earth, are silicate volcanism. Assuming this is the case for exoplanetary bodies, tidally induced volcanism will shift the white dwarf pollutants towards mantle-rich. On the other hand, as is discussed in \citealp{2019GeoRL..46.5055A,2020NatAs...4...41J}, for a differentiated planetesimal with thin/no rocky mantle, a top-down solidification of the iron core can result in significant pressure growth in the iron magma ocean, which in turn induces dike formation and eruptions. The quiescent duration of Iron volcanism is controlled by strain accumulation during solidification. The time interval between two iron volcanism active periods, according to the calculations of \citealp{2019GeoRL..46.5055A} for a 100\,km planetesimal is of the order of magnitude $1000\,\rm yr$, which is unlikely the major cause of successive transit-active periods within a short time-span (unless the volcanism is long-lasting), for instance, those of WD\,1145+017. 

We are yet unable to confirm whether tidally induced volcanism tends to enhance the total mass delivered to the white dwarf. On one hand, tidally induced volcanism expands the pericentre distance range where the planetary material can be accreted by the white dwarf, and save planetesimals from being ejected from the system. On the other hand, the efficiency of mass delivery for tidal disruption and tidally induced volcanism remains unknown, and could be much higher for tidal disruption.

\subsection{The importance of additional processes}  \label{other potential scenarios discussion}

Tidally induced volcanism does not occur in isolation. There are many other accompany physical processes, some of which may enhance the observed signatures, whilst other disrupt them. 

First, general relativistic precession is important both for the planetsimals' orbit and for the orbit of dust released. The precession period of a planetesimal avoiding tidal disruption is usually too long for observations ($\gtrsim 100\,\rm yr$ for the planetesimal around WD\,1145+017, see Appendix Q). On the other hand, a precessing accretion disk much closer to the white dwarf ($\lesssim 0.1\,R_{\odot}$, where sublimation may play a major role) may mimic a $\sim100\,\rm day$ transit period.

Second, collisions among tidally evolving planetesimals may both lead to transit (dust production, trigger volcanism by inducing stress, producing fracture, and supplying heat) and prevent long-term transits (destructive collisions). We estimate the collision rate of tidally evolving planetesimals (see Appendix R), finding out that less than one collision is expected per Myr under our model. Collisions for planetesimals undergoing tidal evolution are mostly destructive (high encounter speed), ruling out collision-induced volcanism. Therefore, although occasional destructive collisions may produce a high level of dust, we do not consider collisions among planetesimals a major cause of transits due to its low frequency. However, we acknowledge the possibility that a planetesimal may interact with an existing disk directly (collisions) or indirectly (gravitational perturbations), triggering disk activities that may be observed as transits, with the periodicity of disk activities potentially correlated with the orbital period of the planetesimal. 

Third, tidally induced volcanism is usually accompanied with planetesimal property variations, potentially altering the strength and rheology of the planetesimal, thus changing the conditions of tidal disruption. As a result, tidal disruption may be re-triggered under the action of tidal interactions and volcanism, due to spin-up and/or weakening strength.

Fourth, if the planetesimal undergoes rapid tidal circularization when the white dwarf remains luminous, the tidal flux and/or stellar irradiation may be strong enough to cause surface melting/sublimation (although not a common representative of the population, see Section \ref{surface sublimation discussion}) \citep{2015Natur.526..546V}. With volcanism changing the surface condition of dust particles (e.g., more loosely detached), sublimation may enhance the escape of volcanic dust from the planetesimal and/or the formation of dust coma. Sublimation and volcanism can equivalently lead to high speed ejecta leaving the planetesimal. 

Notably, assuming a Maxwell-Boltzmann distribution of sublimated forsterite, the speed distribution effectively truncates at $\sim 1000\,\rm m/s$ for the irradiation experienced by the planetesimal around WD\,1145+017, thus making sublimation degenerate with tidally induced volcanism in terms of the ejection speed. The key differences between sublimation and tidally induced volcanism might be the stochasticity (with tidally induced volcanism being more stochastic in the magnitude of emission and ejection velocity while sublimation being more continuous), the correlation with stellar irradiation (with sublimation require strong stellar irradiation and hence strong positive correlation between incident irradiation and the presence of transits/transit depth/activity level), and the lifetime of the secondary transits (distinct dynamics due to different particle sizes). On the other hand, tidally induced volcanism may possess intrinsic periodicity in activities (nearly constant interval between two volcanically active periods) controlled by, for instance, the magma chamber refilling time \citep{DOBRETSOV20151663}. 

The tidally induced volcanism scenario is complex, involving various physics. However, these processes are unlikely to diminish/outweigh the features of tidally induced volcanism.

\subsection{Different models for the origin of transits} \label{other models discussion}

Tidally induced volcanism described in this paper provides a self-consistent explanation for planetesimals to both arrive at the observed transit periods (tidal decay) and to produce the transits (volcanism). 

Meanwhile, there exist other models to explain transits and are not necessarily exclusive to tidally induced volcanism:

\begin{itemize}
    \item Sublimation under sufficient irradiation \citep{2015Natur.526..546V}.
    \item Fragments leaving the planetesimal at L1 point undergoing further sublimation \citep{2016MNRAS.458.3904R}.
    \item Partial tidal disruption of a differentiated planetesimal's mantle \citep{2017MNRAS.465.1008V,2020ApJ...893..166D}.
    \item Rotational fission of triaxial planetesimal due to exchange of spin and orbital angular momentum, potentially undergoing rotational fission \citep{2020MNRAS.492.5291V}.
\end{itemize}

Each theory has their specific constraints on the properties of the planetesimal and/or white dwarf while the probability of lying in the parameter space of a given model remains mystery. For instance, tidally induced volcanism requires a planetesimal composition/rheology where strong tidal interactions are induced and volcanism is favoured, with the conditions essential for other theories, e.g., strong stellar irradiation for sublimation, not necessary. Meanwhile, different theories may lead to distinct interpretations of the transit features. For instance, partial tidal disruption and sublimation are more inclined to continuous processes, thus requiring further modeling for the stochastic nature of the observed transits if not of observational origin (e.g., the large variability, presence and disappearance of transits around WD\,1145+017). Compared to more energetic processes such as sublimation and tidally induced volcanism, partial tidal disruption and rotational fission leads to smaller orbit dispersion  (see \citealp{2017MNRAS.465.1008V,2020MNRAS.492.5291V} for the filling time), hence requiring additional dispersion mechanisms when dealing with large transit depth. On the other hand, similar to the case of Io, high temperature processes such as sublimation and tidally induced volcanism may produce plasma, where the white dwarf magnetic field can alter the subsequent orbital evolution of the ejecta, which must be accounted for when matching the model to observations. Finally, all theories listed above, including tidally induced volcanism, are more inclined to increase the silicate abundances of white dwarf pollutants (a mantle/crust--core mass loss sequence).

There exist various mechanisms to explain the transits around white dwarfs, potentially with distinct parameter space preferences, characteristic features, and may work concurrently (see Section \ref{other potential scenarios discussion}).

\section{Conclusions}\label{conclusions}

Evolution of planetary systems around white dwarfs are crucial for understanding the arrival of planetary material onto the white dwarf. Transits of a handful of polluted white dwarfs reveal the rich dynamics of clos-in planetary material. 

In this work, we propose a scenario where large planetesimals gravitationally perturbed just outside the Roche limit potentially undergo rapid tidal evolution, where orbital decay, circularization and tidal heating take place. Tidal heating may trigger volcanism, on orbits ranging from long-period highly eccentric orbits to short-period near circular orbits. Tidally induced volcanism, although potentially less common than tidal disruption ($\lesssim 10\%$), may lead to dramatic observable signatures like transits, where the ejected dust that is initially optically thick blocks the white dwarf. Tidally induced volcanism may work alongside with tidal disruption in delivering planetary material to the white dwarf, increasing the abundances of mantle-rich material in white dwarf pollutants. This model can broadly match the transit duration, transit depth, $180^{\circ}$ phase shift of the primary transit of WD\,1145+017, as well as the range of secondary transits. On the other hand, to match the predicted mass loss rate required for transits around WD\,1145+017, a large body ($R_p\gtrsim 100\,\rm km$) is required. This model predicts short-period transits (hour--day) are more concentrated on a single orbital period, whilst longer-period (days--hundreds of days) transits could have much higher dispersion in the orbital period. However, more comprehensive modelings for the thermal, rheological, tidal evolution of the planetesimal as well as the evolution of volcanic ejecta are required for a better understanding of this scenario.

\section*{Acknowledgements}

AB acknowledges the support of a Royal Society University Research Fellowship, URF\textbackslash R1\textbackslash 211421. YL acknowledges the support of a STFC studentship. 

\section*{Data Availability}

Codes and data used in this work are available upon reasonable request to the author, Yuqi Li. 



\bibliographystyle{mnras}
\bibliography{example} 




\appendix


\bsp	
\label{lastpage}
\end{document}


\maketitle
\appendix

\section{Orbital energy dissipation}\label{energy comparison}

We expand the tidal power $\dot{E}_t$ at high and low eccentricities for $\frac{q_0}{Q_0}\rightarrow 0$: 

\begin{equation}\label{tidal power high e}
\frac{dE_t}{dt}(e\rightarrow 1) \approx 0.19819G^2M_pM_* T_p q_0^{-7.5}Q_0^{-1.5},
\end{equation}

\begin{equation}\label{tidal power low e}
\frac{dE_t}{dt}(e\rightarrow 0)\approx 0.006836G^2M_pM_* T_p q_0^{-9}e^2.    
\end{equation}

Along the tidal evolution track, the maximum of tidal power ($\frac{dE_t}{dt}$) is determined by that of $e^2(1-e^2)^{\frac{3}{2}}F(e)$, at $e=e_{\mathit{max}}\approx 0.735$:

\begin{equation}\label{tidal power max}
\frac{dE_t}{dt}(e= e_{\mathit{max}})\approx  0.002363 M_pM_* T_p q_0^{-9}. 
\end{equation}

The accumulated tidal energy when a planetesimal starting at $(q_0,Q_0)$ ($e_0=\frac{Q_0-q_0}{Q_0+q_0}$) reaches an eccentricity $e$ is given by:

\begin{equation}\label{accumulated tidal energy}
\Delta E_t =\frac{GM_pM_*\left(q_0+Q_0\right)}{4q_0Q_0}\left[\left(\frac{Q_0-q_0}{Q_0+q_0}\right)^2-e^2\right].
\end{equation}

We approximate the energy required for melting as $c_pM_p\Delta T_{\mathit{c}}$, such that $e_0-e\sim (e_0-e)e_0<e_0^2-e^2=\frac{4q_0Q_0}{q_0+Q_0}\frac{c_p\Delta T_c}{GM_*}\sim 2\times 10^{-4}\frac{q_0}{0.01\,\mathrm{AU}}\frac{c_p}{1000\,\rm J/(K\cdot kg)}\frac{\Delta T_c}{3000\,\rm K}\frac{0.6\,M_{\odot}}{M_*}$. Hence, eccentricity decay is negligible as the accumulated tidal energy reaches the melting energy for the parameters we consider. It is reasonable to approximate the melting time $t_{\mathit{melt}}$ using:

\begin{equation}
t_{\mathit{melt}}\sim\frac{c_pM_p\Delta T_{\mathit{c}}}{\dot{E}_{\mathit{t,0}}}.    
\end{equation}

The cooling timescale $t_{\mathit{cool}}$ is approximated by the diffusion timescale scaled by the Nusselt number ($\mathrm{Nu}$, the ratio of total heat loss rate to conductive heat loss rate):

\begin{equation}
t_{\mathit{cool}}\sim \frac{R_p^2}{\pi^2 \alpha_d \mathrm{Nu}}\sim \frac{R_p^2\rho_pc_p}{\pi^2\kappa \mathrm{Nu}}.    
\end{equation}

$\frac{t_{\mathit{melt}}}{t_{\mathit{cool}}}$ can hence be approximated as:

\begin{equation}\label{t_melt t_cool ratio analytical}
\frac{t_{\mathit{melt}}}{t_{\mathit{cool}}} \sim 208.6\frac{\kappa \mathrm{Nu} \Delta T_{\mathit{c}}q_0^{7.5}Q_0^{1.5}}{G^2K_pM_*^3R_p^4}. 
\end{equation}

\section{Critical energy of melting}\label{critical energy of melting appendix}

In this section, we estimate the temperature representation of the melting energy with respect to a fixed specific heat capacity $c_p$, $\Delta T_c$, such that $\bar{c}_pM_p\Delta T_c$ is the critical energy of melting (we will refer to the constant specific capacity as $\bar{c}_p$ in this section). For a temperature dependent heat capacity, $c_p(T)$, a specific enthalpy of fusion $\Delta h_{\mathit{f}}$, and a critical temperature of melting $T_c$, we have:

\begin{equation}
\Delta T_c= \frac{\int_{T_0}^{T_c}c_p(T)dT+\Delta h_{\mathit{f}}}{\bar{c}_p},
\end{equation}

\noindent where $T_0$ is the initial temperature of the planetesimal. We use the empirical relation of $c_p(T)$ for forsterite \citep{article}:

\begin{equation}
\begin{aligned}
&c_p(T)=-2863+528\ln{T}+\frac{623\times 10^3}{T}-\frac{184\times 10^6}{T^2}+\frac{180\times 10^8}{T^3},     
\end{aligned}
\end{equation}

\noindent with the heat capacity in SI unit ($\rm J/(K\cdot kg)$). Note that the empirical relation has a minimum at $T\approx 206\,\rm K$, which may not be physical. Hence, we assume that $c_p(T<206\,\mathrm{K})=c_p(T=206\,\rm K)$. We find out that $\Delta T_c\sim 3000\,\rm K$ for $T_c\sim 2000\,\rm K$, $50\,\mathrm{K}\lesssim T_0\lesssim 500\,\rm K$ and $\Delta h_f\sim 1000\,\rm kJ/kg$ \citep{LESHER2015113}. The lower and upper limits of investigation for $T_0$ are the equilibrium temperatures a planetesimal would have if it is on a highly eccentric orbit when the host star just enters its white dwarf phase and at a cooling age $\sim 100\,\rm Myr$ (with other parameters identical to those in Table 1 in the main text).

\section{The Nusselt number}\label{parameterized convection}

We use the Nusselt number $\mathrm{Nu}$ to quantify the total heat loss (excluding heat loss during volcanism) to conductive heat loss ratio. We constrain the range of $\mathrm{Nu}$ by considering parameterized convection in a partially melted solid-like mantle whose viscosity is reduced in the presence of liquid melt, with the Nusselt number given by \citep{SOLOMATOV200791,2013EGUGA..15.4653L,2015E&PSL.427...74Z}:

\begin{equation}
\mathrm{Nu}=0.089\mathrm{Ra}^{\frac{1}{3}},    
\end{equation}

\noindent with $\mathrm{Ra}$ the Raleigh number:

\begin{equation}
\mathrm{Ra}=\frac{\beta g\Delta Tl^3}{\alpha_t\nu},
\end{equation}

\noindent where $\beta$ is the coefficient of thermal expansion, $g$ is the gravity, $\Delta T$ is the surface temperature-critical temperature contrast, $l$ is the length scale of convection, $\alpha_t$ is the thermal diffusivity, $\nu$ is the kinematic viscosity. We apply $\beta=5\times 10^{-5}\,\rm K^{-1}$, $\alpha_t\sim 10^{-6}\,\rm m^2/s$, $\Delta T\sim 1000\,\rm K$, $gl^3\sim \frac{4}{3}\pi GR_p^4(1-f_m)^3\rho_p$. We further apply the parameterized viscosity for melt-bearing solid \citep{SOLOMATOV200791,2013EGUGA..15.4653L,2015E&PSL.427...74Z}:

\begin{equation}
\nu=\nu_se^{-\alpha_\nu\phi},    
\end{equation}

\noindent where $\nu_s\sim 10^{15}\,\rm m/s^2$, $\alpha_\nu=26$--31. We consider the limiting case where the melt fraction $\phi$ of the crust just reaches the critical value $\phi\sim 0.4$, with the maximum $\mathrm{Nu}$ given by:

\begin{equation}
\begin{aligned}
\mathrm{Nu}&=100\left(\frac{\beta}{5\times 10^{-5}\,\mathrm{K^{-1}}}\frac{\Delta T}{2000\,\mathrm{K}}\frac{1\times 10^{-6}\,\mathrm{m^2/s}}{\alpha_t}\frac{4\times 10^{9}\,\mathrm{m/s^2}}{\nu}\right)^{\frac{1}{3}}\\&\times \left(\frac{\rho_p}{3000\,\mathrm{kg/m^3}}\right)^{\frac{1}{3}}\left(\frac{R_p}{100\,\mathrm{km}}\right)^{\frac{4}{3}}(1-f_m),    
\end{aligned}
\end{equation}

\noindent which indicates that $\mathrm {Nu}\lesssim 10$ under the fiducial parameters and $\mathrm {Nu}$ can hardly exceed 100 even in the extreme case where the conduction length scale and convection length scale both equal to the size of the planetesimal.

We acknowledge that there exist other scaling relations for the Nusselt number, with which we will compare \citep{1999PEPI..116....1R,2005PEPI..149..361R, 2015A&A...584A..60C}:

\begin{equation}
 \mathrm{Nu}=A\theta^{B}\mathrm{Ra}^{C},   
\end{equation}

\noindent where we apply $\theta\sim \frac{4\times 10^4\,\mathrm{K}}{T_c^2}\Delta T$. Two sets of $(A,B,C)$ are investigated, $(0.67,-\frac{4}{3},\frac{1}{3})$ and $(2.51,-1.2,0.2)$, giving slightly smaller $\mathrm{Nu}$ than its counterpart obtained by parametrized convection. 

Hence, in the majority of the cases we have $\log_{10}\mathrm{Nu}\sim 0$--1.

\section{Alternative models for the condition of melting} \label{melting condition appendix}

\begin{figure}
\begin{center}
\includegraphics[width=0.8\textwidth]{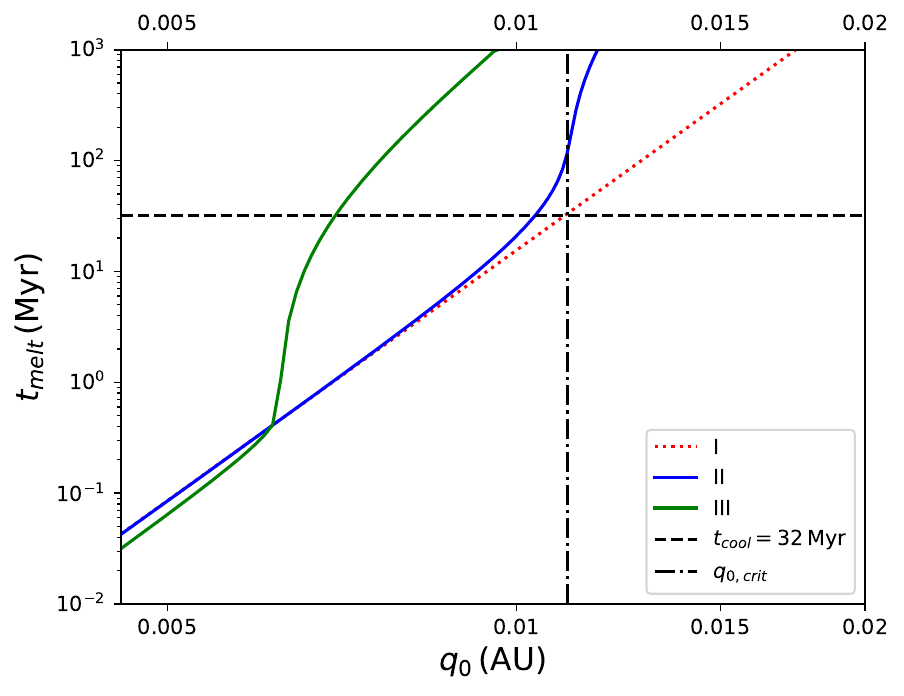}
\caption{A comparison of the predicted melting timescale using the model presented in this section (II: blue solid line, III: green solid line) and the model in Section 2.2.1 (I, red dotted line). The critical initial pericentre distance of melting are given by the $q_0$ at the intersections of the lines with $t=1\,\rm Gyr$ ($\sim 0.0096\,\rm AU$ and $0.0121\,\rm AU$) for model II and III, and the $q_0$ at the intersection of the red dotted line and $t=t_{\mathit{cool}}$ ($\sim 0.011\,\rm AU$, dashdot line) for model I.}
\label{melting model plot}    
\end{center}

\end{figure}

The condition of melting can be quantified as the accumulated heat (tidal heating subtracted by cooling) exceeding the melting energy within the timescale we consider (1\,Gyr). To quantify the cooling rate, we use the simple relation:

\begin{equation}
\dot{E}_{\mathit{acc}}=\dot{E}_t-\frac{E_{\mathit{acc}}}{t_{\mathit{cool}}}. 
\end{equation}

\noindent We will consider two models of $t_{\mathit{cool}}$, model II (maximum diffusion length):

\begin{equation}
 t_{\mathit{cool}}\sim \frac{R_p^2\rho_pc_p}{\pi^2\kappa},   
\end{equation}

\noindent and model III (minimum diffusion length):

\begin{equation}
t_{\mathit{cool}}\sim \frac{\left[R_p-\left(\frac{3}{4\pi\rho_p}\frac{E_{\mathit{acc}}}{c_p\Delta T_c}\right)^{\frac{1}{3}}\right]^2\rho_pc_p}{\pi^2\kappa}.    
\end{equation}

\noindent The model in Section 2.2.1 is model I. 

For melting to occur, we require that $\left(E_{\mathit{acc}}\right)_{\mathit{max}}\geq c_p M_p \Delta T_c$ for model II and $\left(E_{\mathit{acc}}\right)_{\mathit{max}}\geq c_p f_m^3M_p \Delta T_c$ for model III (to avoid singularity). The corresponding melting timescale satisfies $E_{\mathit{acc}}(t_{\mathit{melt}})=c_p M_p \Delta T_c$ and $E_{\mathit{acc}}(t_{\mathit{melt}})=c_p f_m^3M_p \Delta T_c$, respectively. The critical initial pericentre distance of melting $q_{\mathit{0,crit}}$ for model II and III is defined by the $q_0$ where $t_{\mathit{melt}}=1\,\rm Gyr$. 

We compute $t_{\mathit{melt}}$ numerically and compare to model I in Fig.\ref{melting model plot}, showing that $q_\mathit{0,crit}$ computed using model II and III, $\sim 0.0096\,\rm AU$ and $\sim 0.0121\,\rm AU$, are generally consistent with that of model I, $\sim 0.0111\,\rm AU$. However, non-negligible differences in $t_{\mathit{melt}}$ may present, especially towards large $q_0$, where tidal heating is less dominant and the cooling model matters more. Although all of our models are for qualitative purpose only, there is a generally agreement on the critical initial pericentre distance of melting $q_{\mathit{0,crit}}\sim 0.01\,\rm AU$ due to the strong dependence of tidal evolution on $q_0$ (a small change in $q_0$ could move the planetesimal from a cooling-dominant regime to heating dominant regime). 

Model II and III generally predicts a larger parameter space for melting ($K_p\lesssim 1\,\rm s$ and $R_p\lesssim 20\,\rm km$, while much larger $Q_0$ to rule out tidally induced melting when varying single parameter) as we no longer require melting to occur at the early tidal evolution stage before reaching the cooling timescale as is in model I. 

\section{Cooling through a spherical shell}\label{steady state appendix}

Assuming that the temperature profile of the planetesimal reaches a steady state where $T_{\mathit{c}}(x=f_mR_{\mathit{p}})$ at the transition from solidus to liquidus behaviour and $T(x=R_p)=T_s$ at the surface, the radial thermal diffusion equation reduces to:

\begin{equation}
\kappa x^2\frac{dT}{dx}=C,   
\end{equation}

\noindent which, for a constant $\kappa$ (can be understood as using $\bar{\kappa}=\left|\frac{\int_{T_c}^{T_s}\kappa(T)dT}{T_s-T_c}\right|$) gives:

\begin{equation}
T=\frac{f_mR_p}{x(1-f_m)}(T_{\mathit{c}}-T_s)+\frac{T_s-f_mT_{\mathit{c}}}{1-f_m},    
\end{equation}

\noindent and the corresponding heat loss rate is:

\begin{equation}\label{heat loss equation appendix}
\dot{Q}_l=-\kappa \mathrm{Nu'} 4\pi x^2\frac{dT}{dx}=4\pi \kappa \mathrm{Nu'} \frac{f_m}{1-f_m}R_p (T_{\mathit{c}}-T_s),   
\end{equation}

\noindent where $\mathrm{Nu'}\geq1$ is the Nusselt number accounting for additional heat loss (the 'prime' is used to remind that melting and volcanism are distinct thermal processes). 

At $e\rightarrow e_0\rightarrow 1$, the ratio of tidal power to heat loss rate ($\frac{\dot{E}_{\mathit{t}}}{\dot{Q}_l}$) is:

\begin{equation}\label{power ratio large e}
\frac{\dot{E}_{\mathit{t}}}{\dot{Q}_l}\sim 0.01577 \frac{G^2K_pM_*^3R_p^4}{\kappa \mathrm{Nu'} (T_c-T_s)q_0^{7.5}Q_0^{1.5}}\frac{1-f_m}{f_m},   
\end{equation}

\noindent and $\frac{\dot{E}_{\mathit{t}}}{\dot{Q}_l}$ at $e\rightarrow 0$ can be approximated as:

\begin{equation}\label{power ratio small e}
\frac{\dot{E}_{\mathit{t}}}{\dot{Q}_l}\sim 0.000544 \frac{G^2K_pM_*^3R_p^4e^2}{\kappa \mathrm{Nu'} (T_c-T_s)q_0^{9}}\frac{1-f_m}{f_m}.
\end{equation}

More generally, when $\kappa$ is a function of temperature, one need to solve:

\begin{equation}
\int \kappa(T) dT=\int \frac{C}{x^2}dx.   
\end{equation}

For $\kappa=\frac{1}{a+bT}$ \citep{miao2014temperature}, we have:

\begin{equation}
\begin{aligned}
\ln(a+bT)&=\frac{f_mR_p}{x(1-f_m)}\left[\ln{(a+bT_c)}-\ln{(a+bT_s)}\right]
\\&+\frac{1}{1-f_m}\left[\ln{(a+bT_s)}-f_m\ln{(a+bT_c)}\right],      
\end{aligned}
\end{equation}

The corresponding heat loss rate is:

\begin{equation}\label{heat loss T dependent}
\begin{aligned}
\dot{Q}_l&=4\pi \mathrm{Nu'}\frac{f_m}{(1-f_m)}R_p\frac{1}{b}\left[\ln{(a+bT_c)}-\ln{(a+bT_s)}\right]
\\&=4\pi \mathrm{Nu'}\frac{f_m}{(1-f_m)}R_p\frac{1}{b}\ln{\frac{\kappa(T_s)}{\kappa(T_c)}}.   
\end{aligned}   
\end{equation}

\section{Depth of magma reservoir}\label{thermal stress appendix}

\begin{figure}
\begin{center}
 \includegraphics[width=0.95\textwidth]{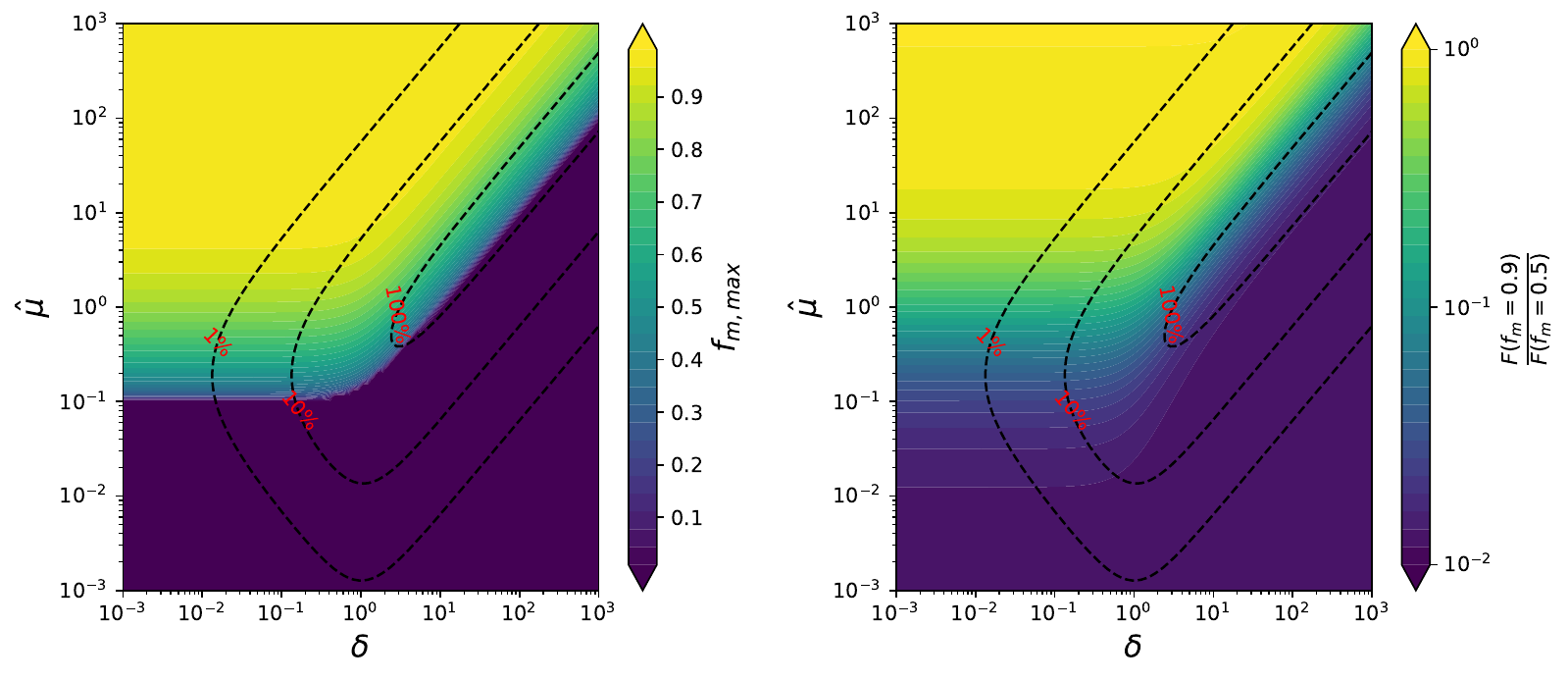}
\caption{The $f_m$ where $\Im(\tilde{k}_2)$ reaches its maximum (left panel), and the ratio of $F\equiv \Im(\tilde{k}_2)\frac{1-f_m}{f_m}$ at $f_m=0.9$ to its counterpart at $f_m=0.5$ (right panel) in $\delta$--$\hat{\mu}$ space. The magnitude of $\Im(\tilde{k}_2)(f_m=0.5)$ is illustrated by the contour lines normalised with respect to the $100\%$ contour line (corresponding to $\Im(\tilde{k}_2)=0.6$).}
\label{k2 plot appendix}   
\end{center}

\end{figure}

One estimation for the depth of magma reservoir is where the thermal stress reaches the material strength, and the magma tend to accumulate instead of propagating in the absence of excess pressure. The thermal stress is given by \citep{GELLER1972213,TAMBOVTSEVA1999319}:

\begin{equation}
\begin{aligned}
\sigma_i(r)=\frac{\alpha E}{1-v}T_i,        
\end{aligned}
\end{equation}

\noindent where $i=r$ or $\phi$, with $\sigma_r$ the radial stress (0 at surface), $\sigma_\phi$ the tangential stress, $\alpha$ is the coefficient of linear expansion, $E$ is the Young's modulus, $\nu$ is the Poisson ratio, $T_i$ is the temperature representations of thermal stress given by:

\begin{equation}
T_r=\frac{2}{R_p^3}\int_{0}^{R_p}T(r)r^2dr-\frac{2}{r^3}\int_{0}^{r}T(r)r^2dr,   
\end{equation}

\begin{equation}
T_\phi=\frac{2}{R_p^3}\int_{0}^{R_p}T(r)r^2dr+\frac{1}{r^3}\int_{0}^{r}T(r)r^2dr-T(r)   
\end{equation}

We apply a simplified temperature profile (see Appendix \ref{steady state appendix}):

\begin{equation}
T(r)=\begin{cases}
T_c,&r<f_mR_p\\
\frac{f_mR_p}{r(1-f_m)}(T_{\mathit{c}}-T_s)+\frac{T_s-f_mT_{\mathit{c}}}{1-f_m},&f_mR_p<r<R_p
\end{cases}, 
\end{equation}

\noindent from which we obtain:

\begin{equation}
T_r=2A-2B(r),    
\end{equation}

\begin{equation}
T_\phi=2A+B(r)-T(r),    
\end{equation}

\noindent where $A$ and $B$ for $f_mR_p\leq r\leq R_p$ are given by:

\begin{equation}
\begin{aligned}
A&=\frac{1}{R_p^3}\int_{0}^{R_p}T(r)r^2dr\\&=\frac{1}{2}(f_m^2+f_m)(T_{\mathit{c}}-T_s)+\frac{1}{3}(T_s-f_mT_{\mathit{c}})(f_m^2+f_m+1)+\frac{1}{3}f_m^3T_c,        
\end{aligned}
\end{equation}

\begin{equation}
\begin{aligned}
B&=\frac{1}{r^3}\int_{0}^{r}T(r)r^2dr\\&=\frac{1}{2}\frac{f_mR_p}{(1-f_m)}(T_{\mathit{c}}-T_s)\left(\frac{1}{r}-\frac{f_m^2R_p^2}{r^3}\right) +\frac{1}{3}\frac{T_s-f_mT_{\mathit{c}}}{1-f_m}\left(1-\frac{f_m^3R_p^3}{r^3}\right)\\&+\frac{1}{3}\frac{f_m^3R_p^3}{r^3}T_c,
\end{aligned}
\end{equation}

We investigate $T_i$ at $r=f_mR_P$, where $B=\frac{1}{3}T_c$ and $T_r=T_\phi$. We acknowledge that there is a wide range for the free parameters: $\alpha\sim 5\times 10^{-6}\,\rm K^{-1}$--$1\times 10^{-5}\,\rm K^{-1}$, $\nu\sim 0.1$--$0.4$, $E\sim 50\,\rm GPa$--$100\,\rm GPa$, a compressive strength $-\sigma_{\mathit{-}}\sim 10\,\rm MPa$--$1000\,\rm MPa$  \citep{1996GeoRL..23.1143B,TAMBOVTSEVA1999319,GERCEK20071,2017SoSyR..51...64S,min9120787,2019A&A...629A.119T,2020M&PS...55..962P,TANG2023107154,10.1093/mnras/stad2006}, allowing $f_m$ to range from 0 to 1. We obtain the fiducial value of $f_m$ using the numerical average of $\alpha$, $\nu$, $E$, together with $-\bar{\sigma}_{\mathit{-}}=200\,\rm MPa$ for L chondrites \citep{2020M&PS...55..962P} (most data available and only data available for dynamic measurements). We obtain $f_m\lesssim 0.9$.

Another estimation for the depth of melt reservoir can be obtained by considering the viscoelastic tidal dissipation as a function of melt fraction. Based on the model in \citealp{2024ApJ...961...22S}, we aim to locate the melt fraction corresponding to the maximum of the imaginary part of the love number ($\Im(\tilde{k}_2)$):

\begin{equation}
\Im(\tilde{k}_2)=\frac{3}{2}\frac{Z(f_m)\delta \hat{\mu}}{|1+Z(f_m)\hat{\mu}|^2+\delta^2},    
\end{equation}

\noindent where $\hat{\mu}=\frac{\mu}{\rho_p gR_p}$, with $\mu$ the rigidity, $\delta$ is the ratio of maxwell frequency to tidal forcing frequency. $Z(f_m)$ is given by:

\begin{equation}
Z(f_m)=12\left(\frac{19-75f_m^3+112f_m^5-75f_m^7+19f_m^{10}}{24+40f_m^3-45f_m^7-19f_m^{10}}\right),    
\end{equation}

\noindent which decreases monotonically with $f_m$ from the maximum, $Z(f_m=0)=9.5$, to, e.g.,  $Z(f_m=0.99)=0.055$.

If $\hat{\mu}$ is small, or if $\delta$ is large and $\delta \gg \hat{\mu}$, $\Im(\tilde{k}_2)$ is proportional to $Z(f_m)$ where tidal dissipation efficiency monotonically decreases with $f_m$. If $\hat{\mu}$ is large and $\hat{\mu}\gg \delta$, $\Im(\tilde{k}_2)$ is inversely proportional to $Z(f_m)$ where tidal dissipation efficiency monotonically increases with $f_m$. In these cases, there is no maximum for $\Im(\tilde{k}_2)$. Furthermore, $\Im(\tilde{k}_2)(f_m)$ is greatly suppressed if: 1. either $\hat{\mu}$ or $\delta$ is large and orders of magnitude larger than the other and 2. either $\hat{\mu}$ or $\delta$ is small. Hence, we are mainly interested in the cases where $\hat{\mu}\sim \delta\gtrsim 1$ (see Fig.\ref{k2 plot appendix} for an example). We consider the special case where $\hat{\mu}= \delta$, finding out that the maximum of $\Im(\tilde{k}_2)$ occurs at $f_m\sim 0.8$, which leads to more optimistic estimation of the parameter space of volcanism compared to $f_m\sim0.9$ obtained above. The maximum of $\Im(\tilde{k}_2)$ will shift towards larger $f_m$ with the increase in $\frac{\hat{\mu}}{\delta}$ and vice versa, before reaching the monotonic regions. 

The ascending rate of the melt frontier is proportional to $\Im(\tilde{k}_2)\frac{1}{f_m^2}$. Meanwhile, the descending rate of the melt frontier is proportional to (using Eq.\ref{heat loss equation appendix}) $\frac{1}{f_m(1-f_m)}$. Therefore, the evolution of $F\equiv \Im(\tilde{k}_2)\frac{1-f_m}{f_m}$ with $f_m$ indicates the evolution towards heating/melting dominated regime or cooling/crystallization dominated regime. In the parameter space of large $ \Im(\tilde{k}_2)$, we find order(s) of magnitude decay in $F$ when the melt frontier ascends from $f_m=0.5$ to $f_m=0.9$, such that the propagation rate of the melt frontier towards the surface is significantly suppressed.

In summary, we consider $f_m=0.9$ as a reasonable trial value for the depth of melt frontier.

\section{The ejection speed of volcanic flow}\label{ejection speed}

We estimate the volcanic ejection speed under the 1-D conduit model and choked flow condition \citep{1980GeoJ...63..117W,WOODS1995189,KOYAGUCHI200529,Gonnermann_Manga_2013}.

The continuity equation assuming a constant vent cross-sectional area is:

\begin{equation}\label{continuity equation}
\frac{d(\rho u)}{dz}=0,
\end{equation}

\noindent where $u$ is magma velocity. By using the fact that $\frac{d\rho}{dz}=\frac{d\rho}{dP}\frac{dP}{dz}$, Eq.\ref{continuity equation} can be written as:

\begin{equation}\label{continuity equation 1}
u\frac{d\rho}{dP}\frac{dP}{dz}+\rho\frac{du}{dz}=0,   
\end{equation}

\noindent where $\frac{d\rho}{dP}=\frac{1}{c_s^2}$ with $c_s$ the isothermal sound speed.

The momentum equation is:

\begin{equation}\label{momentum equation}
\rho u \frac{du}{dz}=-\frac{dP}{dz}-\rho g-f,
\end{equation}

\noindent with $f$ the friction force. Eq.\ref{momentum equation} can be combined with Eq.\ref{continuity equation 1} to give:

\begin{equation}\label{continuity-momentum equation 1}
(1-\frac{u^2}{c_s^2})\frac{dP}{dz}=-\rho g-f,    
\end{equation}

\noindent from which one can deduce that $u\leq c_s$ for $\frac{dP}{dz}\leq 0$. Therefore, it is reasonable to approximate the maximum vent exit speed using the sound speed (choked flow condition).

To derive the isothermal speed of sound, we combine Eq.\ref{continuity equation} and Eq.\ref{momentum equation} as:

\begin{equation}\label{continuity-momentum equation}
-u^2 \frac{d\rho}{dz}=-\frac{dP}{dz}-\rho g-f.
\end{equation}

The density of gas-magma mixture $\rho$ can be expressed as:

\begin{equation}\label{density equation}
\frac{1}{\rho}=\frac{1-n}{\sigma}+\frac{nRT}{P},   
\end{equation}

\noindent where $\sigma$ is magma density, $P$ is pressure, $R$ is specific gas constant, $n$ is the gas mass fraction, which can be related to the gas mass fraction at zero pressure $n_0$ as:

\begin{equation}
n=n_0-C,    
\end{equation}

\noindent where the mass fraction of dissolved gas $C$ is:

\begin{equation}
C=sP^{\beta}    
\end{equation}

\noindent with $s$ the saturation constant. $\beta$ and $s$ depends on the type of dissolved gas.

Differentiating Eq.\ref{density equation} gives: 

\begin{equation}
 -\frac{1}{\rho^2}\frac{d \rho}{d z}=-\frac{1}{\sigma}\frac{dn}{dz}+\frac{RT}{P}\frac{dn}{dz}-\frac{nRT}{P^2}\frac{dP}{dz},  
\end{equation}

\noindent where

\begin{equation}
\frac{d n}{d z}=-\frac{dC}{dz}= -\beta\frac{C}{P}\frac{dP}{dz}, 
\end{equation}

\noindent and hence

\begin{equation}
\frac{d \rho}{d z}=-\rho^2\left(\frac{\beta C}{P \sigma}-\frac{\beta C RT}{P^2}-\frac{nRT}{P^2}\right)\frac{dP}{dz}.   
\end{equation}

Eq.\ref{continuity-momentum equation} can be written as:

\begin{equation}
\begin{aligned}
\left[1-\frac{u^2\rho^2}{P^2}\left(n_0RT+(\beta-1) CRT-\frac{\beta CP}{\sigma}\right)\right]\frac{dP}{dz}=-\rho g-f,    
\end{aligned}
\end{equation}

\noindent which, by comparing with Eq.\ref{continuity-momentum equation 1}, indicates that the sound speed $c_s$ is:

\begin{equation}\label{sound speed equation}
\begin{aligned}
c_s=\sqrt{n_0RT}\left[1+\frac{(1-n)P}{nRT\sigma}\right]\frac{n}{\sqrt{n_0}\sqrt{{n_0+(\beta-1) C-\frac{\beta CP}{\sigma RT}}}}.
\end{aligned}
\end{equation}

One may notice that for the last term in the square root, $\frac{\beta CP}{\sigma RT}=\frac{\beta C \rho_{\mathit{gas}}}{\sigma}\ll \beta C \leq \beta n_0$ should generally hold. As we do not expect $\beta$ to be a large number (where solubility boosts with pressure), $\frac{\beta CP}{\sigma RT}\ll n_0$ should hold and one can neglect this term. We expect that the majority of the gas is exsolved at vent exit ($C\ll n_0$), and hence $\frac{n}{\sqrt{n_0}\sqrt{{n_0+(\beta-1) C}}}\sim 1$ (vent exit not at high pressure). On the other hand, the term $\frac{(1-n)P}{nRT\sigma}$ is equivalent to $\frac{V_{\mathit{magma}}}{V_{\mathit{gas}}}$, which is supposed to be small at vent exist \citep{2010GeoJI.182..843B,b8c5fb8f5396422290efe8786fef7d8b}. Therefore, $c_s\sim \sqrt{n_0RT}$ should generally hold.

After vent exit, the excess pressure of the flow relative to the ambient leads to decompression \citep{WOODS1995189}. The continuity equation implies that:

\begin{equation}
\rho_v u_vA_v=\rho_au_aA_a,
\end{equation}

\noindent where the subscript '$v$' represents the vent exit and '$a$' represents decompression to the ambient, $A$ is the cross-sectional area. The momentum equation becomes:

\begin{equation}
\rho_au_a^2A_a-\rho_vu_v^2A_v=p_vA_v+p_a(A_a-A_v)-p_aA_a.
\end{equation}

The speed after decompression can then be expressed as:

\begin{equation}\label{ua equation}
\begin{aligned}
u_a&=u_v+\frac{P_v-P_a}{\rho_vu_v}\sim 2\sqrt{n_0RT},   
\end{aligned}   
\end{equation}

\noindent where we apply $u_v\sim c_s\sim \sqrt{n_0RT}$ and $\rho_v\sim \frac{P_v}{n_0RT}$ at vent exit. 

To sum up, the volcanic ejection speed is of the order of magnitude $\sqrt{n_0RT}$, or: 

\begin{equation}
u_{\mathit{eject}}\sim 3000\sqrt{\frac{n_0}{\mu_m}\frac{T}{1000\,\mathrm{K}}}\,\mathrm{m/s},    
\end{equation}

\noindent where $\mu_m$ is the (relative) mean molecular mass of gas in the flow.

\section{Computations of orbit dispersion}\label{near circular appendix}

\begin{figure}
\begin{center}
 \includegraphics[width=0.8\textwidth]{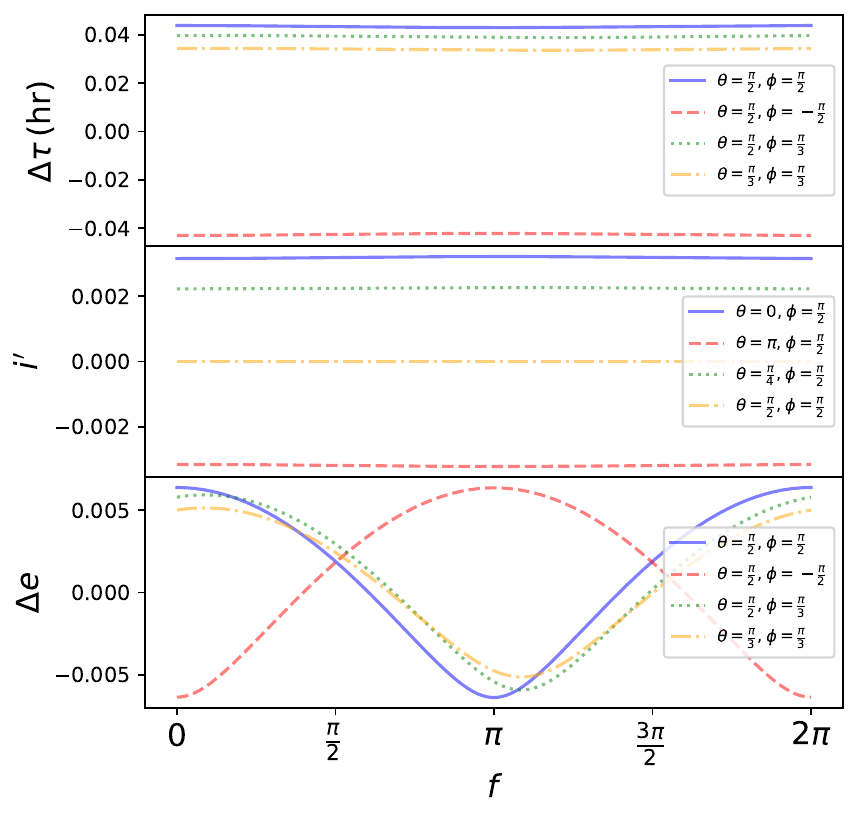}
\caption{The orbital period (upper panel), orbital inclination (middle panel) and eccentricity dispersion (lower panel) relative to the the planetesimal with $R_p=100\,\rm km$ and $\rho_P=6000\,\rm kg/m^3$ on a 4.5\,hr period with $e=0.01$ around WD\,1145+017, as a function of true anomaly ($f$) for four sets of ejection angles (note the different choices for $i'$, the middle panel). The ejection speed is assumed to be 1000\,m/s.}
\label{eccentricity plot appendix}   
\end{center}

\end{figure}

An order of magnitude approximation for $\Delta \bar{e}=\bar{e}'-\bar{e}$ assuming a near circular orbit is:

\begin{equation}
\begin{aligned}
\Delta \bar{e}&\approx \sin\theta \left[(S+2)\frac{R_p}{a}\cos\phi+\frac{2u_{\mathit{eject}}}{v_{\mathit{orb}}}\sin\phi\right]\hat{i}\\ &+\sin\theta\left[(S-1)\frac{R_p}{a}\sin\phi-\frac{u_{\mathit{eject}}}{v_{\mathit{orb}}}\cos\phi\right]\hat{j}, 
\end{aligned}
\end{equation} 

\noindent which approaches 0 when ejection occurs at the poles of the planetesimal ($\sin\theta=0$, where only second-order effects remain). We consider the maximum variation in $e$, $\pm|\Delta \bar{e}|$:

\begin{equation}
\begin{aligned}
\left|\Delta e\right|^2&\lesssim \left(\frac{5}{2}S^2+S+1\right)\frac{R_p^2}{a^2}+\frac{5}{2}\frac{u_{\mathit{eject}}^2}{v_{\mathit{orb}}^2} +\left\{\left[\left(\frac{3S^2}{2}+3S\right)\frac{R_p^2}{a^2}-\frac{3}{2}\frac{u_{\mathit{eject}}^2}{v_{\mathit{orb}}^2}\right]^2\right.\\ &\left.+\left[\left(3S+3\right)\frac{R_p}{a}\frac{u_{\mathit{eject}}}{v_{\mathit{orb}}}\right]^2\right\}^{\frac{1}{2}}. 
\end{aligned}
\end{equation}  

Taking WD\,1145+017 as an example, and for a planetesimal with $R_p=100\,\rm km$, $u_{\mathit{eject}}\lesssim 1000\,\rm m/s$, we find out that $|\Delta e|\lesssim 0.0064$, such that the ejecta remains on near circular orbit, as is verified in the lower panel of Fig.\ref{eccentricity plot appendix}.

More importantly, as is shown in Fig.\ref{eccentricity plot appendix}, although $\Delta e$ is sensitive to the true anomaly $f$ (vector addition), $\Delta \tau$ and $i'$ almost remain constant, nearly equal to their circular-orbit counterparts. Hence, when considering $\Delta \tau$ and $i'$ on a near circular orbit, one can effectively simplify the geometry to a circular orbit. 

We generalize the approximations for near circular orbits to moderately eccentric orbits, where the local keplerian speed and semi-major axis, is of the same order of magnitude as the orbital speed and distance, respectively, throughout the orbit, such that we may assume:

\begin{itemize}
    \item $\frac{v_{\mathit{spin}}}{v_k}\ll 1$,
    \item $\frac{u_{\mathit{eject}}}{v_k} \ll 1$
    \item $\frac{v_{\mathit{orb}}}{v_k}\sim 1$
    \item $\frac{R_p}{r}\ll 1$,
    \item $\frac{a}{r}\sim 1$,
    \item $\frac{2M_pa}{M_*R_p}=\frac{v_{\mathit{esc}}^2}{v_{\mathit{k}}^2} <\frac{u_{\mathit{eject}}^2}{v_{\mathit{k}}^2}$,
\end{itemize}

\noindent where $v_k=\sqrt{\frac{GM_*}{a}}$ is the keplerian speed.

The first-order approximation for semi-major axis of the ejecta is:

\begin{equation}
\begin{aligned}
\frac{1}{a'}&\approx\frac{1}{a}-\frac{1}{a}\frac{2R_p}{r}\frac{a}{r}\sin\theta\cos\phi\\&-\frac{1}{a}\left[ 2\frac{\omega_pR_p}{v_k}\frac{v_{\mathit{orb}}}{v_k}(-\cos\gamma\sin\theta\sin\phi+\sin\gamma\sin\theta\cos\phi) \right. \\
&\left. + 2\frac{u_{\mathit{eject}}}{v_k}\frac{v_{\mathit{orb}}}{v_k}(\cos\gamma\sin\theta\cos\phi+\sin\gamma\sin\theta\sin\phi)\right],
\end{aligned}
\end{equation}

\noindent from which we can deduce that:

\begin{equation}
\begin{aligned}
\frac{\Delta\tau}{\tau}&\approx3\sin\theta\left[\frac{R_p}{r}\frac{a}{r}\cos\phi+  \frac{\omega_pR_p}{v_k}\frac{v_{\mathit{orb}}}{v_k}(-\cos\gamma\sin\phi+\sin\gamma\cos\phi) \right. \\
&\left. +\frac{u_{\mathit{eject}}}{v_k}\frac{v_{\mathit{orb}}}{v_k}(\cos\gamma\cos\phi+\sin\gamma\sin\phi)\right],
\end{aligned}
\end{equation}

\noindent which increases with $v_{\mathit{orb}}$ and decreases with $r$, while further correlated with $\gamma$. Since the term $\frac{u_{\mathit{eject}}}{v_k}$ usually dominates for the parameters we consider, $\Delta \tau$ is maximized near the pericentre.

The first-order approximation for the orbital angular momentum is:

\begin{equation}
\begin{aligned}
\bar{J}'&=-R_pv_{\mathit{orb}}\sin\gamma\cos\theta\hat{i}\\ &+(R_pv_{\mathit{orb}}\cos\gamma\cos\theta-u_{\mathit{eject}}r\cos\theta)\hat{j}\\&+(rv_{\mathit{orb}}\sin\gamma+\omega_pR_pr\sin\theta\cos\phi+ru_{\mathit{eject}}\sin\theta\sin\phi\\&+R_pv_{\mathit{orb}}\sin\gamma\sin\theta\cos\phi-R_pv_{\mathit{orb}}\cos\gamma\sin\theta\sin\phi)\hat{k}, 
\end{aligned}
\end{equation} 

\noindent which gives the inclination:

\begin{equation}
i'\approx  \frac{\cos\theta}{\sin\gamma}\sqrt{\frac{R_p^2}{r^2}-2\frac{R_p}{r}\frac{u_{\mathit{eject}}}{v_{\mathit{orb}}}\cos\gamma+\frac{u_{\mathit{eject}}^2}{v_{\mathit{orb}}^2}},   
\end{equation}

\noindent whose maximum (although unclear analytically, but can be deduced based on the order of magnitude of the terms) is usually reached when $v_{\mathit{orb}}$ is minimized, i.e., the apocentre.

\section{Back reaction of volcanic ejecta on the planetesimal}\label{back reaction appendix}

In this section we investigate the effect of non-isotropic volcanism ejection on the orbit of the planetesimal. The induced impulse velocity of the planetesimal $\delta \bar{ v}$ due to ejecta mass $\delta m$ with ejection velocity $\bar{u}_{\mathit{eject}}$ satisfies:

\begin{equation}
\delta \bar{v}(M_p-\sum \delta m)+\sum \left(\bar{u}_{\mathit{eject}}\delta m\right)=0,  
\end{equation}

\noindent and hence:

\begin{equation}
\delta\bar{v}=-\frac{1}{M_p-\sum \delta m}\sum \left(\delta m u_{\mathit{eject}}\begin{bmatrix}\sin\theta\cos{\phi}\\\sin{\theta}\sin{\phi}\\\cos{\theta}\end{bmatrix}\right).   
\end{equation}

For simplicity, we consider single ejection case (alternatively, a weighted average), such that the summation can be omitted. The speed of the planetesimal after volcanism ejection is:

\begin{equation}\label{vtot appendix}
\begin{aligned}
v_{\mathit{tot}}&=\left| \begin{bmatrix}v_{\mathit{orb}}\cos\gamma-\Delta u_{\mathit{eject}} \sin\theta\cos\phi\\ v_{\mathit{orb}}\sin\gamma-\Delta u_{\mathit{eject}} \sin\theta\sin\phi\\-\Delta u_{\mathit{eject}} \cos \theta\end{bmatrix} \right|, 
\end{aligned}
\end{equation}

\noindent where $\Delta=\frac{\delta m }{M_p- \delta m}$.

We then solve the equation:

\begin{equation}
\frac{1}{2}v_{\mathit{tot}}^2-\frac{GM_*}{r}=-\frac{GM_*}{2a'},   
\end{equation}

\noindent with $\frac{1}{2}v_{\mathit{orb}}^2-\frac{GM_*}{r}=-\frac{GM_*}{2a}$, and get (to the leading order):

\begin{equation}\label{a' equation appendix}
\begin{aligned}
\frac{1}{a'}&\approx\frac{1}{a}+\frac{2\Delta v_{\mathit{orb}}u_{\mathit{eject}}(\cos\gamma\sin\theta\cos\phi+\sin\gamma\sin\theta\sin\phi)}{GM_*}.
\end{aligned}
\end{equation}

Assuming that $\Delta v_{\mathit{orb}}u_{\mathit{eject}}\ll v_{k}^2$, Eq.\ref{a' equation appendix} can be approximated as:

\begin{equation}
\frac{\delta a}{a} \approx -2\Delta \frac{v_{\mathit{orb}}}{v_k}\frac{u_{\mathit{eject}}}{v_k}(\cos\gamma\sin\theta\cos\phi+\sin\gamma\sin\theta\sin\phi).   
\end{equation}

For a near circular orbit Eq.\ref{a' equation appendix} becomes:

\begin{equation}\label{delta a cir appendix}
\frac{\delta a}{a} \approx -2\Delta \frac{  u_{\mathit{eject}} }{v_{\mathit{orb}}}\sin\theta\sin\phi,   
\end{equation}

\noindent which, since $\Delta \ll 1$, is much smaller compared to the orbit dispersion of the ejecta and is usually negligible. 

The differential form of Eq.\ref{delta a cir appendix} can be obtained by substituting $\Delta=-\frac{dM_p}{M_p}$:

\begin{equation}
\frac{\dot{a}}{a}\approx\frac{2\dot{M}_p}{M_p}\frac{u_{\mathit{eject}} }{v_{\mathit{orb}}}\sin\theta\sin\phi,
\end{equation}

\noindent whose magnitude is negligible over an observational timescale of years ($\frac{\delta a}{a}\lesssim 10^{-8}\,\rm yr^{-1}$ for a planetesimal with a radius of 100\,km, density $6000\,\rm kg/m^3$ around WD\,1145+017 with a mass loss rate of $8\times 10^{9}\, \rm g/s$ \citealp{2015Natur.526..546V} and ejection speed of 500\,m/s). On the other hand, the maximum orbital perturbations due to the back reactions of non-spherically symmetric volcanism may outweighs the orbital evolution under tide on a near circular orbit. Therefore, there may exist a scenario where volcanism raises orbital eccentricity of the planetesimal and help sustain volcanism.

Similarly, mass loss may exert torque on the planetesimal, altering its spin. The torque $\bar{\tau}$ is given by:

\begin{equation}
\bar{\tau}=\frac{d(I_p\bar{\omega}_p)}{dt}=\sum\dot{M}_p\bar{r}\times\bar{u}_{\mathit{eject}},
\end{equation}

\noindent where $\bar{r}$ is the position vector of the ejection relative to the planetesimal's centre of mass. Meanwhile, the spin of the planetesimal is also altered by tidal evolution \citep{1981A&A....99..126H,2007A&A...462L...5L,2010A&A...516A..64L,2010ApJ...725.1995M,2011A&A...535A..94B,2011A&A...528A..27H,2012ApJ...751..119B,2012ApJ...757....6H,2022ApJ...931...11G,2022ApJ...931...10R,2023ApJ...948...41L}:

\begin{equation}\label{tidal w equation}
\begin{aligned}
&\frac{d\omega_p}{dt}|_{\mathit{tide}}=K_pn^2\frac{M_*^2}{(M_p+M_*)M_p}\frac{R_p^3}{a^3}\frac{1}{C_p}(1-e^2)^{-6}\\&\times\left[f_2(e)n\cos \epsilon_p-\frac{1}{2}(1+\cos^2 \epsilon_p)(1-e^2)^{\frac{3}{2}}f_5(e)\omega_p\right],
\end{aligned}
\end{equation}

\noindent where $\epsilon_p$ is obliquity, which evolves towards 0 over a timescale similar to the pseudo-synchronization timescale. $f_2$ and $f_5$ are given by:

\begin{equation}\label{eccentricity functions}
\begin{aligned}
&f_2(e)=1+\frac{15}{2}e^2+\frac{45}{8}e^4+\frac{5}{16}e^6,\\
&f_5(e)=1+3e^2+\frac{3}{8}e^4.
\end{aligned}
\end{equation}

Assuming that the surface of the planetesimal is strength-less, the breakup limit is:

\begin{equation}
\omega_{\mathit{break}}=\sqrt{\frac{GM_p}{R_p^3}-\frac{2GM_*}{q_0^3}},    
\end{equation}

The maximum spin-up rate due to volcanism can be approximated as:

\begin{equation}
\begin{aligned}
\frac{d\omega_{\mathit{p}}}{dt}|_{\mathit{volc}}=-\frac{\dot{M}_pu_{\mathit{eject}}}{M_pR_pC_p},    
\end{aligned}
\end{equation}

\noindent And the equilibrium spin rate (volcanism-induced spin-up rate matches the tidally induced spin-down rate) is determined by setting $\frac{d\omega_{\mathit{p}}}{dt}|_{\mathit{volc}}+\frac{d\omega_p}{dt}|_{\mathit{tide}}=0$:

\begin{equation}
\begin{aligned}
\omega_{\mathit{eq}}= \frac{2f_2(e)n\cos\epsilon_p-2\frac{\dot{M}_pu_{\mathit{eject}}}{GK_pM_*^2R_p^4}\left(\frac{2q_0Q_0}{q_0+Q_0}\right)^{6}}{(1+\cos^2\epsilon_p)(1-e^2)^{\frac{3}{2}}f_5(e)},       
\end{aligned}
\end{equation}

\noindent which for circular orbit and spin-orbit alignment is:

\begin{equation}
\omega_{\mathit{eq}}=n-\frac{\dot{M}_pu_{\mathit{eject}}a^3(M_p+M_*)}{K_pn^2M_*^2R_p^4}, 
\end{equation}

\noindent and is effectively identical to the synchronous rotation rate ($n$) for a planetesimal around WD\,1145+017 assuming a mass loss rate of $8\times 10^{9}\, \rm g/s$ \citep{2015Natur.526..546V} (volcanism-induced break-up timescale $\gtrsim 3000\,\rm yr$). One can show that this is also true for eccentric orbits. Therefore, under the CTL model, volcanism-induced break-up is rare.

\section{Energy for fracture}\label{conduit energy appendix}

To cause fracture in the solid mantle for volcanic eruption, tidal energy needs to support the strain energy corresponding to the strength of the planetesimal, leading to an additional energy loss $Q_s$:

\begin{equation}
Q_s=\frac{1}{2}\frac{V\sigma_{\mathit{-}}^2}{E},    
\end{equation}

\noindent where $V$ is the conduit volume, $E$ is the Young's modulus of the solid layer and we consider the compressive strength $\sigma_{-}$. We estimate $Q_s$ assuming $V=\frac{4}{3}\pi R_p^3(1-f_m^3)$, $\sigma_{\mathit{-}}\sim -100\,\rm MPa$, $E\sim 100\,\rm GPa$ \citep{2017SoSyR..51...64S,min9120787,2019A&A...629A.119T,2020M&PS...55..962P,TANG2023107154}. We find out that $\frac{Q_s}{c_pM_p}\sim 0.005\,\mathrm {K}\ll \Delta T_c$ and can hence be neglected. 

Furthermore, the conduit is not necessarily in the radial direction, hence inducing additional compressive stress due to gravity (hydrostatic stress) and friction (deviatoric stress) \citep{Turcotte_Schubert_2014}:

\begin{equation}\label{compressive stress}
\begin{aligned}
& \sigma_x=-\rho_pgh\left[1\pm\frac{2f_s}{(1+f_s^2)^{\frac{1}{2}}\mp f_s}\right], \\&\sigma_{y}=-\rho_pgh,
\end{aligned}
\end{equation}

\noindent where we assume $g=\frac{GM_p}{R_p^2}$, $h=(1-f_m)R_p$, $f_s\sim 0.85$ (coefficient of friction \citealp{Byerlee1978FrictionOR,Turcotte_Schubert_2014}). We find out that $\sigma=\sqrt{\sigma_x^2+\sigma_y^2} \sim \rho_pgh\lesssim 10\,\rm MPa\ll \sigma_{-}$. The corresponding strain energy is negligible compared to that corresponds to the compressive strength and can hence be neglected. 

Finally, additional thermal energy is lost to the planetesimal's interior cooler than the magma before eruption \citep{2019GeoRL..46.5055A}. For a magma refilling timescale of $t_{\mathit{refill}}$ and assuming that the magma chamber diffuses heat to its ambient from a semi-sphere with radius $\frac{1}{2}(1-f_m)R_p$, the ratio of heat lost to the ambient to the thermal energy of the magma can be estimated as:

\begin{equation}
\frac{Q_{\mathit{diff}}}{E_{\mathit{magma}}}\sim \frac{6\pi\sqrt{t_{\mathit{refill}}\alpha_t}}{(1-f_m)R_p}.
\end{equation}

Assuming $t_{\mathit{refill}}\sim 1\,\rm yr$ (constrained by WD\,1145+017 observational time span), we have $\frac{Q_{\mathit{diff}}}{E_{\mathit{magma}}}\sim 0.06$. Hence, the majority of the thermal energy is retained in the magma.

\section{Thermal properties of planetesimal}\label{thermal properties appendix}

\begin{table}
\begin{center}
\begin{tabular}{|c|c|c|c|c|c|} 
 \hline
Type & $T_c\,\rm (K)$ & $T_0\,\rm (K)$& $\left(\Delta T_c-\frac{\Delta h_f}{c_p}\right)\,\rm (K)$& $f_Q$\\
\hline
Forsterite& 2000&50& 2296&0.73\\
&2000&500& 1921&0.62\\
&1500&50&1607&0.81\\
&1500&500&1231&0.69\\
&1000&50&961&0.92\\
&1000&500&586&0.78\\
 
 \hline
Granite& 2000&50& 2009& 0.55\\
&2000&500& 1784& 0.44\\
&1500&50&1406&0.63\\
&1500&500&1180&0.49\\
&1000&50&802&0.74\\
&1000&500&577&0.58\\
 
 \hline
Granodiorite& 2000&50& 2165&0.46\\
&2000&500& 1791&0.40\\
&1500&50&1559&0.51\\
&1500&500&1185&0.44\\
&1000&50&953&0.57\\
&1000&500&580&0.49\\

 \hline
Gabbro& 2000&50& 2046&0.57\\
&2000&500& 1696&0.50\\
&1500&50&1417&0.63\\
&1500&500&1067&0.55\\
&1000&50&856&0.70\\
&1000&500&507&0.61\\

 \hline
Garnet amphibolite& 2000&100& 2388&0.56\\
&2000&500& 2011&0.50\\
&1500&50&1631&0.61\\
&1500&500&1255&0.55\\
&1000&50&943&0.67\\
&1000&500&567&0.60\\

\hline
\end{tabular}
\end{center}
\caption{The temperature representation of critical energy of melting excluding the contributions from the enthalpy of fusion, $\Delta T_c-\frac{\Delta h_f}{c_p}$, as well as the ratio of steady state heat loss rate under time dependent $\kappa (T)$ to its counterpart of fixed $\kappa$, $f_Q$, for different types of rocks, different critical temperature of melting $T_c$ and different initial temperature $T_0$. The data for forsterite is from \citealp{article,LESHER2015113}, with the remaining from \citealp{miao2014temperature}.}
\label{heat capacity table}
\end{table}

In this section, we examine the variations in the conditions for melting and volcanism using the temperature dependent thermal properties for different types of rocks.

We investigate the empirical relations of $c_p(T)$ for different types of rocks \citep{miao2014temperature}, assuming that below/above the numerical minimum/maximum of the empirical relations, $c_p$ stays constant, equal to the minimum/maximum. The empirical relations are of the form of:

\begin{equation}
c_p=a+\frac{b}{T^\frac{1}{2}}+\frac{c}{T}+\frac{d}{T^2}+\frac{e}{T^3},    
\end{equation}

\noindent with the constants $a$--$e$ dependent on the types of rocks. We consider three critical temperature of melting $T_c\sim 1000\,\rm K$, 1500\,K and 2000\,K, accounting for the values of different substances \citep{LESHER2015113,2015aste.book..533S,2023A&A...671A..74J} together with two initial temperatures, $T_0\sim 50\,\rm K$ and $500\,\rm K$, corresponding to the equilibrium temperature of a planetesimal on a highly eccentric orbit when the host star enters its white dwarf phase, and with a cooling age of $\sim 100\,\rm Myr$, respectively. The corresponding $\Delta T_c$ excluding the contributions of the heat of fusion is shown in Table \ref{heat capacity table}. The range of the enthalpy of fusion, is $\sim 300\,\rm kJ/kg$--1000\,kJ/kg \citep{1985E&PSL..73..407F,LESHER2015113}. In this case, $\Delta T_c$ ranges from $\sim 800\,\rm K$ to $\sim 3000\,\rm K$. Compared to the results obtained using the fiducial values, melting is possible with a lower tidal evolution rate. For instance, $q_{\mathit{0,crit}}$ (Eq.11 in the main text) may maximally increase by $\sim 20\%$.

We consider the empirical relation in \citealp{miao2014temperature} for $\kappa(T)$:

\begin{equation}
\kappa=\frac{1}{a+bT},    
\end{equation}

\noindent with $a$ and $b$ compositional dependent constants. The corresponding heat loss rate is given by Eq.\ref{heat loss T dependent}.

With other parameters remain identical, the ratio of heat loss rate under time dependent $\kappa (T)$ to its counterpart of fixed $\kappa$, $f_Q$ (the ratio of Eq.\ref{heat loss T dependent} to Eq.\ref{heat loss equation appendix}) is:

\begin{equation}
f_Q=\frac{\ln{\frac{\kappa(T_s)}{\kappa(T_c)}}}{b\kappa(T_c-T_s)},    
\end{equation}

As is shown in Table \ref{heat capacity table}, we find out that $f_Q\lesssim 1$. Melt can hence be maintained at a smaller tidal power than the case of constant $\kappa$. For instance, $q_{\mathit{0,crit}}'$ (Eq.15 in the main text) may maximally increase by $\sim 10\%$.

\section{Surface heating of the planetesimal}\label{surface heating appendix}

We investigate the case where the sum of tidal power and stellar irradiation balances the radiative cooling:

\begin{equation}
\begin{aligned}
4\pi R_p^2\sigma <T_s^4>=\dot{E}_t+\frac{R_p^2L_*}{4a^2\beta_r\sqrt{1-e^2}}    
\end{aligned} 
\end{equation}

\noindent where the factor $\frac{1}{\sqrt{1-e^2}}$ originates from temporal average of stellar irradiation and $<T_s^4>^{\frac{1}{4}}\neq <T_s>$ (Eq.13 in the main text) is an radiative power related time-averaged equilibrium temperature and $\beta_r$ accounts for the fraction of re-radiation (1 for a fast rotator and 0.5 for a tidally-locked body).

Numerical simulations show that the maximum $T_s$ is usually dominated by either tidal power or stellar flux, with cases with comparable contributions from both simultaneously being rare (because of their distinct dependence on orbital eccentricity). Hence, we may consider tidal heating and stellar irradiation dominated cases separately. In the cases where tidal power dominates, the maximum surface temperature of the planetesimal is reached at the maximum of tidal power, $e\approx 0.735$:

\begin{equation}
\begin{aligned}
(<T_s^4>)^{\frac{1}{4}}_{\mathit{max,tide}}&\sim 800\,\mathrm{K} \left(\frac{R_p}{100\,\mathrm{km}}\right)^{\frac{3}{4}} \left(\frac{M_*}{0.6\,M_{\odot}}\right)^{\frac{3}{4}}\left(\frac{0.005\,\mathrm{AU}}{q_0}\right)^{\frac{9}{4}},    
\end{aligned}
\end{equation}

\noindent which, assuming a planetesimal with a radius of 100\,km, density and tensile strength being the average of the measured Solar System ordinary chondrites, $\rho_p=3500\,\rm kg/m^3$, $\sigma_{+}=25\,\rm MPa$ \citep{2020M&PS...55..962P}, can only exceed 1400\,K (solidus temperature of silicates \citealp{2015aste.book..533S,2023A&A...671A..74J}) if the planetesimal is perturbed near/within its Roche limit ($q_0\lesssim 0.004\,\rm AU\approx r_{\mathit{Roche}}$). 

Meanwhile, the irradiation dominated surface temperature given by:

\begin{equation}
\begin{aligned}
(<T_s^4>)^{\frac{1}{4}}_{\mathit{irr}}&\sim 900\,\mathrm{K}  \left(\frac{L_*}{0.01\,L_{\odot}}\right)^{\frac{1}{4}}\left(\frac{0.005\,\mathrm{AU}}{q_0}\right)^{\frac{1}{2}}\frac{(1-e^2)^{\frac{3}{8}}}{\beta_r^{\frac{1}{4}}},       
\end{aligned} 
\end{equation}

\noindent where we assume that $q_0\ll Q_0$. Surface melting/sublimation usually occurs in the rapid circularization case where the planetesimal is tidally locked on a close-in near circular orbit with the white dwarf remaining hot. In this case, for $T_s$ to exceed 1400\,K, we require (using Eq.14 in the main text):

\begin{equation}
\left(\frac{M_*}{0.6\,M_{\odot}}\right)^{\frac{1}{4}}\left(\frac{0.005\,\mathrm{AU}}{q_0}\right)^{\frac{1}{2}}\left(0.1+\frac{t_{\mathit{cooling-age}}}{\mathrm{Myr}}\right)^{-0.295}\gtrsim 0.3. 
\end{equation}

Taking $M_*=0.6\,\rm M_{\odot}$ and using the fact that white dwarf with observed transits all have cooling ages above $100\,\rm Myr$, we obtain $q_0\lesssim 0.004\,\rm AU$.

\section{Orbit dispersion under additional perturbations}\label{radiation pressure appendix}

\begin{figure}
\begin{center}
\includegraphics[width=0.95\textwidth]{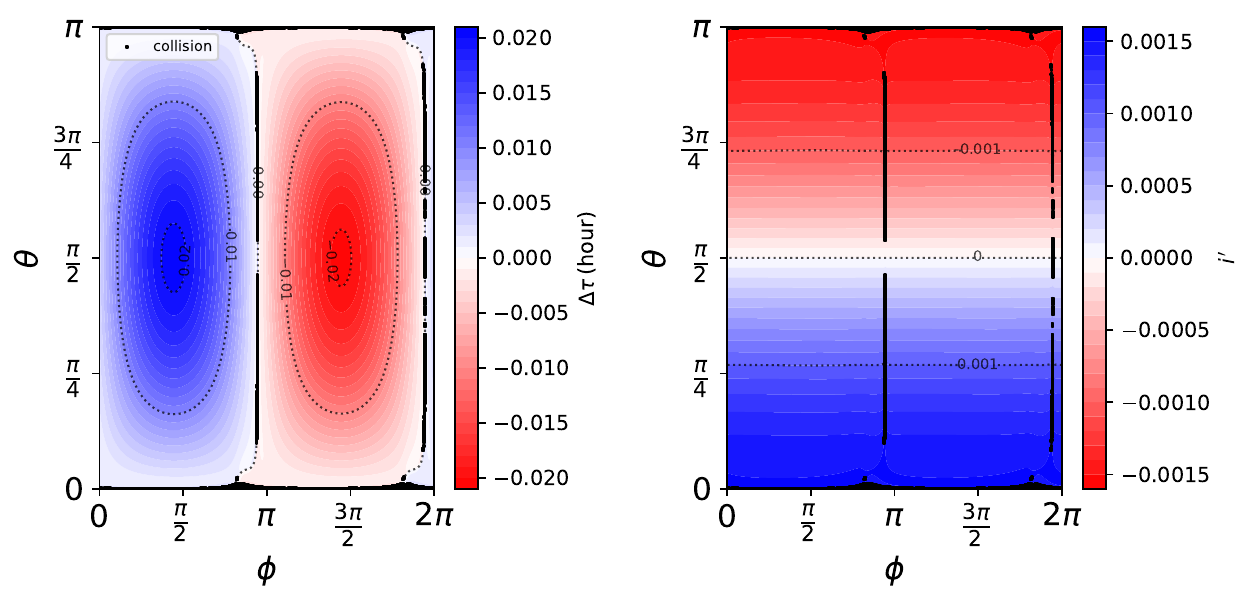}
\caption{Same plot as Fig.4 in the main text, but simulated using Rebound \citep{2012A&A...537A.128R} for an integration time of one orbital period, including the gravitational force from the planetesimal and general relativistic effect. The black pixels represent where collisions between the ejecta and the planetesimal occur.}
\label{period dispersion rebound}    
\end{center}

\end{figure}

\begin{figure}
\begin{center}
 \includegraphics[width=0.95\textwidth]{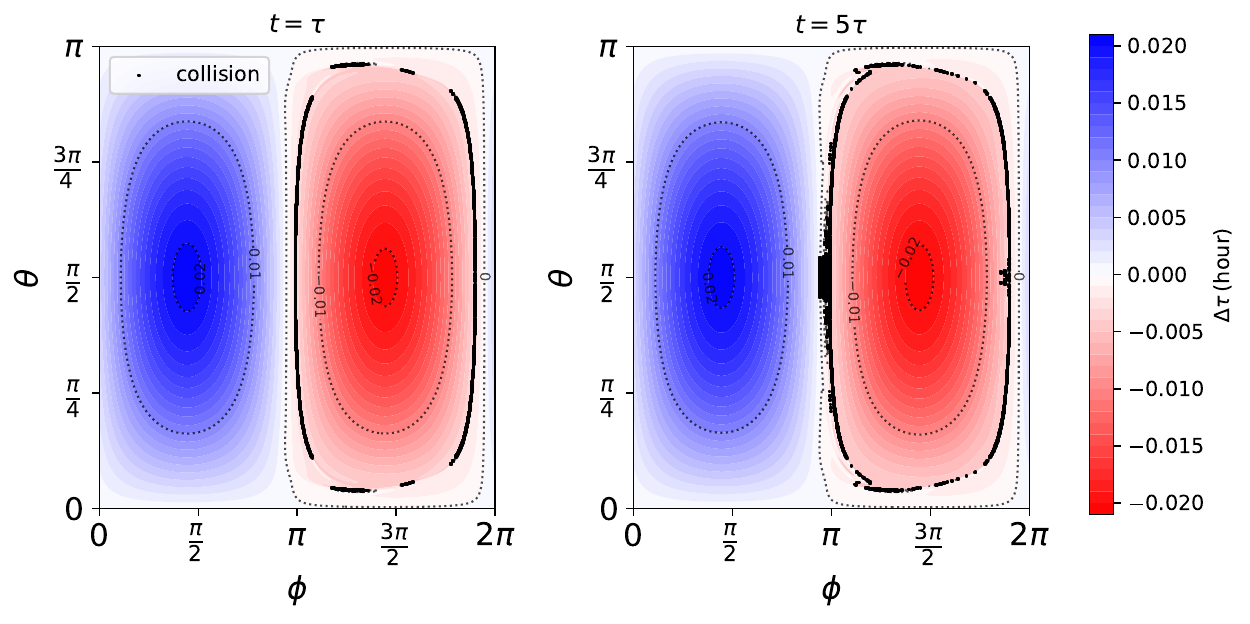}
\caption{Same plot as the left panel of Fig.4 in the main text (period dispersion), but simulated using Rebound \citep{2012A&A...537A.128R} for an integration time of 1 (left panel) and 5 (right panel) orbital periods, including the gravitational force from the planetesimal, general relativistic effect and radiation perturbations. The ejected dust is assumed to have a radius of $10^{-5}\,\rm m$ and a density of $3000\,\rm kg/m^3$. The black pixels represent where collisions between the ejecta and the planetesimal occur.}
\label{period dispersion radiation rebound}   
\end{center}

\end{figure}

We investigate the effects of the planetesimal on the ejecta using Rebound \citep{2012A&A...537A.128R}. For identical parameters as Fig.4 in the main text, we plot the predicted orbital period dispersion and inclination accounting for the planetesimal after 1 orbital period in Fig.\ref{period dispersion rebound}. The black pixels indicate the collisions of the ejecta with the planetesimal. In general, the simulations show consistency with the analytical results, both qualitatively (the dependence on $\theta$ and $\phi$) and qualitatively (fractional error $\lesssim 5\%$ for $u_{\mathit{eject}}=500\,\rm m/s$, which decreases with the increase in $u_{\mathit{eject}}$). The exceptions occur around the regions where $\Delta \tau\rightarrow 0$ under the analytical model, where the ejecta remains close to the planetesimal, with their orbits strongly torqued by the planetesimal and a higher risk of collision with the planetesimal. The analytical model remains a reasonable approximation in the majority of the parameter space, for instance, when considering the maximum orbital period dispersion.

We further include the radiation force. Analytically, the inclusion of radiation pressure is equivalent to a reduced mass of the host star ($(1-\beta)M_*$ with $\beta$ the ratio of radiation to gravitational force), with the semi-major axis perturbation for a near circular orbit given by:

\begin{equation}
\begin{aligned}
\frac{\Delta a}{a} &\approx 2\frac{R_p}{a}\sin\theta\cos\phi +\frac{\beta}{1-\beta}+2\frac{\omega_p R_p}{v_{\mathit{orb}}(1-\beta)}\sin\theta \cos\phi\\&+2\frac{u_{\mathit{eject}}}{v_{\mathit{orb}}(1-\beta)}\sin\theta \sin\phi,
\end{aligned}    
\end{equation}

\noindent where the term $\frac{\beta}{1-\beta}$ corresponds to the expanding semi-major axis under radiation pressure, and the contributions of spin and ejection velocities to $\frac{\Delta a}{a}$ are enlarged by a factor of $\frac{1}{1-\beta}$.

We simulate the orbital period dispersion ($\Delta \tau$) including both PR drag and radiation pressure, as is shown in Fig.\ref{period dispersion radiation rebound}, where we plot the resultant period dispersion after 1 (left panel) and 5 (right panel) orbital periods together with the parameter space where collisions occur (black pixels). In general, by comparing Fig.\ref{period dispersion radiation rebound} to Fig.\ref{period dispersion rebound}/Fig.4 in the main text the inclusion of radiation perturbations does not alter the orbit dispersion qualitatively shortly after ejection. Two major effects of radiation force are: 1. the parameter space of collisions and the behaviours of the ejecta with small $\Delta \tau$ are altered and 2. the ejecta is drifted towards shorter periods with time.

\section{Maximum dust production rate on a near circular orbit}\label{mass loss rate appendix}

\begin{figure}
\begin{center}
\includegraphics[width=0.8\textwidth]{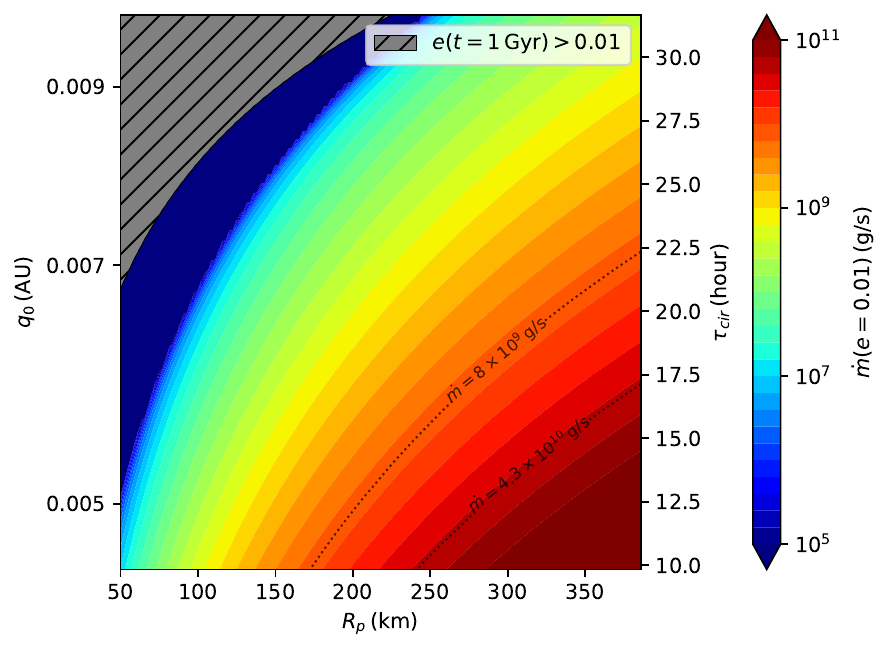}
\caption{The numerically computed maximum mass loss rate allowed by tidal heating for different-sized planetesimals starting at different pericentre distances when the planetesimal reaches $e=0.01$ (other free parameters identical to those in Table 1 in the main text, Roche limit 0.00456\,AU) for a volcanism ejection speed of 500\,m/s. The contour lines are the predicted mass loss rate to explain the transit around WD\,1145+017 \citep{2015Natur.526..546V} and the accretion rate \citep{2016ApJ...816L..22X}.}
\label{mass loss e plot}    
\end{center}

\end{figure}

We compute the mass loss rate for a sample planetesimal under our fiducial parameters in Table 1 in the main text at $e=0.01$ in Fig.\ref{mass loss e plot}, assuming a fixed volcanic ejection speed of 500\,m/s (excluding the planetesimal with a radius larger than $\sim 400\,\rm km$ where the escape speed exceeds 500\,m/s). To reach the mass loss rate of $8\times10^{9}\,\rm g/s$ ($4.3\times10^{10}\,\rm g/s$), the sample planetesimal should be larger than $\sim 170\,\rm km$ ($260\,\rm km$) and starts at a pericentre distance below $0.007\,\rm AU$/orbital period after circularization $\tau_{\mathit{cir}}\lesssim 20\,\rm hr$ ($q_0\lesssim 0.006\,\rm AU$, $\tau_{\mathit{cir}}\lesssim 17\,\rm hr$). Hence, rocky planetesimals on short-period near circular orbits are able to produce the mass loss rate at the same level as that required to explain the transits around WD\,1145+017 provided that they are large and perturbed close enough to the white dwarf. These rocky planetesimal, potentially associated with large mass loss rate and deep transit depth, should circularize to an orbital period $\lesssim 20\,\rm hr$.

\section{Orbital evolution of the ejecta}\label{evolution of ejecta appendix}

\begin{figure}
\begin{center}
\includegraphics[width=0.95\textwidth]{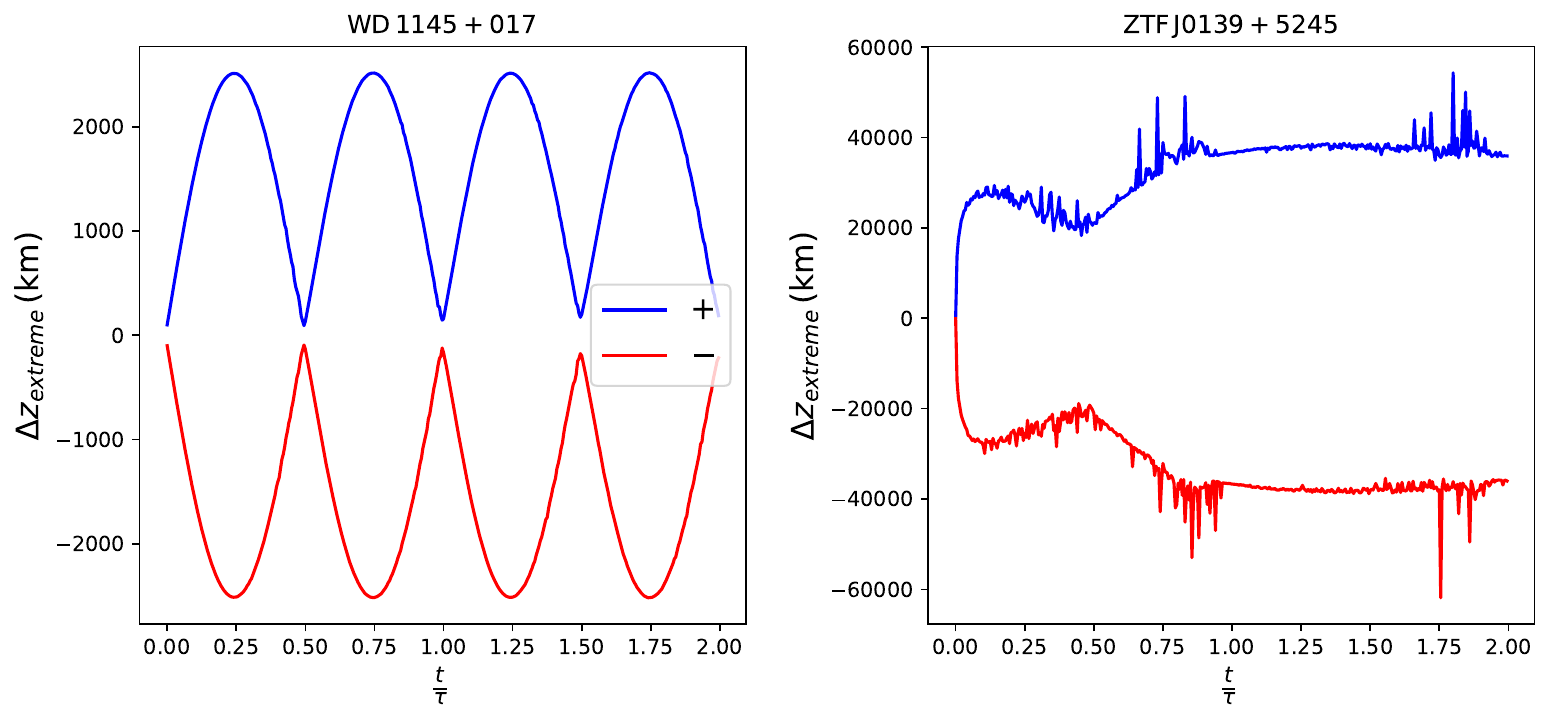}
\caption{The maximum z-span of volcanism ejecta for two systems, WD\,1145+017 ($0.6\,M_{\odot}$, $\rho_p=6000\,\rm kg/m^3$, $\tau=4.5\,\rm hr$, $e=0.01$) and ZTF\,J0139+5245 ($0.52\,M_{\odot}$, $\rho_p=3000\,\rm kg/m^3$, $\tau=107.2\,\rm day$, $e=0.98$) at two different true anomalies ($f=\frac{\pi}{2}$ and $f=\pi$). Other free parameters are identical to those in Table 1 in the main text.}
\label{z max plot}    
\end{center}

\end{figure}

\begin{figure}
\begin{center}
\includegraphics[width=0.95\textwidth]{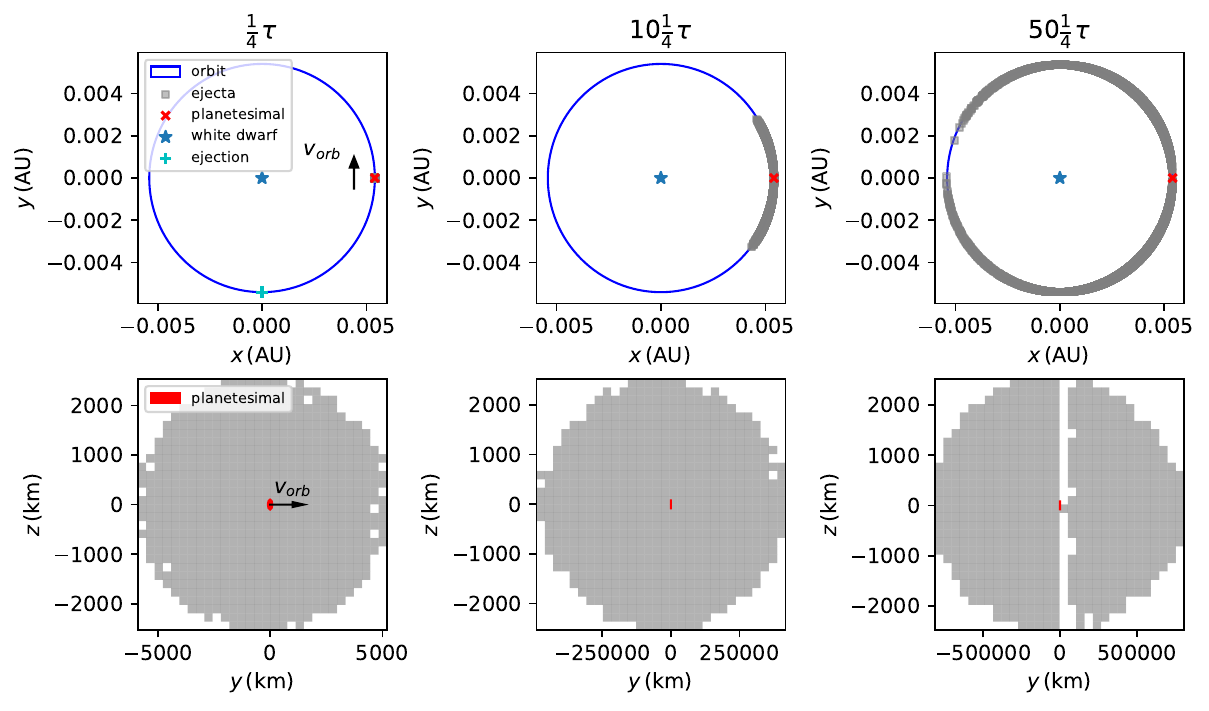}
\caption{Rebound simulation of the positions of 10000 ejected particles (not colliding with the planetesimal) with size $10^{-5}\,\rm m$ and density of $3000\,\rm kg/m^3$ in the orbital plane (upper panels) and along the x-axis (lower panels) when the planetesimal is at $y=0$. See the text for other settings.}
\label{volc 2d density plot}    
\end{center}

\end{figure}

We simulate the position of volcanic ejecta for 10000 particles erupted at the same time that avoid collision with the planetesimal. We focus on the span of the ejecta. The direction of the ejection velocity is assumed to follow a normal distribution with the mean at $\theta$ and $\phi$ and a standard deviation of $\frac{\pi}{10}$. 

The maximum z span ($\Delta z_{\mathit{extreme}}$) of the volcanism ejecta which corresponds to the maximum transit depth is shown in Fig.\ref{z max plot} for two systems, planetesimal on a near circular orbit around WD\,1145+017 and planetesimal on a highly eccentric orbit around ZTF\,J0139+5245. $\Delta z_{\mathit{extreme}}$ for WD\,1145+017 is almost periodic for single ejection, with $\Delta z_{\mathit{extreme}}$ maximized at $\frac{(2n+1)}{4}\tau$. For a highly eccentric orbit, $\Delta z_{\mathit{extreme}}$ is dependent on the true anomaly and can be orders of magnitude larger than its counterpart for a near circular orbit, corresponding to the possibility of a much larger transit depth (but note a higher probability of optically thin ejecta). 

In Fig.\ref{volc 2d density plot}, we plot the orbital evolution of the ejecta's span for the planetesimal around WD\,1145+017. We assume that the volcanism ejection occurs at $x=0$ (cyan plus) while we always observe the ejecta at $y=0$ (to illustrate the maximum span along the z axis). Due to the orbital period dispersion, the leading (ejecta with shorter orbital period) and trailing tails (ejecta with longer orbital period) diffuse away from the planetesimal with time. The leading tail will catch up the trailing tail at $\sim 50\tau$.

\section{Tidal and PR escape timescale}\label{escape timescale appendix}

One can show that for a planetesimal/disk starting on an extremely eccentric orbit, when the apocentre decays by one Hill radius ($r_H$) of the planet (defined as the condition for escape), the planetesimal/disk remains on a highly eccentric orbit with little circularization. Hence, the tidal escaping timescale is:

\begin{equation}
\tau_{\mathit{e,tide}}\approx 5\frac{r_HM_pq_0^{7.5}}{GK_pM_*^2R_p^5Q_0^{0.5}},    
\end{equation}

The PR escaping timescale is:

\begin{equation}
\begin{aligned}
\tau_{\mathit{e,PR}}\approx 7\frac{ r_Hc^2 D\rho q_0^{1.5}}{L_*Q_0^{0.5}}.
\end{aligned}
\end{equation}

\section{General relativistic precession} \label{GR precession appendix}

\begin{figure}
\begin{center}
\includegraphics[width=0.8\textwidth]{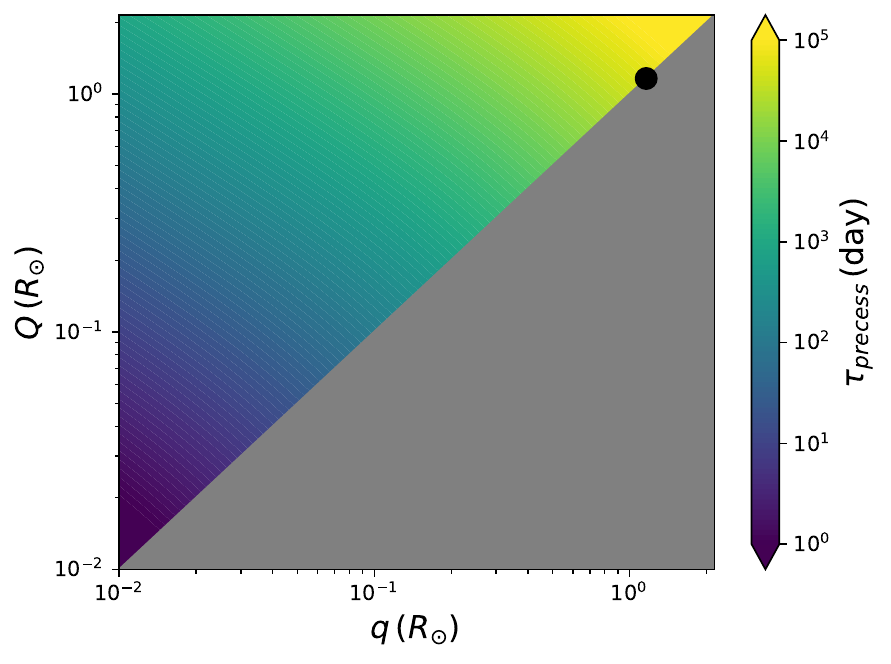}
\caption{The precession period as a function of pericentre distance and apocentre distance. The black dot is the position of the transiting body on a 4.5\,hr near circular orbit around WD\,1145+017.}
\label{precession plot}    
\end{center}

\end{figure}

The GR precession of a disk may mimic periodic signals. The period of this signal is given by the precession period \citep{2008MNRAS.389..191P,2022Symm...15...39Y,2024ApJ...962...77C}:

\begin{equation}\label{precession approximate equation}
\begin{aligned}
\tau_{\mathit{precess}}=\frac{2^{\frac{1}{2}}\pi c^2qQ(q+Q)^{\frac{1}{2}}}{3(1-\beta)(GM_*)^{\frac{3}{2}}},
\end{aligned}
\end{equation}

\noindent where the factor $1-\beta$ accounts for radiation pressure \citep{1950ApJ...111..134W}. We estimate $\tau_{\mathit{precess}}$ (with $\beta\sim 0$) in $q$--$Q$ space, which is shown in Fig.\ref{precession plot}. The precession period of the transiting body on a 4.5\,hr near circular orbit around WD\,1145+017, $\sim 150\,\rm yr$ is represented by the black dot. 

\section{Collisions among tidally evolved planetesimals}\label{collision appendix}

\begin{figure}
\begin{center}
 \includegraphics[width=0.8\textwidth]{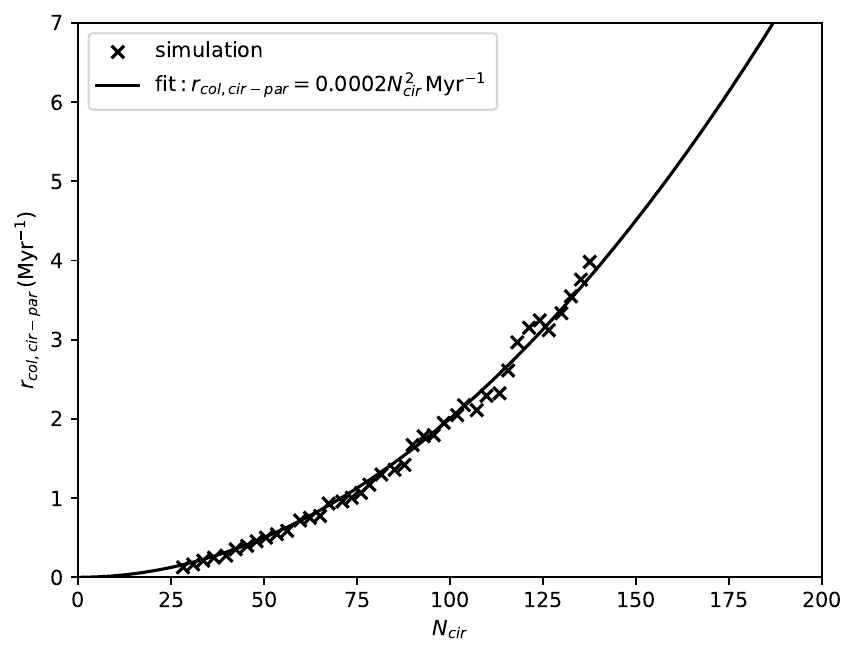}
\caption{The predicted collision rate between nearly circularized planetesimals $e\leq 0.01$ with their partially circularized counterpart as a function of the number of circularized planetesimals based on the same orbit distribution used for Fig.11 of \citealp{10.1093/mnras/staf182}. We assume a truncated planetesimal size distribution of $P(R_p)\propto R_p^{-4}$ from $50\,\rm km$ to $1000\,\rm km$, together with a uniform size distribution between $3000\,\rm kg/m^3$ to $6000\,\rm kg/m^3$.}
\label{collision rate cir par}   
\end{center}

\end{figure}

For a spherical shell of tidally circularized planetesimals, the collision rate between the circularized planetesimals (which are assumed to not collide with each other) and partially circularized planetesimals $r_{\mathit{collision}}$ can be estimated as:

\begin{equation}
\begin{aligned}
r_{\mathit{collision}}\sim n_{\mathit{cir}}\sum_i w_i P_{i} \sigma_iv_{\mathit{i}},
\end{aligned}
\end{equation}

\noindent where $n_{\mathit{cir}}$ is the number density of (nearly) circularized planetesimals (with $e\leq 0.01$), $\sigma_i$ is the collisional cross-section of partially circularized planetesimal $i$ with the circularized planetesimals:

\begin{equation}\label{sigma equation}
\sigma_i=\pi(R_i+R_{\mathit{cir}})^2\left(1+\frac{v_{\infty}^2}{v_{i}^2}\right),  
\end{equation}

\noindent with the term $\frac{v_{\infty}^2}{v_{i}^2}$ accounting for the gravitational focusing effect. $v_{\infty}$ is given by:

\begin{equation}
v_{\infty}=\sqrt{\frac{2G(M_i+M_{\mathit{cir}})}{R_i+R_{\mathit{cir}}}}.    
\end{equation}

\noindent $v_{i}$ is the encounter speed of the partially circularized planetesimal $i$ with the circularized planetesimal:

\begin{equation}
v_{i}=\left|v_{\mathit{orb}}(r)\begin{bmatrix}\cos\gamma\\\sin\gamma\\0\end{bmatrix}-{v_k}(r)\begin{bmatrix}0\\\cos\theta'\\\sin\theta'\end{bmatrix}\right|,
\end{equation}

\noindent where $v_{\mathit{orb}}(r)=\sqrt{G(M_*+M_{i})(\frac{2}{r}-\frac{1}{a_i})}$ is the speed of the partially circularized planetesimal $i$ during orbital crossing with a circularized planetesimal with keplerian speed $v_k=\sqrt{\frac{G(M_*+M_p)}{r}}$. $\theta'$ ($0\leq \theta'\leq \pi$) is the angle between the two orbital planes (the angle between the orbital angular momentum of the partially circularized planetesimal $i$ and the circularized planetesimal it collides with). $\gamma$ is the angle between the radial vector and velocity vector of the partially circularized planetesimal. $P_{\mathit{i}}$ accounts for the fact that not all circular orbits within the spherical shell will intersect the orbit of the partially circularized planetesimal, leading to a reduced $n_{\mathit{cir}}$. Consider a random point on a circle with radius $r'$ and a random point on a sphere with radius $r$. We define the coordinate system such that they are described by $(r'\cos\phi,r'\sin\phi,0)$ and $(r\sin\theta,0,r\cos\theta)$, respectively. The distance squared between these two points is given by:

\begin{equation}
d^2=r^2+r'^2-2rr'\sin\theta \cos\phi.   
\end{equation}

And hence:

\begin{equation}
\theta=\sin^{-1}\left(\frac{r^2+r'^2-d^2}{2rr'\cos\phi}\right),    
\end{equation}

\noindent where $\theta$ only has solution if $\frac{r^2+r'^2-d^2}{2r'r}\leq 1$. As $d\ll r'$ and $d\ll r$, this condition is only fulfilled as $r'\rightarrow r$, where $\theta\rightarrow \frac{\pi}{2}$. Then, we aim to minimize $\theta$ for $0<\theta<\frac{\pi}{2}$, which is reached at the maximum of $d$, $d=R_i+R_{\mathit{cir}}$ (the gravitational focusing term is negligible due to the high encounter speed), the maximum of $\cos\phi$ ($\cos\phi=1$, $\phi=0$), and $r'=\sqrt{r^2-d^2}$. And hence:

\begin{equation}
\theta\approx \sin^{-1}\left(1-\frac{d^2}{2r^2}\right)\approx \frac{\pi}{2}-\frac{d}{r}  
\end{equation}

For a random $\theta$ between 0 and $\frac{\pi}{2}$, the probability of lying in an interval between $\frac{\pi}{2}-\frac{d}{r}$ and $\frac{\pi}{2}$ is:

\begin{equation}
P_{\mathit{i}}\sim\frac{2\left(R_i+R_{\mathit{cir}}\right)}{\pi r},    
\end{equation}

\noindent which leads to a significant reduction in the effective number density of circularized planetesimals. $w_i$ is the effective number of the partially circularized planetesimal $i$, describing the fraction of time that the partially circularized planetesimal spends in the spherical shell containing the circularized planetesimals. $w$ is given by:

\begin{equation}\label{w equation}
 \begin{aligned}
 w&=\frac{2}{\tau}\int_{\mathrm{max}(r_{\mathit{min}},q)}^{\mathrm{min}(r_{\mathit{max}},Q)}\frac{dt}{dr}dr=\frac{(1-e^2)^{\frac{1}{2}}}{\pi ae}\int_{\mathrm{max}(r_{\mathit{min}},q)}^{\mathrm{min}(r_{\mathit{max}},Q)}\frac{dr}{\sin f},
 \end{aligned}   
\end{equation}

\noindent with $\sin f$ being $\sqrt{1-\left[\frac{a(1-e^2)}{re}-\frac{1}{e}\right]^2}$, the factor of 2 representing entering and exiting the shell. For a narrow shell, Eq.\ref{w equation} reduces to:

\begin{equation}\label{w dr equation}
\begin{aligned}
w(r|q<r<Q)&=\frac{2}{\tau}\frac{\delta r}{v_r}=\frac{(1-e^2)^{\frac{1}{2}}\delta r}{\pi a e \sin f}= \frac{\delta r r}{\pi a \sqrt{(r-q)(Q-r)}},  
\end{aligned}
\end{equation}

\noindent where $v_r=\frac{dr}{dt}=v_{\mathit{orb}}\cos\gamma$.

If the probability ratio of being on a (nearly) circularized orbit to that of being on a partially circularized orbit is fixed, we have $\sum_i\propto n_{\mathit{cir}}\propto N\propto N_{\mathit{cir}}$ and $r_{\mathit{collision}}\propto N^2\propto N_{\mathit{cir}}^2$. Hence, as is verified in Fig.\ref{collision rate cir par}, the collision rate between partially circularized planetesimals and circularized planetesimals can be fitted to the relation (assuming a fixed orbit distribution):

\begin{equation}\label{collision rate equation}
r_{\mathit{collision}}=AN_{\mathit{cir}}^2\,\rm Myr^{-1},  
\end{equation}

\noindent with $A$ a fitting constant. Assuming all collisions are catastrophic (see below for a justification of this assumption), a constant rate of replenishment for circularized planetesimals $r_{\mathit{rep}}$, and neglecting the potential variations of Eq.\ref{collision rate equation} due to collisions, the number of circularized planetesimals can be approximated as:

\begin{equation}
\begin{aligned}
&\frac{dN_{\mathit{cir}}}{dt}=r_{\mathit{rep}}-\frac{AN_{\mathit{cir}}^2}{\mathrm{Myr}},\\
&N_{\mathit{cir}}(t)=\sqrt{\frac{r_{\mathit{rep}}}{A}}\tanh\left(\sqrt{Ar_{\mathit{rep}}}\frac{t}{\mathrm{Myr}}\right),     
\end{aligned}
\end{equation}

\noindent which converges to $\sqrt{\frac{r_{\mathit{rep}}}{2A}}$ at $t\gtrsim \frac{1}{\sqrt{2Ar_{\mathit{rep}}}}$. We estimate $A$ numerically, finding out that $A\sim 0.0002\,\rm Myr^{-1}$ (Fig.\ref{collision rate cir par}). Assuming a mass reservoir for potentially tidally circularized planetesimals each with $R_p=100\,\rm km$ and $\rho_p=3500 \,\rm kg/m^3$ comparable to the mass of Solar System main asteroid belt, the number of potentially circularized planetesimals is approximately 100. Assuming that these circularized planetesimals arrive at a constant rate within 1\,Gyr such that $r_{\mathit{rep}}=0.1\,\rm Myr^{-1}$, the number of circularized planetesimals converge to 15 at $t \gtrsim 150\,\rm Myr$, where 1 collision occurs every 20\,Myr on average, indicating a low probability of observing collisions.

Similarly, one can estimate the collision rate among partially circularized planetesimals by accounting for orbital crossing probability at $r$ ($P\sim \frac{\Sigma \sigma}{4\pi r^2}$) as:

\begin{equation}
r_{\mathit{collision}}\sim \sum_{i} \int _{q_i}^{Q_i}n_i\sigma_{\mathit{i}}^2v_{\mathit{i}}\sum _{i\neq j} n_j dr,
\end{equation}

\noindent where the effective number density $n$ is 0 for $r\leq q$ or $r\geq Q$ and otherwise given by:

\begin{equation}
n(r)=\frac{w(r)}{4\pi r^2dr}=\frac{1}{4\pi^2 r a \sqrt{(r-q)(Q-r)}}. 
\end{equation}

We find out that the collision rate among partially circularized planetesimals is much smaller than their counterpart between circularized and partially circularized planetesimals. This is expected as the partially circularized planetesimals are much more dispersed in space and orbital crossings have stricter constraint on the orientations of orbits.

Collisional outcomes are investigated in \citealp{2009ApJ...691L.133S,2012ApJ...745...79L,2017ApJ...844..116K,2017A&A...605A...7L}. We summarize the scenarios relevant to this study below:

\begin{itemize}
    \item dust production (mass ejection),
    \item merge/accretion
    \item triggering of volcanism (weakening stress, providing heat),
    \item remove close-in planetesimals (catastrophic collisions).
\end{itemize}

We apply the model and the free parameters in \citealp{2017ApJ...844..116K} and \citealp{2017A&A...605A...7L}, where the specific impact energy $Q_c$ and the specific binding energy $Q_B$ are given by:

\begin{equation}
Q_{\mathit{C}}=\frac{1}{2}\frac{M_iM_j}{(M_i+M_j)^2}v_{\mathit{ij}}^2,    
\end{equation}

\begin{equation}
\begin{aligned}
Q_{\mathit{B}}&=500\,\mathrm{J/kg}\left[\left(\frac{s_{\mathit{ij}}}{\mathrm{m}}\right)^{-0.37}+\left(\frac{s_{\mathit{ij}}}{\mathrm{km}}\right)^{1.38}\right]\left(\frac{v_{\mathit{ij}}}{3\,\mathrm{km/s}}\right)^{0.5}\\&+\frac{3G(M_i+M_j)}{5s_{\mathit{ij}}},
\end{aligned}
\end{equation}

\noindent where $s_{\mathit{ij}}=\left(R_i^3+R_j^3\right)^{\frac{1}{3}}$. The mass escaping from the colliding planetesimals can be estimated via:

\begin{equation}
M_{\mathit{esc}}= \frac{1}{2}\left(M_i+M_j\right)\frac{Q_C}{Q_B}.   
\end{equation}

Catastrophic collisions are defined as $\frac{Q_C}{Q_B}\geq 1$ \citep{2012ApJ...745...79L}. We compute $\frac{Q_C}{Q_B}$ numerically, finding out that collisions among partially circularized and circularized planetesimals are predominantly catastrophic. The dominance of catastrophic outcomes is expected due to the high local keplerian speed of circularized planetesimals ($\sim 100\,\rm km/s$) and hence the impact speed/energy. Non-catastrophic collisions require that the partially circularized planetesimals to be small ($R_p\lesssim 10\,\rm km$ for a target circularized planetesimal with radius $R_p\sim 100\,\rm km$), whose orbital evolution is usually dominated by gravitational instability instead of tidal evolution.

\section{Dynamical tide}\label{dynamical tide}

Under the assumption that dynamical tidal theory applied to gas giants/stars equivalently applies to rocky planetesimals, for the dominant mode of (non-chaotic) dynamical tide, the quadrupole fundamental mode ($l=2$ f-mode), the orbital energy damped in forced oscillations excited during each pericentre cross can be approximated as \citep{1977ApJ...213..183P,2018ApJ...854...44M,2022ApJ...931...10R,2022ApJ...931...11G}:

\begin{equation}
\begin{aligned}
\Delta E&= f_{\mathit{dyn}}\frac{G(M_p+M_*)M_*^2}{M_p}R_p^{8}q^{-9}, 
\end{aligned}
\end{equation}

\noindent where $f_{\mathit{dyn}}$ is a constant quantifying the dissipation efficiency (unconstrained despite using the assumption that tidal dissipation under dynamical tide is usually more efficient than its counterpart under equilibrium tide). The corresponding tidal power is:

\begin{equation}\label{dynamical tide E equation}
\begin{aligned}
\frac{dE}{dt}&=\frac{\Delta E}{\tau}=\frac{n}{2\pi}\Delta E. 
\end{aligned}
\end{equation}

Expanding Eq.\ref{dynamical tide E equation} at highly eccentric orbit, where dynamical tidal evolution is usually relevant (if triggered) we have:

\begin{equation}\label{tidal power high e dynamical}
\frac{dE_t}{dt}(e\rightarrow 1) \approx 0.45f_{\mathit{dyn}}G^{1.5}M_*^{3.5}M_p^{-1}R_p^{8} q_0^{-9}Q_0^{-1.5}, 
\end{equation}

\noindent which has a steeper dependence on $q_0$ compared to Eq.\ref{tidal power high e}.

The corresponding semi-major axis evolution is:

\begin{equation}\label{dynamical tide a equation}
\begin{aligned}
\frac{da}{dt}&=\frac{a}{E}\frac{dE}{dt}=-\frac{na^2}{\pi}\frac{\Delta E}{GM_pM_*}, 
\end{aligned}
\end{equation}

\noindent with the eccentricity evolution constrained by conservation of orbital angular momentum ($\frac{de}{dt}=\frac{(1-e^2)}{2ae}\frac{da}{dt}$).

We then consider the point where the transition from dynamical to equilibrium tidal evolution occurs. \citealp{2022ApJ...931...10R} and \citealp{2022ApJ...931...11G}, despite using the criteria that the transition between dynamical and equilibrium tide happens at the eccentricity where the ratio $\frac{\dot{a}_{\mathit{dynamical}}}{\dot{a}_{\mathit{equilibrium}}} \equiv \beta=1$, also introduce an artificial cut-off at $e_{\mathit{cut}}$ at the minimum of $\beta$ to avoid the divergence of dynamical tidal evolution towards circularization, with $\beta$ given by:

\begin{equation}
\begin{aligned}
&\beta= \frac{R_p^3}{7\pi GM_p}\frac{f_{\mathit{dyn}}}{K_p}n\frac{(1-e^2)^{\frac{15}{2}}}{e^2(1-e)^9F(e)},\\&F(e)=\frac{1+\frac{45}{14}e^2+8e^4+\frac{685}{224}e^6+\frac{255}{448}e^8+\frac{25}{1792}e^{10}}{1+3e^2+\frac{3}{8}e^4}, 
\end{aligned}
\end{equation}

\noindent where we assume pseudo-synchronization and spin-orbit alignment. 

$\beta$ can be further split into physical property part and orbital parameter part as:

\begin{equation}\label{beta equation}
\begin{aligned}
&\beta=\Tilde{T}_p\Tilde{F}(e|q_0,Q_0), \\&\Tilde{T}_p=\frac{R_p^3\sqrt{GM_*}}{7\pi G M_p}\frac{f_{\mathit{dyn}}}{K_p},\\& \Tilde{F}(e|q_0,Q_0)=\left(\frac{2q_0Q_0}{q_0+Q_0}\right)^{-\frac{3}{2}}\frac{(1+e)^{9}}{e^2F(e)},
\end{aligned}
\end{equation}

\noindent where the minimum of $\Tilde{F}(e|q_0,Q_0)$ is at $e\approx0.34$ (note that this is a mathematical cut-off instead of a physical cut-off; see \citealp{2018ApJ...854...44M} for a different cutoff $e_{\mathit{cut}}\sim0.8$).

\bibliography{appendix}